\documentclass[twocolumn]{aastex63}

\newcommand\csh{University of Bern, Center for Space and Habitability, Gesellschaftsstrasse 6, 3012 Bern, Switzerland }
\newcommand\obs{ASD IMCCE, CNRS-UMR 8028, Observatoire de Paris, PSL, UPMC, 77 avenue Denfert-Rochereau, 75014 Paris, France}
\newcommand\geo{University of Bern, Institute of Geology, Baltzerstrasse 1+3, 3012 Bern, Switzerland }
\newcommand\zur{University of Z\"{u}rich, Institute for Computational Science, Winterthurerstrasse 190, 8057 Z\"{u}rich, Switzerland }
\newcommand\war{University of   Warwick, Department of Physics, Astronomy and Astrophysics Group, Coventry CV4 7AL, UK}

\usepackage{ulem}
\usepackage{mathtools}

\newcolumntype{P}{>{$}p{5.5cm}<{$}} 

\received{August 26, 2020}
\accepted{January 29, 2021; The Planetary Science Journal}
\shorttitle{Lithologic Controls on Silicate Weathering}
\shortauthors{Hakim et al.}
\graphicspath{{./}}

\begin{document}

\title{Lithologic Controls on Silicate Weathering Regimes of Temperate Planets}

\correspondingauthor{Kaustubh Hakim}
\email{kaustubh.hakim@csh.unibe.ch}

\author[0000-0002-0786-7307]{Kaustubh Hakim}
\affiliation{\csh}

\author[0000-0002-0673-4860]{Dan J. Bower}
\affiliation{\csh}

\author[0000-0002-7384-8577]{Meng Tian}
\affiliation{\csh}

\author[0000-0001-9423-8121]{Russell Deitrick}
\affiliation{\csh}

\author[0000-0002-9577-2489]{Pierre Auclair-Desrotour}
\affiliation{\obs}
\affiliation{\csh}

\author[0000-0003-4269-3311]{Daniel Kitzmann}
\affiliation{\csh}

\author[0000-0001-6110-4610]{Caroline Dorn}
\affiliation{\zur}

\author[0000-0002-2443-8539]{Klaus Mezger}
\affiliation{\geo}

\author[0000-0003-1907-5910]{Kevin Heng}
\affiliation{\csh}
\affiliation{\war}



\begin{abstract}

Weathering of silicate rocks at a planetary surface can draw down CO$_2$ from the atmosphere for eventual burial and long-term storage in the planetary interior.  This process is thought to provide an essential negative feedback to the carbonate-silicate cycle (carbon cycle) to maintain clement climates on Earth and potentially similar temperate exoplanets. We implement thermodynamics to determine weathering rates as a function of surface lithology (rock type). These rates provide upper limits that allow estimating the maximum rate of weathering in regulating climate.  This modeling shows that the weathering of mineral assemblages in a given rock, rather than individual minerals, is crucial to determine weathering rates at planetary surfaces. By implementing a fluid-transport controlled approach, we further mimic chemical kinetics and thermodynamics to determine weathering rates for three types of rocks inspired by the lithologies of Earth's continental and oceanic crust, and its upper mantle. We find that thermodynamic weathering rates of a continental crust-like lithology are about one to two orders of magnitude lower than those of a lithology characteristic of the oceanic crust. We show that when the CO$_2$ partial pressure decreases or surface temperature increases, thermodynamics rather than kinetics exerts a strong control on weathering. The kinetically- and thermodynamically-limited regimes of weathering depend on lithology, whereas, the supply-limited weathering is independent of lithology. Our results imply that the temperature-sensitivity of thermodynamically-limited silicate weathering may instigate a positive feedback to the carbon cycle, in which the weathering rate decreases as the surface temperature increases. 

\end{abstract}



\section{Introduction} \label{sec:intro}

Greenhouse gases such as CO$_{2}$ are essential in raising Earth's surface temperature \citep{1993Icar..101..108K,2013ApJ...765..131K}. To regulate the amount of CO$_{2}$ in the atmosphere, processes such as the weathering of rocks, degassing and regassing are necessary \citep{1845AdM..book.....B,1952ptod.conf.....U}; but see \citet{2018ApJ...864...75K}. One of the basic assumptions for the definition of the classical habitable zone is therefore the presence of the carbonate-silicate cycle (carbon cycle) that regulates the long-term climate \citep{1993Icar..101..108K}. However, it remains unclear how this cycle may operate on rocky exoplanets where surface conditions could depart from those on modern Earth. The essence of the inorganic carbon cycle is captured by the Ebelman-Urey reaction involving the conversion of a silicate mineral (e.g., wollastonite CaSiO$_{3}$) to a carbonate mineral (e.g., calcite CaCO$_{3}$) in the presence of atmospheric CO$_2$, 

\begin{equation} \label{eq:silWeathering}
   \mathrm{CaSiO_{3} + CO_{2} \xleftrightarrow{} CaCO_{3} + SiO_{2} }.
\end{equation} 
The reverse of reaction (\ref{eq:silWeathering}), metamorphism, converts carbonates back to silicates and releases CO$_2$ to the atmosphere. In addition to metamorphic decarbonation, the degassing of the mantle at mid-ocean ridges and at volcanic arcs also contributes to the CO$_2$ supply to the atmosphere. Although wollastonite is used to exemplify the carbon cycle, in reality its contribution is insignificant. In this study, we evaluate the contribution of silicate minerals and rocks on the weathering component of the carbon cycle. 

An important feature of the carbon cycle on Earth is the negative feedback of silicate weathering \citep[e.g.,][]{1981JGR....86.9776W,1983AmJS..283..641B,2000AREPS..28..611K,2001JGR...106.1373S,2012ApJ...756..178A,2015ApJ...812...36F,2017NatCo...815423K,2020ApJ...896..115G}. This feedback buffers the climate against changes in stellar luminosity and impacts the extent of the habitable zone \citep[e.g.,][]{1993Icar..101..108K,2013ApJ...765..131K}. This feedback is sensitive to partial pressure of CO$_2$ in the atmosphere, surface temperature, and runoff which facilitates weathering reactions and transport of aqueous chemical species from continents to oceans \citep[e.g.,][]{1981JGR....86.9776W,1983AmJS..283..641B}. In oceans, cations and bicarbonate or carbonate ions react to form carbonates that are deposited on the seafloor, thereby removing CO$_2$ from the atmosphere-ocean system \citep[e.g., calcite,][]{2005E&PSL.234..299R}. High concentrations of CO$_2$ in the atmosphere elevate the surface temperature due to the greenhouse effect of CO$_2$. High temperatures give rise to high evaporation and high precipitation (rainfall) rates. As precipitation intensifies, runoff also intensifies. As a result, silicate weathering intensifies, reducing the abundance of atmospheric CO$_2$ which in turn decreases the temperature. When temperatures become low, evaporation and precipitation rates become low thereby decreasing runoff and weathering. In the absence of intense weathering, volcanic degassing increases the abundance of CO$_2$ in the atmosphere sufficiently to increase the temperature again. In the absence of a self-regulating mechanism, the fate of Earth's atmosphere may have been like that of Venus \citep{2014E&PSL.403..307G}. 

In addition to silicate weathering on continents, silicate weathering on the seafloor also has the potential to provide an equivalent negative feedback \citep[e.g.,][]{1992AmJS..292...81F,1997GeCoA..61..965B,2001JGR...106.1373S,2013GGG....14.1771C,2017NatCo...815423K,2017E&PSL.474...97C}. Seafloor weathering is the low-temperature (\textless 313~K) carbonation of the basaltic oceanic crust facilitated by the circulation of seawater through hydrothermal systems \citep{2018AREPS..46...21C}. Seafloor weathering reactions, like continental weathering reactions, dissolve silicate minerals constituting rocks in the presence of water and CO$_2$. Most studies modeling the carbon cycle expect the contribution of seafloor weathering to be smaller than continental weathering by up to a few orders of magnitude and thus neglect it \citep[e.g.,][]{1981JGR....86.9776W,1983AmJS..283..641B,1995AmJS..295.1077C,2001AmJS..301..182B}. However, recent studies have claimed that seafloor weathering is likely a significant component of global weathering during Earth's history \citep[e.g.,][]{2013GGG....14.1771C,2017E&PSL.474...97C,2018PNAS..115.4105K}; but see \citet{2018Natur.560..471I}. Seafloor weathering may be of critical importance on oceanworlds \citep[e.g.,][]{2012ApJ...756..178A,2015ApJ...812...36F,2019A&A...627A..48H}; but see \citet{2018ApJ...864...75K}. 

A prevalent assumption among studies modeling silicate weathering is that kinetics of mineral dissolution reactions determines the weathering flux

\begin{equation} \label{eq:w_pCO2}
    w =  w_0 \left( \frac{P_{\mathrm{CO}_{2}}}{P_{\mathrm{CO}_{2},0}} \right)^{\beta} \exp{\left[ - \frac{E}{R} \left(\frac{1}{T} -\frac{1}{T_0}\right) \right] } 
\end{equation} 
where the subscript `0' denotes reference values, $P_{\mathrm{CO}_{2}}$ is the partial pressure of CO$_2$ in the  gaseous state, and $\beta$ is the kinetic power-law constant constrained empirically from laboratory or field data   \citep[e.g.,][]{1981JGR....86.9776W,1983AmJS..283..641B,1993Icar..101..108K,2001JGR...106.1373S,2015ApJ...812...36F,2017NatCo...815423K}. The activation energy $E$ is also determined empirically from the dependence of kinetic rate coefficients on the Arrhenius law with $T$ as the surface temperature and $R$ as the universal gas constant \citep{palandri2004compilation,2008kwri.book.....B}. \citet{1981JGR....86.9776W} adopt $\beta=0.3$ based on kinetic rate measurements of feldspar weathering in a laboratory \citep{1965BdM...88.....2}. Later studies adjusted the value of $\beta$ using more laboratory and field measurements or balancing carbon fluxes \citep{2008kwri.book.....B}. A Bayesian inversion study performed by \citet{2017NatCo...815423K} for the carbon cycle on Earth based on data from the past 100~Myr shows that the value of $\beta$ for continental weathering is largely unconstrained: between 0.21$-$0.48 (prior: 0.2$-$0.5) for their nominal model and between 0.05$-$0.95 (prior: 0$-$1) for their Michaelis-Menten model \citep{1987AmJS..287..763V,2004tpcc.book.....B}. Moreover, the power-law exponent of seafloor weathering in Equation~(\ref{eq:w_pCO2}) is either assumed to be equal to 0.23 \citep{1997GeCoA..61..965B} or varied between 0$-$1 \citep[e.g.,][]{2001JGR...106.1373S,2015E&PSL.415...38C}. More sophisticated formulations of seafloor weathering assume a dependence on the oceanic crustal production rate with yet another power-law exponent between 0$-$2 \citep{2018PNAS..115.4105K}. 

The flow of fluids such as rainwater through the pore-space of soils facilitates silicate weathering reactions on continents. If the fluid residence time is shorter than the time needed for a reaction to attain chemical equilibrium, the reaction becomes rate limiting and reaction kinetics govern the amount of solute released \citep[e.g.,][]{1982AmJS..282..237A,1984GeCoA..48.2405H}. Then the weathering is kinetically-limited. In contrast, at long fluid residence times, silicate weathering reactions reach chemical equilibrium resulting in a thermodynamic limit \citep[also known as the runoff limit,][]{2000AREPS..28..611K,2011E&PSL.312...48M}. The thermodynamic limit of weathering is referred to as the transport limit in recent literature \citep[e.g.,][]{2018E&PSL.485..111W}; however, the transport limit is traditionally used to describe weathering limited by the supply of fresh minerals, i.e., supply-limited weathering \citep{thompson1959local,1983JGR....88.9671S,2011E&PSL.312...48M,2000AREPS..28..611K}. In this study, the usage of the term transport limited is avoided. Observations that regional weathering fluxes on Earth depend on runoff \citep[e.g.,][]{2004E&PSL.224..547R} in addition to reaction kinetics suggest that global weathering fluxes are a mixture of kinetically- and thermodynamically-limited regimes. The fluid-transport approach \citep{2005E&PSL.240..539S,2010E&PSL.294..101M,2014Sci...343.1502M,2017ESRv..165..280L} allows the modeling of both thermodynamic and kinetic limits of weathering on temperate planets \citep[][]{2018E&PSL.485..111W,2020ApJ...896..115G}. 

The lithology (rock type) is anticipated to play a role in the intensity of the silicate weathering feedback \citep[e.g.,][]{1981JGR....86.9776W,1983JGR....88.9671S,2000AREPS..28..611K}. At the thermodynamic limit of weathering, \citet{2018E&PSL.485..111W} demonstrate that minerals in the feldspar mineral group exhibit feedback that is weaker to stronger than the kinetic weathering feedback. For example, their calculations show that the thermodynamic $\beta$ (or $\beta_{\rm th}$) is 0.25 for albite and K-feldspar and 0.67 for anorthite, compared to the classic kinetic $\beta = 0.3$ from \citet{1981JGR....86.9776W}. The kinetic weathering models are based on reaction rates of feldspars, common silicate minerals comprising granitic rocks on modern continents \citep[e.g.,][]{1981JGR....86.9776W}. However, rocks on the Hadean and early Archean Earth were more mafic than present-day Earth and therefore poorer in SiO$_2$ and feldspar content \citep[e.g.,][]{2020GeCoA.278...16C}; but see \citet{2020Keller}. The modern seafloor mostly contains basaltic rocks. Observations of clay minerals on Mars suggest past weathering processes \citep{2018SciA....4.3330B}. Clues from stellar elemental abundances point to diverse planetary compositions \citep[e.g.,][]{2019MNRAS.482.2222W,2020arXiv200709021S}. Consequently, the surface lithologies (and therefore weathering fluxes) on rocky exoplanets are not necessarily similar to modern Earth.

In this study, we develop a silicate weathering model of the inorganic carbon cycle with key inclusion of lithology by applying the fluid-transport model of \citet{2014Sci...343.1502M} to fluid-rock reactions: \texttt{CHILI} (CHemical weatherIng model based on LIthology). This model tracks the aqueous carbon reservoir assimilating three weathering regimes as a function of lithology. In addition to continental weathering, this model is applied to seafloor weathering. The primary goal is to determine lithology-based weathering fluxes on the surface of temperate exoplanets by mitigating the impact of present-day Earth calibrations. The key philosophy behind this study is to investigate the extent to which the silicate weathering model may be generalized, beyond its Earth-centric origins, in order to apply it to rocky exoplanets with secondary atmospheres.

\begin{figure*}[!ht]
  \centering
  \medskip
  \includegraphics[width=\textwidth]{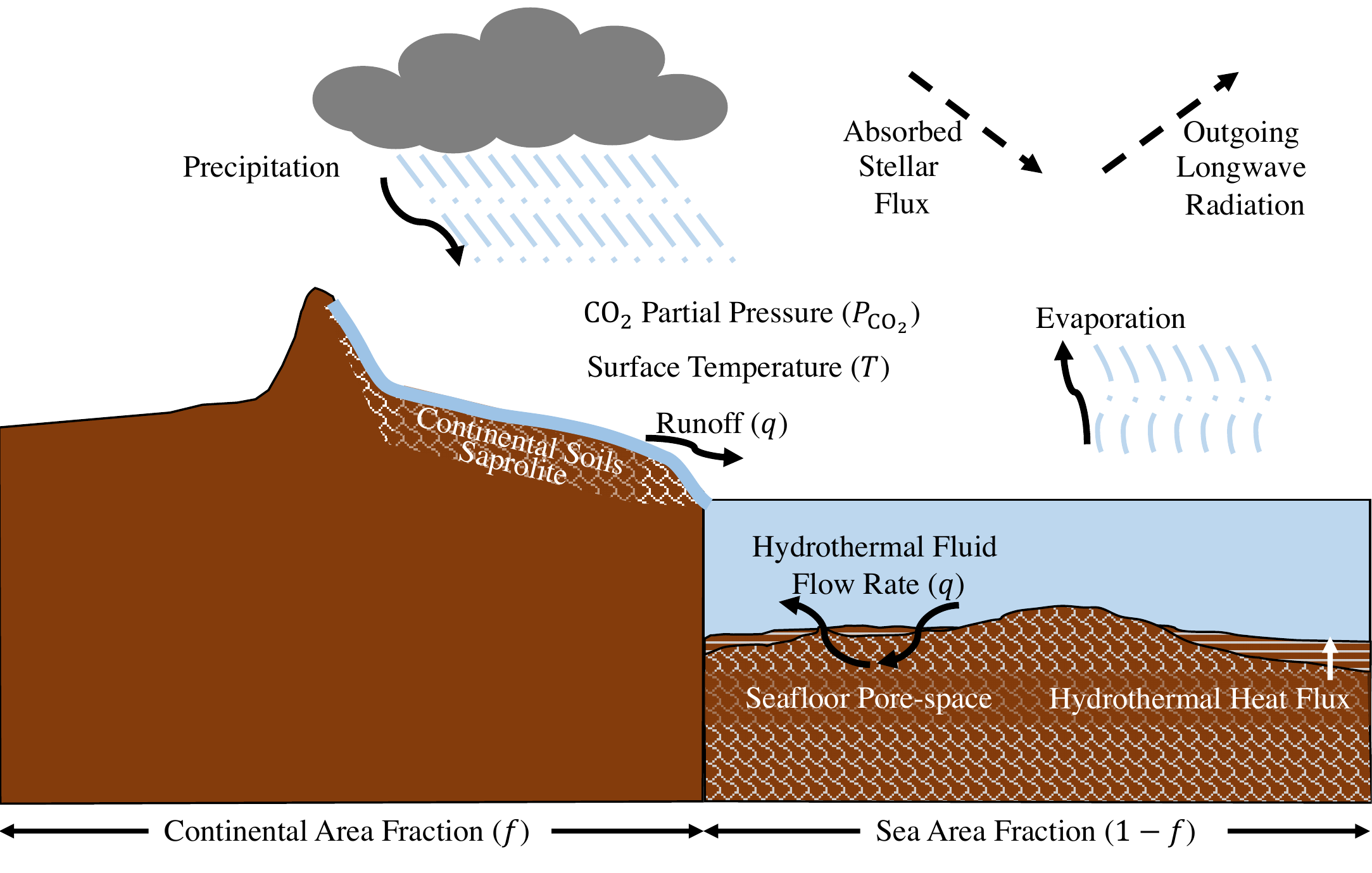}
  \caption{ Silicate weathering model based on weathering reactions and fluid flow rates. The continental runoff and the seafloor fluid flow rate enable weathering reactions in continental soils and saprolite, and seafloor pore-space, respectively. Key parameters and processes are labeled (see Tables \ref{tab:params}, \ref{tab:variables} and \ref{tab:f_params} for all parameters and computed quantities). }
  \label{fig:illustration}
\end{figure*}

\section{Weathering Model} \label{sec:methods}

\subsection{Proxies for Weathering} \label{sec:methodsWeatheringFlux}

Continental weathering occurs in continental soils and saprolite where runoff (water discharge per unit surface area) facilitates fluid-rock weathering reactions (Figure~\ref{fig:illustration}). Seafloor weathering occurs in pores and veins of the oceanic crust (hereafter, pore space) where seawater reacts with basalts. Analogous to runoff, the fluid flow in off-axis low-temperature hydrothermal systems facilitates seafloor weathering reactions \citep{1994JGR....99.3081S,2018AREPS..46...21C}. The fluid flow rate $q$ (either runoff or hydrothermal fluid flow rate) is a free parameter in \texttt{CHILI} (Table~\ref{tab:params}). On Earth, the present-day continental runoff varies between 0.01$-$3~m~yr$^{-1}$ with a mean of approximately 0.3~m~yr$^{-1}$ \citep{1999ChGeo.159....3G,2002GBioC..16.1042F}. On the seafloor, the non-porosity corrected hydrothermal fluid flow rates are between 0.001$-$0.7~m~yr$^{-1}$ with a mean of about 0.05~m~yr$^{-1}$ \citep{1994JGR....99.3081S,2003E&PSL.216..565J,2013E&PSL.380...12H}. 

In the fluid transport-controlled model of continental weathering, the weathering flux $w$ (mol~m$^{-2}$~yr$^{-1}$) is the product of the concentration of a solute of interest $C$ and runoff $q$  \citep[e.g.,][]{2014Sci...343.1502M},

\begin{equation} \label{eq:w}
    w = C \; q.
\end{equation}
We also apply the same approach to seafloor weathering. The total silicate weathering rate\footnote{The weathering rate has dimensions of moles per unit time, whereas the weathering flux has dimensions of moles per unit area per unit time (Table~\ref{tab:variables}).} $W$ (mol~yr$^{-1}$) is the sum of the continental ($W_{\mathrm{cont}}$) and seafloor weathering rates ($W_{\mathrm{seaf}}$). These rates are given by product of the continental ($w_{\mathrm{cont}}$) or seafloor weathering flux ($w_{\mathrm{seaf}}$) and the continental ($f \, A_{\rm s}$) or seafloor surface area ($(1 -  f) A_{\rm s}$), where $A_{\rm s}$ is the planet surface area and $f$ is the continental area fraction: 

\begin{equation} \label{eq:w_rate}
    W = w_{\mathrm{cont}} \, f \, A_{\rm s} + w_{\mathrm{seaf}} \, (1-f) \, A_{\rm s}.
\end{equation}

To model silicate weathering, the aqueous bicarbonate ion concentration [HCO$_3^-$] is normally used as a proxy for weathering because it is a common byproduct of silicate weathering \citep[e.g., anorthite weathering,][]{2018E&PSL.485..111W},

\begin{equation} \label{eq:sil_weath}
\begin{split}
    \mathrm{CaAl_2Si_2O_8} + 2 \mathrm{CO}_{2} + 3 \mathrm{H_2O} \xrightarrow{} 2 \mathrm{HCO}_3^{-} + \mathrm{Ca}^{2+} \\ 
    + \mathrm{Al_2Si_2O_5(OH)_4}.
\end{split}
\end{equation}
This is a good assumption under modern Earth conditions since HCO$_3^-$ is the primary CO$_2$-rich product of continental silicate weathering that is carried to oceans by rivers and reacts with Ca$^{2+}$ to precipitate Ca-rich carbonates on the seafloor. We are interested in modeling weathering under conditions more diverse than those on modern Earth. In highly alkaline conditions, the carbonate ion CO$_{3}^{2-}$ is produced in amounts similar or exceeding [HCO$_{3}^{-}$]. Both CO$_{3}^{2-}$ and HCO$_{3}^{-}$ have the potential to  contribute to Ca-Mg-Fe carbonate precipitation on the seafloor \citep[e.g., calcite precipitation,][]{1978AmJS..278..179P},

\begin{equation}\label{eq:carb_prec}
\begin{split}
   2 \mathrm{HCO}_3^{-} + \mathrm{Ca^{2+}} & \xrightarrow{} \mathrm{CaCO}_3 +  \mathrm{CO}_{2} + \mathrm{H_2O}, \\
   \mathrm{CO}_3^{2-} + \mathrm{Ca^{2+}}  & \xrightarrow{} \mathrm{CaCO}_3.
\end{split} 
\end{equation}
Equation~(\ref{eq:carb_prec}) exemplifies that, in addition to bicarbonate and carbonate ions, divalent cations are required to drive the flux of CO$_2$ out of the atmosphere-ocean system. Nonetheless, since twice as many HCO$_{3}^{-}$ ions as CO$_{3}^{2-}$ ions are needed to precipitate one mole of carbonate \citep{2017NatCo...815423K}, in addition to HCO$_{3}^{-}$ ($C = \mathrm{[HCO_3^-]}$ in Equation~\ref{eq:w}), we consider carbonate alkalinity $A$ resulting from reactions between silicate rocks and fluids as a proxy for weathering ($C = A$ in Equation~\ref{eq:w}) where

\begin{equation} \label{eq:ALK}
    A = \mathrm{[HCO_3^-] + 2 [CO_3^{2-}]}.
\end{equation}

\begin{deluxetable}{llc}
\tablecaption{Control parameters and their reference values for modern Earth. \label{tab:params}}
\tablehead{
\colhead{Symbol} & \colhead{Description} & \colhead{Reference}
}
\startdata
$P_{\mathrm{CO}_2}$ & CO$_2$ partial pressure & 280 $\mu$bar \\
$T$    & Surface temperature &  288 K      \\
$P$    & Surface pressure & 1 bar    \\
$q$    & Runoff or fluid flow rate  & 0.3 m~yr$^{-1}$  \\
$t_{\mathrm{s}}$  & Soil or pore-space age & 10$^5$ yr \\
$\psi$ & Dimensionless pore-space parameter & 222\,750 \\
\enddata
\tablecomments{ $T$ is not a control parameter for some calculations ($T = f(P_{\mathrm{CO}_2})$, Section~\ref{sec:methodsClimate}).  }
\end{deluxetable}

\begin{deluxetable}{llc}
\tablecaption{Computed quantities. \label{tab:variables}}
\tablehead{
\colhead{Symbol} & \colhead{Description} & \colhead{Units}
}
\startdata
$w$    & Weathering flux & mol~m$^{-2}$~yr$^{-1}$  \\
$W$    & Weathering rate & mol~yr$^{-1}$ \\
$T'$   & Seafloor pore-space temperature & K      \\
$A$    & Carbonate alkalinity  & mol~dm$^{-3}$  \\
pH     & Negative logarithm of H$^+$ activity & $-$ \\
$C$    & Concentration  & mol~dm$^{-3}$ \\
$[\mathrm{C}]$  & Concentration of a species C  & mol~dm$^{-3}$ \\
$a_{\mathrm{C}}$ & Activity of a species C & $-$ \\
$D_w$  & Damk\"{o}hler coefficient  & m~yr$^{-1}$ \\
$K$ & Equilibrium constant & $-$ \\
$k_{\mathrm{eff}}$ & Kinetic rate coefficient & mol~m$^{-2}$~yr$^{-1}$  \\
$E$ & Activation energy &  kJ~mol$^{-1}$ \\
$E_{\rm th}$ & Thermodynamic activation energy &  kJ~mol$^{-1}$ \\
$\beta$ & Power-law exponent &  $-$ \\
$\beta_{\rm th}$ & Thermodynamic power-law exponent &  $-$ \\
\enddata
\end{deluxetable}

\begin{figure*}[!ht]
  \centering
  \medskip
  \includegraphics[width=\textwidth]{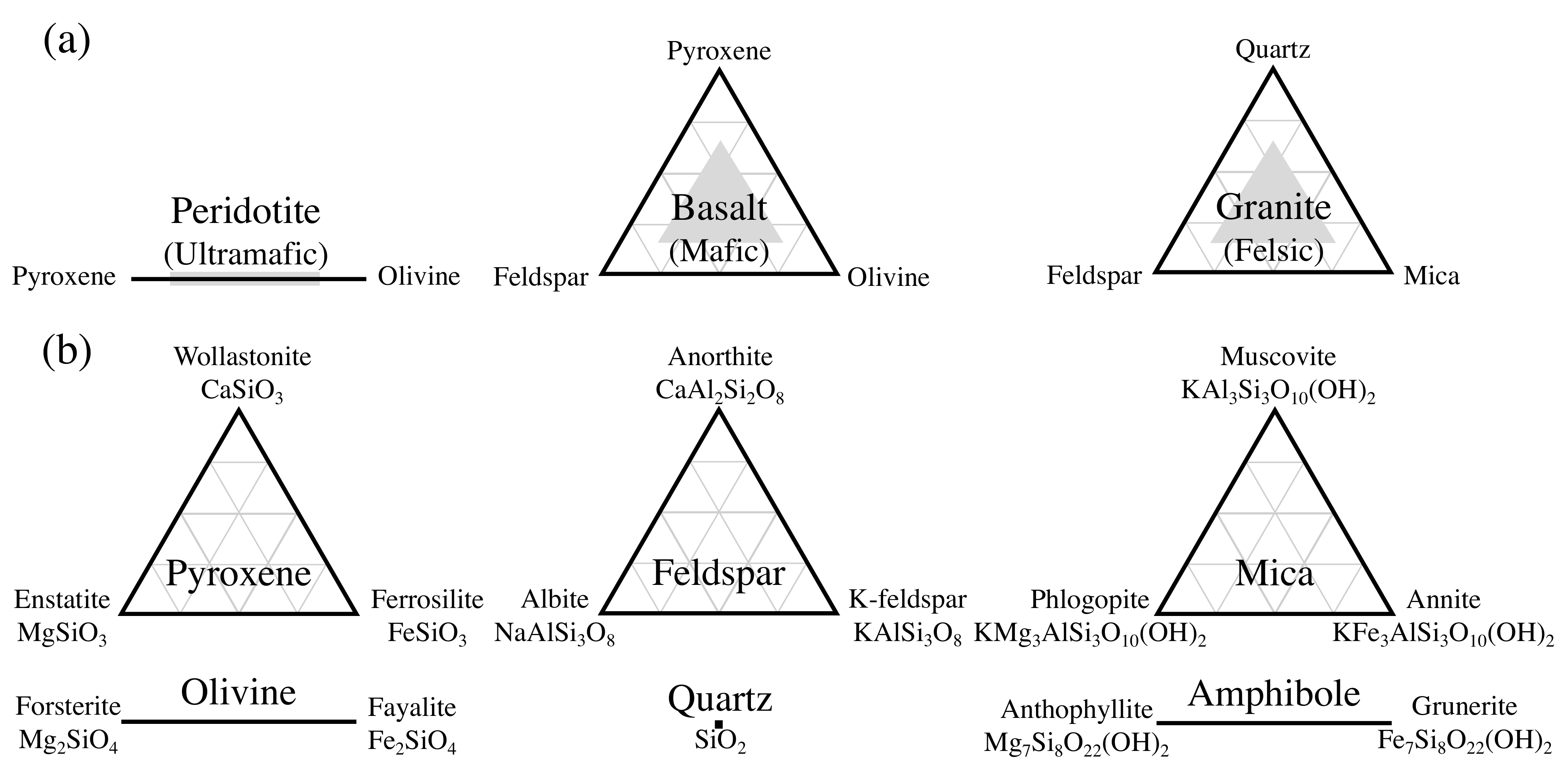}
  \caption{(a) The composition of peridotite, basalt and granite considered in this study in terms of major endmember mineral groups. All mineral groups for a given rock are assumed to be present in significant proportions (shaded regions). (b) The major weatherable silicate minerals constituting silicate rocks: ternary endmembers of pyroxene, feldspar and mica, binary endmembers of olivines and amphibole, and quartz. Our choice of rock compositions is limited to certain endmember minerals in a mineral group. Mica and amphibole exhibit a large number of endmembers \citep{2011JMetG..29..333H}. The endmembers of mica and amphibole chosen here are merely illustrative. }
  \label{fig:mineral_types}
\end{figure*}

\subsection{Major Silicate Lithologies} \label{sec:methodsRocks}

Previous studies applying the fluid transport-controlled model to continental weathering limit the lithology to monomineralic \citep[e.g., oligoclase feldspar,][]{2020ApJ...896..115G} or one rock type \citep[e.g., granite, ][]{2014Sci...343.1502M,2018E&PSL.485..111W}. However, the chemistry of common silicate rocks ranges from ultramafic to felsic, with increasing SiO$_2$ content. To model the surface mineralogy of terrestrial exoplanets, we approximate an ultramafic, a mafic and a felsic lithology with major mineralogical compositions of peridotite, basalt and granite, respectively. Peridotites are upper mantle rocks rich in MgO and FeO and poor in SiO$_2$ relative to basalts and granites. Basalts are igneous rocks that are common examples of rocks present in the oceanic crust of Earth, lunar mare on the Moon, and the crust of Mars. Granites are present on the modern continental crust of Earth. These are highly differentiated rocks that are rich in SiO$_2$, Na$_2$O and K$_2$O due to partial melting and crystallization processes.

The three rock types (peridotite, basalt and granite considered in this study are assumed to be composed of 2$-$3 major mineral groups, and can thus be projected onto binary or ternary diagrams spanned by the major mineral groups considered (Figure~\ref{fig:mineral_types}a). Each mineral group is similarly defined by 1$-$3 endmember minerals (Figure~\ref{fig:mineral_types}b). For instance, olivine is a solid solution of two endmember minerals, forsterite and fayalite.  In the reduced set of mineral groups with only endmember minerals, we represent peridotite using pyroxenes (wollastonite and enstatite) and olivines (forsterite and fayalite), basalt using plagioclase feldspars (anorthite and albite) and pyroxenes (wollastonite, enstatite and ferrosilite), and granite using alkali feldspars (K-feldspar and albite), quartz and biotite micas (phlogopite and annite). Halloysite, a phyllosilicate mineral, is a byproduct (secondary mineral) of the weathering of basalt and granite but not of peridotite (see Appendix~\ref{app:Generalized}). Kaolinite, another phyllosilicate secondary mineral is also considered but no secondary minerals containing divalent and monovalent cations are modeled. The choice of endmember primary minerals to define these rocks makes them idealized compared to natural rocks. Moreover, for a rock, not all endmember minerals of a mineral group can be modeled due to the unity activity assumption for endmember minerals (Section~\ref{sec:methodsMaximum}). Only major minerals in typical rock types are considered and minor minerals such as magnetite, hematite and pyrite are not discussed. Since the net contribution of carbonate weathering on the carbon cycle is small on timescales of the order of 100~kyr for Earth \citep[e.g.,][]{2001JGR...106.1373S}, the weathering of carbonate minerals is neglected.

\subsection{Maximum Weathering Model} \label{sec:methodsMaximum}

Concentrations of products of weathering reactions at chemical equilibrium allow one to calculate a thermodynamic upper limit to weathering (maximum weathering). To calculate concentrations of silicate weathering products, a number of chemical reactions need to be considered. Reactions between water, CO$_2$ and silicate minerals produce the bicarbonate ion HCO$_{3}^{-}$. The bicarbonate ion further dissociates into the carbonate ion CO$_{3}^{2-}$ and H$^+$. Water dissociates into H$^+$ and OH$^-$. The relations between equilibrium constants of these reactions and thermodynamic activities of reactants and products are given in Appendix~\ref{app:Generalized}. Thermodynamic activities quantify the energetics of mixing of constituent components in solid or aqueous solutions \citep[][]{2001tac..book.....P}. Since no non-ideal solid-solution behaviors are considered, the endmember compositions of the minerals are used for calculations and thus the activities of minerals are set at unity. Furthermore, in a dilute solution, the activity of liquid water is approximately unity.

The activity of an aqueous species (e.g., HCO$_3^-(aq)$) is given by the product of its concentration [HCO$_3^-(aq)$] and the activity coefficient $\gamma$ normalized to the standard state of concentration ($C_0$ = 1~mol~dm$^{-3}$), $a_{\mathrm{HCO}_3^-(aq)} = \frac{\gamma \; [\mathrm{HCO}_3^-(aq)]}{C_0}$ \citep[][]{2001tac..book.....P}. In an ideal, dilute solution, $\gamma \to 1$ and $[\mathrm{HCO}_3^-(aq)] = a_{\mathrm{HCO}_3^-(aq)} \, C_0$, an assumption made throughout the study for all aqueous species. The activity of a gaseous species such as CO$_2(g)$ is given by the ratio of its fugacity in the gas mixture to the fugacity of pure CO$_2(g)$ at the total pressure $P$ (i.e., surface pressure in this study). Thus, $a_{\mathrm{CO}_2(g)} = \frac{f_{\mathrm{CO_{2}}} }{f^{\mathrm{tot}}_{\mathrm{CO_{2}}}} = \frac{\Gamma P_{\mathrm{CO}_2}}{\Gamma^{\mathrm{tot}} P}$, where $P_{\mathrm{CO}_2}$ is the CO$_2$ partial pressure, $P$ is the total surface pressure, $\Gamma$ and $\Gamma^{\mathrm{tot}}$ are fugacity coefficients for the CO$_2$ component and pure CO$_2$, respectively. The fugacity coefficients give a correction factor for the non-ideal behavior due to mixing and/or pressure effects. For CO$_2(g)$, the fugacity coefficient varies between 0.5 and 1 up to pressures of 200~bar \citep{1988spycher}. Our assumptions of unity fugacity coefficients and $P = 1$~bar lead to $a_{\mathrm{CO}_2(g)} = P_{\mathrm{CO}_2}$.

Equilibrium constants depend on temperature and pressure (see Appendix~\ref{app:data}). Chemical  reactions on continents are characterized by the surface temperature $T$ and surface pressure $P$ (Table~\ref{tab:params}). Seafloor weathering reactions are characterized by the seafloor pore-space temperature $T'$ and pressure $P'$. In the temperature range 273$-$373~K, data suggest that $T'$ is within 1\% of $T$ \citep[][and references therein]{2017NatCo...815423K} and hence $T'= T$ is assumed. Moreover, pressure in the range of 0.01$-$1000~bar has a negligible effect on equilibrium constants and resulting concentrations (see Figure~\ref{fig:K} and Figure~\ref{fig:HCO3eq_all_peri}). However, seafloor pressure affects carbonate stability. Nonetheless, $P'$ is fixed to 200~bar which is the pressure at approximately 2~km depth in Earth's present-day oceans. When extending the continental weathering model to seafloor weathering, we consider a fresh water ocean, in which ocean chemistry does not limit the production of cations and carbonate alkalinity. 

To demonstrate the computation of thermodynamic solute concentrations resulting from fluid-rock reactions, weathering of peridotite is considered. Since two pyroxene endmembers (wollastonite and enstatite) and two olivine endmembers (forsterite and fayalite) are considered to constitute  peridotite, dissolution reactions of these four minerals are of interest (Appendix~\ref{app:Generalized}, Table~\ref{tab:reactions}, rows: (a), (b), (d), (e)). Besides, reactions in the water-bicarbonate system are needed (Appendix~\ref{app:Generalized}, Table~\ref{tab:reactions}, rows: (o), (p), (q), (r)). These eight chemical reactions result in eight equations and ten unknowns. The eight equations are given by the relations between activities and equilibrium constants (Appendix~\ref{app:Generalized}). The ten unknowns are the activities of CO$_2(g)$, CO$_2(aq)$, SiO$_2(aq)$, Ca$^{2+}$, Mg$^{2+}$, Fe$^{2+}$, H$^+$, OH$^-$, HCO$_3^-$ and CO$_3^{2-}$. An additional equation is given by balancing the charges of all cations and anions present in the solution. These nine equations are further reduced to one polynomial equation with two unknowns, activities of HCO$_3^-$ and CO$_2(g)$ (see Appendix~\ref{app:Generalized}, Table~\ref{tab:poly_weath}). If ferrosilite, the third endmember of pyroxene, is also considered, an additional equation is introduced but the unknowns remain the same and the system is over-determined. This scenario stems from the assumption of endmember minerals with unity activities instead of solid solutions. In the case of solid solutions, the activities of endmember minerals become unknowns rather than being fixed at unity. This increases the number of unknowns for a given set of equations, and requires more equations that can be derived from additional reactions \citep[e.g., see][for a treatment of solid solutions]{2015E&PSL.430..486G}. Although mineral solid solutions are commonplace in rocks, endmember considerations allow us to establish a simple framework in which the effects of various lithologies can be explored on the weathering of exoplanets. Therefore, the presence of ferrosilite in peridotite is ignored. Moreover, the presence of K-feldspar and quartz in basalt and anorthite in granite is not considered. For a given $P_{\mathrm{CO}_2}$, $a_{\mathrm{HCO}_{3}^{-}}$ is calculated by finding the sole physical root of such a polynomial equation. The bicarbonate ion concentration at chemical equilibrium is obtained from the standard concentration, [HCO$_3^-$]$_{\mathrm{eq}}$ = $a_{\mathrm{HCO}_{3}^{-}} \times$ 1~mol~dm$^{-3}$. Similarly, activities of all aqueous species are converted to concentrations using the standard concentration. Subsequently, all unknowns are calculated from the relations between activities and equilibrium constants (Table~\ref{tab:reactions}).

\begin{figure}[!ht]
  \centering
  \medskip
  \includegraphics[width=\linewidth]{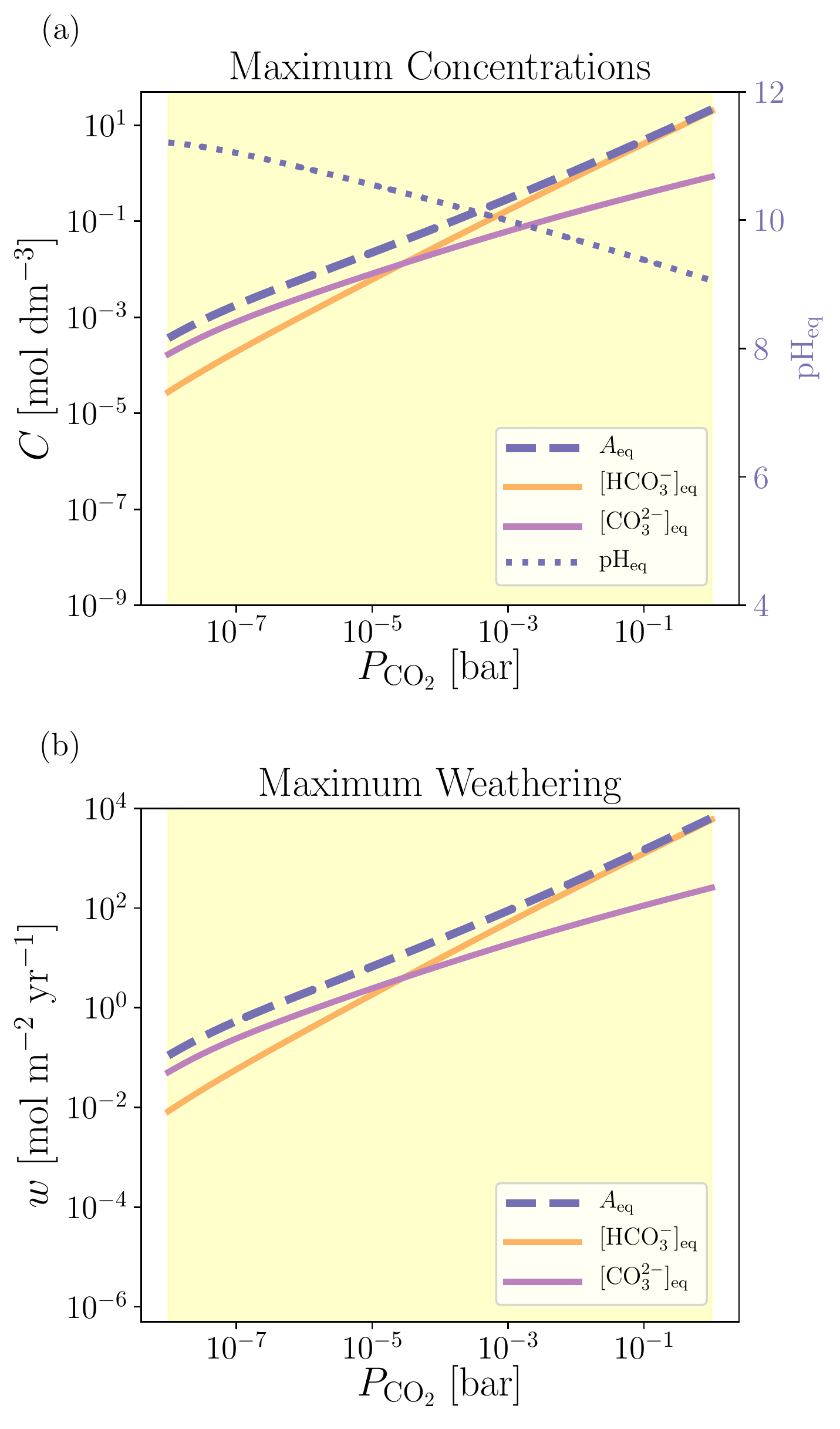}
  \caption{ (a) Maximum concentrations of the components of carbonate alkalinity $A$ and the solution pH of the network of reactions defining peridotite weathering assuming chemical equilibrium as a function of $P_{\mathrm{CO}_2}$ at $T = 288$~K. (b) Maximum weathering flux corresponding to $A$, [HCO$_3^-$] and [CO$_3^{2-}$] at modern mean runoff of $q$ = 0.3~m~yr$^{-1}$ (other parameters take reference values). Yellow background denotes the thermodynamic regime of weathering. The scales on vertical and horizontal axes are equal to those in Figure~\ref{fig:CA_peri}.}
  \label{fig:CAeq_peri}
\end{figure}

At chemical equilibrium, the carbonate alkalinity increases and the pH of solution decreases with an increase in $P_{\mathrm{CO}_2}$ at $T$ = 288~K (Figure~\ref{fig:CAeq_peri}a). As $P_{\mathrm{CO}_2}$ increases, [HCO$_3^-$]$_{\mathrm{eq}}$ and [CO$_3^{2-}$]$_{\mathrm{eq}}$ increase monotonically, thereby increasing $A_{\mathrm{eq}}$, whereas pH$_{\mathrm{eq}}$ decreases monotonically. For all except $P_{\mathrm{CO}_2} < 20$~$\mu$bar, [HCO$_3^-$]$_{\mathrm{eq}}$ contributes significantly to $A_{\mathrm{eq}}$. The weathering flux (Equation~\ref{eq:w}) resulting from thermodynamic solute concentrations gives the maximum weathering flux. Since no secondary minerals are modeled for peridotite weathering, the weathering flux for $P_{\rm CO_2} > 0.1$~bar is likely an overestimate \citep[e.g., aqueous alteration of olivine produces secondary minerals,][]{2019E&PSL.52415718K}. Figure~\ref{fig:CAeq_peri}b shows that the thermodynamic weathering flux calculations using [HCO$_3^-$]$_{\mathrm{eq}}$ and $A_{\mathrm{eq}}$ increase monotonically with $P_{\mathrm{CO}_2}$. We follow the same procedure as for peridotite to compute the maximum solute concentrations and the maximum weathering flux for basalt, granite or individual endmember silicate minerals by solving polynomial equations given in Appendix~\ref{app:Generalized} and corresponding charge balance equations.

\subsection{Generalized Weathering Model} \label{sec:methodsGeneralized}

In natural environments, fluid-rock reactions may not have the time to attain chemical equilibrium due to high fluid flow rates. The maximum weathering model (Section~\ref{sec:methodsMaximum}) does not accurately represent the weathering flux when chemical equilibrium for mineral dissolution reactions is not attained. We present the generalized weathering model by extending the fluid transport-controlled approach of \citet{2014Sci...343.1502M}. Mineral dissolution reactions, being rate limiting, essentially regulate the concentration of reaction products including [HCO$_3^-$]. In this limit, kinetics plays a more dominant role than thermodynamics. \citet{2014Sci...343.1502M} provide a solute transport equation to calculate such a transport-buffered (dilute) solute concentration from its value at chemical equilibrium, fluid flow rate (runoff) $q$ and the Damk\"{o}hler coefficient $D_w$ which gives the `net reaction rate' (see below). The solute transport equation with HCO$_3^-$ is

\begin{equation} \label{eq:solute_eq}
    [\mathrm{HCO}_3^-] = \frac{[\mathrm{HCO}_3^-]_{\mathrm{eq}}}{1 + \frac{q}{D_{w}}}.
\end{equation}
In their formulation, \citet{2014Sci...343.1502M} multiply $D_w$ by an arbitrary scaling constant $\tau = e^2 \approx 7.389$ in order to scale up the solute concentration to 88\% of its equilibrium value at $q = D_w$. \citet{2016GeCoA.190..265I} when applying the solute transport equation of \citet{2014Sci...343.1502M}, use $\tau = 1$ without any scaling. For simplicity, we use $\tau = 1$, implying $[\mathrm{HCO}_3^-] = 0.5 \; [\mathrm{HCO}_3^-]_{\mathrm{eq}}$ at $q = D_w$. The resulting [HCO$_3^-$] of our model at $q = D_w$ is about 43\% lower than that of the \citet{2014Sci...343.1502M} model, which is a much smaller difference than the 5$-$10 orders of magnitude range of solute concentrations explored in this study. 

\begin{figure}[!ht]
  \centering
  \medskip
  \includegraphics[width=0.5\textwidth]{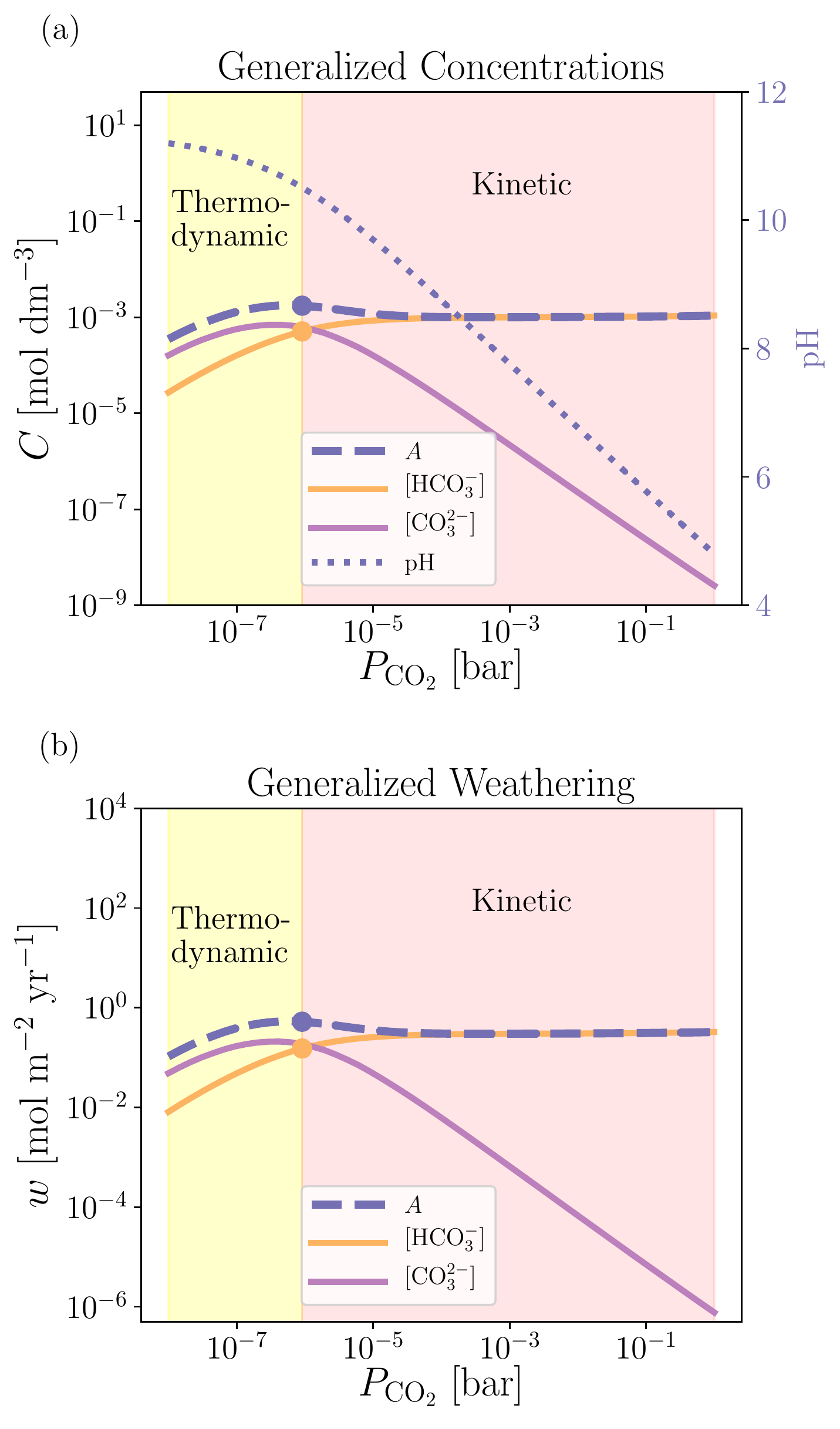}
  \caption{ (a) Generalized concentrations of carbonate alkalinity and the solution pH of peridotite weathering as a function of $P_{\mathrm{CO}_2}$ at $T = 288$~K and modern mean runoff of $q = 0.3$~m~yr$^{-1}$ (other parameters take reference values). From left to right, colored disks mark the transition between thermodynamic and kinetic regimes. (b) The corresponding weathering flux of carbonate alkalinity, [CO$_3^{2-}$] and [HCO$_3^-$]. The scales on vertical and horizontal axes are equal to those in Figure~\ref{fig:CAeq_peri}.}
  \label{fig:CA_peri}
\end{figure}

The quantity $\frac{D_w}{q}$ in Equation~(\ref{eq:solute_eq}) is the Damk\"{o}hler number, the ratio of fluid residence time and chemical equilibrium timescale \citep{2009RvMG...70..485S}. The $\frac{D_w}{q}$ ratio is essentially the ratio of the `net reaction rate' and the fluid flow rate. When the fluid residence time exceeds the chemical equilibrium timescale, or the net reaction rate exceeds the fluid flow rate ($q < D_w$), $[\mathrm{HCO}_3^-] \to [\mathrm{HCO}_3^-]_{\mathrm{eq}}$ and the weathering flux reaches its maximum value for a given $q$, $w = [\mathrm{HCO}_3^-]_{\mathrm{eq}} \; q$ (Figure~\ref{fig:CA_peri}). This weathering regime is called the thermodynamically-limited weathering (hereafter, thermodynamic regime), also known as runoff-limited weathering. When the chemical equilibrium timescale exceeds the fluid residence time, or the  fluid flow rate exceeds the net reaction rate ($q > D_w$), $[\mathrm{HCO}_3^-] \to [\mathrm{HCO}_3^-]_{\mathrm{eq}} \frac{D_w}{q}$ and $w = [\mathrm{HCO}_3^-]_{\mathrm{eq}} \; D_w$, making $w$ independent of [HCO$_3^-$]$_{\mathrm{eq}}$ because  $D_w$ is modeled to be inversely proportional to [HCO$_3^-$]$_{\mathrm{eq}}$ (see below). This regime is known as kinetically-limited weathering (hereafter, kinetic regime). The transition between these two regimes occurs at $q = D_w$. The comparison of timescales described here is conceptually identical to the `quenching approximation' employed in atmospheric chemistry \citep{1977Sci...198.1031P,2017ApJS..228...20T}.  

Rewriting the formulation of $D_w$ from \citet{2014Sci...343.1502M},

\begin{equation} \label{eq:Dw}
    D_{w} = \frac{\psi} {[\mathrm{HCO_3^-}]_{\mathrm{eq}} ( k_{\mathrm{eff}}^{-1} + m A_\mathrm{sp} t_{\mathrm{s}}) }
\end{equation} 
where [HCO$_3^-$]$_{\mathrm{eq}}$ is the equilibrium solute concentration, $k_{\mathrm{eff}}$ is the effective kinetic rate coefficient given by  kinetics data (Appendix~\ref{app:data}), $t_{\mathrm{s}}$ is the age of soils or pore-space, $A_\mathrm{sp}$ is the specific reactive surface area per unit mass of the rock, $m$ is the mean molar mass of the rock, $\psi = L (1 - \phi) \rho X_r A_\mathrm{sp}$ is a dimensionless pore-space parameter that combines five parameters including flowpath length $L$, porosity $\phi$, rock density $\rho$ and fraction of reactive minerals in fresh rock $X_r$. Although \citet{2014Sci...343.1502M} parameterize $D_w$ using nine quantities, in Appendix~\ref{app:Dw}, we show that $D_w$ is mostly sensitive to four of these quantities given their plausible ranges:  [HCO$_3^-$]$_{\mathrm{eq}}$, $k_{\mathrm{eff}}$, $t_{\mathrm{s}}$ and $L$ ($L$ is absorbed in $\psi$). The remaining five parameters are fixed to reference values (Table~\ref{tab:f_params}). The parameter $\psi$ scales $k_{\mathrm{eff}}$ with dimensions of moles per unit reactive surface area of rocks per unit time to $w$ with dimensions of moles per unit exposed continental or seafloor area for a given lithology per unit time. In Equation~(\ref{eq:Dw}), the solid mass to fluid volume ratio $\rho_{\mathrm{sf}}$ given in the $D_w$ formulation of \citet{2014Sci...343.1502M} is rewritten in terms of solid density and porosity using $\rho_{\mathrm{sf}} = \rho (1 - \phi) / \phi$. The kinetic rate coefficients of mineral dissolution reactions are obtained from \citet[][]{palandri2004compilation} as a function of $T$ and pH (see Appendix~\ref{app:data}). The solution pH at chemical equilibrium is used to calculate $k_{\mathrm{eff}}$. For minerals with no $k_{\mathrm{eff}}$ data, $k_{\mathrm{eff}}$ of a corresponding endmember mineral from the same mineral group is adopted. For rocks, we adopt $k_{\mathrm{eff}}$ given by the minimum $k_{\mathrm{eff}}$ among constituent minerals since the slowest reaction is normally rate limiting; but see, e.g., \citet{2009GeoRL..3611202M} where the fastest dissolving mineral sets the pace of rock dissolution.

\begin{deluxetable}{llc}
\tablecaption{Fixed Parameters. \label{tab:f_params}}
\tabletypesize{\small}
\tablehead{
\colhead{Symbol} & \colhead{Description} & \colhead{Value}
}
\startdata
$\phi$   & Porosity  & 0.175 $^\dagger$ \\
$L$      & Flowpath length & 1~m $^\dagger$ \\
$X_r$    & Fraction of reactive minerals  & 1 $^\ast$ \\
& in fresh rock & \\
$A_\mathrm{sp}$ & Specific surface area & 100 m$^2$~kg$^{-1}$ $^\dagger$ \\
& of mineral or rock  \\
$\rho$   & Density of mineral or rock & 2700 kg~m$^{-3}$ $^\dagger$ \\
$m$      & Mean molar mass of rock & 0.27 kg~mol$^{-1}$ $^\dagger$ \\
$A_{\rm s}$      & Planet surface area  & 510.1 Mm$^{2}$ $\ddagger$ \\
$f$      & Continental area fraction  & 0.3 $\ddagger$ \\
$S$      & Stellar flux & 1360 W~m$^{-2}$ $\ddagger$ \\
$\alpha$ & Planetary albedo & 0.3 $^\ddagger$ \\
$P'$     & Seafloor pore-space pressure & 200~bar $\ddagger$  \\
\enddata
\tablecomments{$^\dagger$\citet{2014Sci...343.1502M}, $^\ddagger$Present-day Earth, $^\ast$All considered minerals are reactive.}
\end{deluxetable}

In Equation~(\ref{eq:Dw}), when $t_{\mathrm{s}} = 0$ (young soils), $D_w = \frac{k_{\mathrm{eff}} \psi}{[\mathrm{HCO_3^-}]_{\mathrm{eq}}}$ and $w = k_{\mathrm{eff}} \psi$. In this `fast kinetic' regime, the weathering flux is directly proportional to the kinetic rate coefficient, as assumed in traditional kinetic weathering models \citep[e.g.,][]{1981JGR....86.9776W}. When $t_{\mathrm{s}} \gg \frac{1}{k_{\mathrm{eff}} m A_\mathrm{sp}}$ (old soils), $D_w = \frac{\psi} {[\mathrm{HCO_3^-}]_{\mathrm{eq}} m A_\mathrm{sp} t_{\mathrm{s}}}$ and $w = \frac{\psi}{m A_\mathrm{sp} t_{\mathrm{s}}}$. This regime may be termed as the `slow kinetic' regime, however an accepted terminology is supply-limited weathering (hereafter, supply regime) which is limited by the supply of fresh rocks in the weathering zone \citep{2003GeCoA..67.4411R,2005E&PSL.235..211W}. An increase in $t_{\mathrm{s}}$ decreases the influence of chemical kinetics on weathering. Although $t_{\mathrm{s}}$ depends on the soil production and physical erosion rates which in turn are sensitive to climate, topography and fluid flow properties, there is no consensus on the formulation of soil production and physical erosion rates \citep{2003GeCoA..67.4411R,2005E&PSL.235..211W,2009Geo....37..151G,2012Geo....40..811W,2014Sci...343.1502M,2015ApJ...812...36F}. Treating $t_{\mathrm{s}}$ as a free parameter makes it possible to model both the age of continental soils and seafloor pore-space. The transition between the kinetic and supply regimes can be defined at $t_{\mathrm{s}} = \frac{1}{k_{\mathrm{eff}} m A_\mathrm{sp}}$ where the kinetic reaction rate equals the supply rate of fresh rocks (see Equation~\ref{eq:Dw}). 

In the fluid transport-controlled model, the mineral dissolution reactions (Appendix~\ref{app:Generalized}, rows (a$-$n) in Table~\ref{tab:reactions}) do not attain chemical equilibrium and their reaction products are given by the solute transport equation (Equation~\ref{eq:solute_eq}). On the other hand, reactions in the water-bicarbonate system (Appendix~\ref{app:Generalized}, rows (o$-$r) in Table~\ref{tab:reactions}), are considered to reach chemical equilibrium. This assumption is justified because the equilibrium timescale of chemical reactions in the water-bicarbonate system is of the order of milliseconds to days, unlike mineral dissolution reactions that may take months to thousands of years \citep[e.g.,][]{palandri2004compilation,schulz2006determination}. Thus, the generalized concentrations of aqueous species in the water-bicarbonate system such as [CO$_3^{2-}$], [H$^+$] and [OH$^-$] are calculated from the transport-buffered [HCO$_3^-$] and equilibrium constants (see Appendix~\ref{app:Generalized}, Figure~\ref{fig:methodology} for the flow chart of methodology). Rewriting [CO$_3^{2-}$] in terms of [HCO$_3^-$], $P_{\mathrm{CO}_2}$ and equilibrium constants in Equation~(\ref{eq:ALK}),

\begin{equation} \label{eq:ALK2}
    A = [\mathrm{HCO}_3^-] + 2 \frac{K_{\mathrm{Car}} [\mathrm{HCO}_3^-]^2}{K_{\mathrm{Bic}} P_{\mathrm{CO}_2}}.
\end{equation}

Figure~\ref{fig:CA_peri}(a) shows the carbonate alkalinity and the solution pH for the generalized model of peridotite weathering as a function of $P_{\mathrm{CO}_2}$ at a constant surface temperature ($T$ = 288~K). For $P_{\mathrm{CO}_2} < 1$~$\mu$bar, [HCO$_3^-$], [CO$_3^{2-}$] and pH in Figure~\ref{fig:CA_peri}(a) exhibit the same behavior as in Figure~\ref{fig:CAeq_peri}(a) because $q < D_w$, reiterating that chemical equilibrium calculations are valid in the thermodynamic regime resulting in the maximum weathering flux. The maximum ([HCO$_3^-$]$_{\mathrm{eq}}$ and $A_{\mathrm{eq}}$, Figure~\ref{fig:CAeq_peri}b) and generalized ([HCO$_3^-$] and $A$, Figure~\ref{fig:CA_peri}b) weathering fluxes diverge from each other beyond the thermodynamic to kinetic regime transition that occurs at $P_{\mathrm{CO}_2} \sim 1$~$\mu$bar. The choice of $q$ determines this transition. At a higher $q$, the regime transition would shift to a lower $P_{\mathrm{CO}_2}$ and vice versa. For $P_{\mathrm{CO}_2} > 1$~$\mu$bar, [HCO$_3^-$] becomes independent of $P_{\mathrm{CO}_2}$ in the kinetic weathering regime. This is because the kinetic rate coefficient of the slowest reaction (fayalite dissolution) is constant at a fixed $T$ and does not vary when pH is basic (see Appendix~\ref{app:data}). However, the results shown in Figure~\ref{fig:CA_peri} are expected to change when temperature is not held constant and depends on $P_{\mathrm{CO}_2}$ where the relation between the two is given by a climate model (Section~\ref{sec:methodsClimate}).

In the kinetic weathering regime, [CO$_3^{2-}$] decreases with $P_{\mathrm{CO}_2}$ (Figure~\ref{fig:CA_peri}a) because [CO$_3^{2-}$] is directly proportional to [HCO$_3^-$] and inversely proportional to $P_{\mathrm{CO}_2}$ (Equation~\ref{eq:ALK2}). Since [HCO$_3^-$] is constant in the kinetic regime, [CO$_3^{2-}$] shows a strong decrease because of its sole dependence on $P_{\mathrm{CO}_2}$. We follow the same procedure as for peridotite to compute the generalized concentrations for the weathering of basalt, granite or individual minerals and find that lithology strongly impacts the occurrence of weathering regimes (Appendix~\ref{app:Generalized}, Fig.~\ref{fig:CA_mine}).

\subsection{Climate Model} \label{sec:methodsClimate}

The greenhouse effect of CO$_2$ exerts a strong control on the planetary surface temperature. A climate model enables one to express the surface temperature $T$ as a function of the CO$_2$ partial pressure $P_{\mathrm{CO}_2}$ for a given planetary albedo $\alpha$ and top-of-atmosphere stellar flux $S$. Such a climate model is essential to assess the role of climate in weathering on temperate planets. \texttt{CHILI} provides the functionality to couple $T$ and $P_{\mathrm{CO}_2}$ using any climate model. Previous studies provide formulations of climate models \citep[e.g.,][]{1981JGR....86.9776W,1993Icar..101..108K}. Recently, \citet{2013ApJ...765..131K,2014ApJ...787L..29K} performed 1D radiative-convective calculations to obtain $T$ as a function of $P_{\mathrm{CO}_2}$, $\alpha$ and $S$. Studies such as \citet{2016ApJ...827..120H} and \citet{2019ApJ...875....7K} provide fitting functions to the models of \citet{2013ApJ...765..131K,2014ApJ...787L..29K}. We use the fitting function provided by \citet{2019ApJ...875....7K} to couple $T$ in the range 150$-$350~K with $P_{\mathrm{CO}_2}$ in the range $10^{-5}-10$~bar. For $\alpha = 0.3$ (present-day albedo of Earth) and $S = 1360$~W~m$^{-2}$ (present-day solar flux), the \citet{2019ApJ...875....7K} fitting function results in $T$ between 280$-$350~K for $P_{\mathrm{CO}_2}$ between 10~$\mu$bar and 0.5~bar (see Appendix~\ref{app:climate} for details).

\section{Weathering on Temperate Planets} \label{sec:results}

\subsection{Maximum Weathering for Various Lithologies} \label{sec:resultsThermo}

\begin{figure}[!ht]
  \centering
  \includegraphics[width=\linewidth]{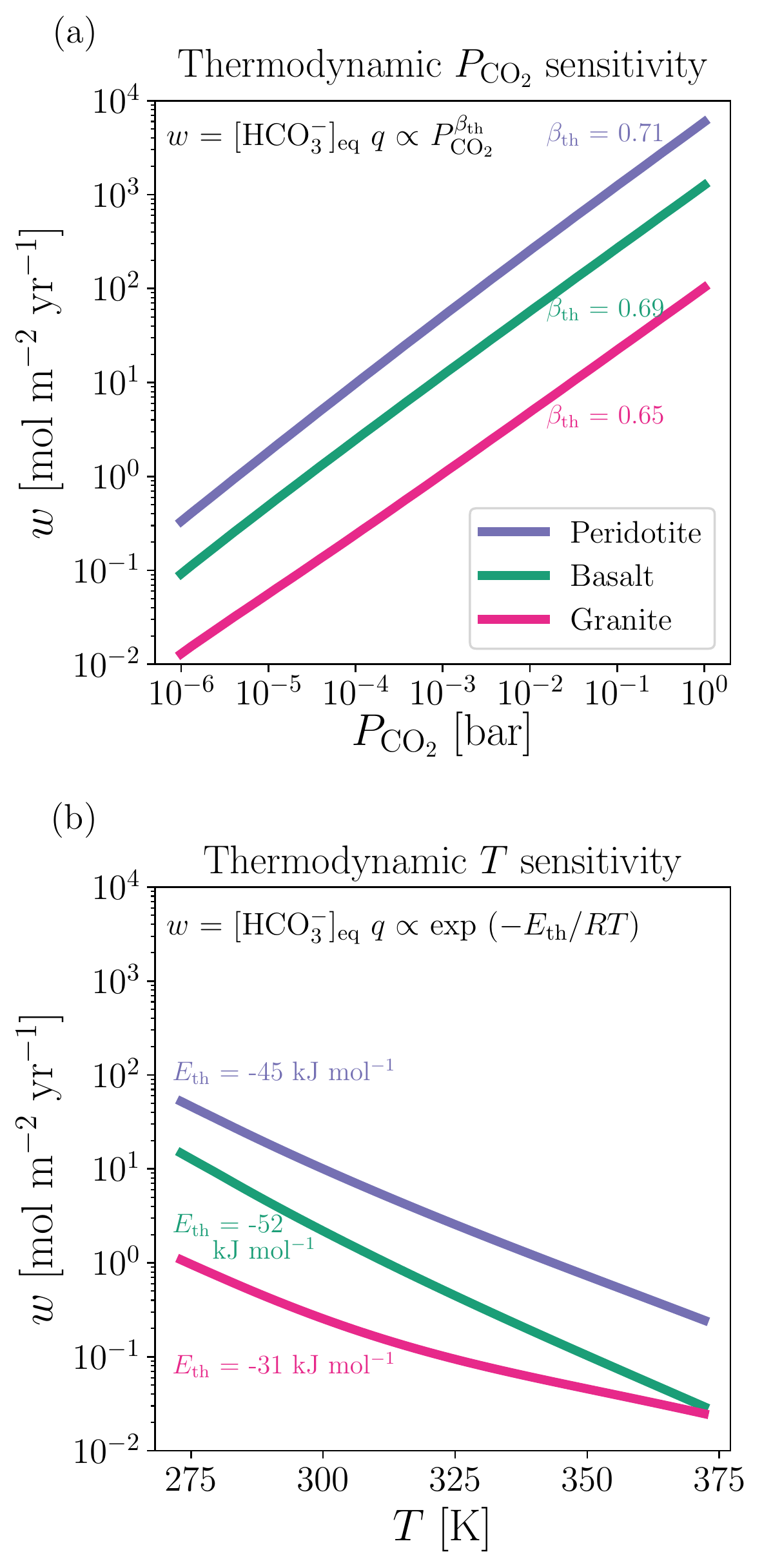}
  \caption{ Sensitivity of thermodynamic [HCO$_3^-$]$_{\mathrm{eq}}$ flux (maximum weathering) of rocks to (a) $P_{\mathrm{CO}_2}$ at $T = 288$~K and (b) $T$ at $P_{\mathrm{CO}_2} = 280$~$\mu$bar, at present-day mean runoff $q$ = 0.3~m~yr$^{-1}$. The labeled fitting parameters are obtained by fitting the thermodynamic weathering flux to the kinetic weathering expression (Equation~\ref{eq:w_pCO2}).}
  \label{fig:w_rock_therm}
\end{figure}

The thermodynamic solute concentrations provide an upper limit to weathering (e.g., peridotite weathering, Figure~\ref{fig:CAeq_peri}). To evaluate the case of maximum weathering on temperate planets, in Figure~\ref{fig:w_rock_therm}, we provide the [HCO$_3^-$]$_{\mathrm{eq}}$ weathering flux of three common rocks (basalt, peridotite, granite) as a function of climate properties, $P_{\mathrm{CO}_2}$ and $T$, at present-day mean runoff of 0.3~m~yr$^{-1}$. The dependence of weathering on total surface pressure is negligible (see Appendix~\ref{app:Generalized}). The [HCO$_3^-$]$_{\mathrm{eq}}$ weathering flux for all rocks increases monotonically with $P_{\mathrm{CO}_2}$ at a constant temperature because weathering intensifies as $P_{\mathrm{CO}_2}$ increases (Figure~\ref{fig:w_rock_therm}a). This is a direct consequence of the calculations of solute concentrations at chemical equilibrium (Section~\ref{sec:methodsMaximum}). Table~\ref{tab:therm_fit} gives fitting parameters of the thermodynamic weathering flux of [HCO$_3^-$]$_{\mathrm{eq}}$ to the kinetic weathering expression (Equation~\ref{eq:w_pCO2}) for silicate rocks and minerals considered in this study. Such a fit, although not the best approximation of the calculated values, provides a way to compare the sensitivity of thermodynamic weathering to climate properties with studies assuming kinetic weathering \citep[e.g.,][]{1981JGR....86.9776W,2001JGR...106.1373S,2015ApJ...812...36F}.

Fluid-rock reactions produce both monovalent (e.g., K$^+$, Na$^+$) and divalent cations (e.g., Ca$^{2+}$, Mg$^{2+}$). The sensitivity of [HCO$_3^-$]$_{\mathrm{eq}}$ flux to $P_{\mathrm{CO}_2}$ depends on the capacity of a rock to produce divalent cations. More divalent cations in the solution require more HCO$_3^-$ and CO$_3^{2-}$ ions to balance the charges. Consequently, the higher the fraction of divalent cations in a rock, the higher the thermodynamic $P_{\mathrm{CO}_2}$ sensitivity ($\beta_{\rm th}$). For instance, peridotite, which produces only divalent cations, exhibits the highest $\beta_{\rm th}$ among the three rocks (Figure~\ref{fig:CAeq_peri}). Granite produces more monovalent cations than peridotite and basalt, and therefore has the lowest $\beta_{\rm th}$. This effect has been discussed for feldspar minerals by \citet{2016GeCoA.190..265I,2018E&PSL.485..111W}. 

\begin{deluxetable*}{lcccccc}
\tablecaption{Parameters for fitting thermodynamic weathering flux to kinetic weathering expression (Equation~\ref{eq:w_pCO2}). \label{tab:therm_fit}}
\tabletypesize{\small}
\tablehead{
\colhead{Rock/Mineral} & \multicolumn{3}{c}{Halloysite or no secondary mineral} & \multicolumn{3}{c}{Kaolinite as a secondary mineral} \\
\colhead{Rock/Mineral} & \colhead{$\beta_{\rm th}$} & \colhead{$E_{\rm th}$} & \colhead{$w_0$} & \colhead{$\beta_{\rm th}$} & \colhead{$E_{\rm th}$} & \colhead{$w_0$} \\
& & (kJ mol$^{-1}$) & (mol~m$^{-2}$~yr$^{-1}$) & & (kJ mol$^{-1}$) & (mol~m$^{-2}$~yr$^{-1}$)
}
\startdata
$Rocks$ \\
Peridotite    & 0.71 & $-$45 & 20.5  \\
Basalt        & 0.69 & $-$52 & 5.0  & 0.65 & $-$57 & 125  \\
Granite       & 0.65 & $-$31 & 0.5  & 0.54 & $-$26 & 1.5  \\
\hline
$Pyroxenes$ \\
Wollastonite  & 0.53 & $-$27 & 2.74  \\
Enstatite     & 0.52 & $-$29 & 0.99  \\
Ferrosilite   & 0.51 & $-$24 & 0.10  \\
\hline
$Olivines$ \\
Forsterite    & 0.59 & $-$19 & 2.99  \\
Fayalite      & 0.58 & $-$16 & 0.70  \\
\hline
$Feldspars$ \\
Anorthite     & 0.69 & $-$52.0 & 5.0  & 0.72 & $-$55.0 & 50.1  \\
Albite        & 0.26 & $+$2.0  & 0.04 & 0.26 & $+$0.04 & 0.12 \\
K-feldspar    & 0.26 & $+$8.0  & 0.01 & 0.26 & $+$5.9  & 0.03 \\
\hline
$Micas$ \\
Muscovite     & 0.50 & $-$7 & 0.001 & 0.50 & $-$18 & 0.076 \\
Phlogopite    & 0.54 & $-$27 & 0.21 & 0.55 & $-$27 & 0.28  \\
Annite        & 0.54 & $-$23 & 0.04 & 0.54 & $-$23 & 0.057 \\
\hline
$Amphiboles$ \\
Anthophyllite & 0.50 & $-$8 & 0.001 \\
Grunerite     & 0.50 & $-$8 & 0.001 \\
\enddata
\tablecomments{ $w_0$ is not a fitting parameter but the value of $w$ at $P_{\mathrm{CO}_2}$ = 280~$\mu$bar, $T$ = 288~K and $q$ = 0.3~m~yr$^{-1}$. No secondary minerals are assumed to be produced during the weathering of peridotite, pyroxenes, olivines and amphiboles. }
\end{deluxetable*}

The $\beta_{\rm th}$ values for endmember minerals within the same mineral group (pyroxene, olivine, mica and amphibole), with the exception of feldspars, are similar to each other (Table~\ref{tab:therm_fit}). This result is again attributed to the presence of monovalent or divalent cations. The divalent cation-producing pyroxene, olivine, mica and amphibole endmembers exhibit $\beta_{\rm th} \sim 0.5$ and monovalent cation-producing albite and K-feldspar exhibit $\beta_{\rm th} \sim 0.25$. The deviation from these ideal values of 0.5 and 0.25 is a result of the simultaneous consideration of the mineral dissolution reaction and the water-bicarbonate reactions. For instance, \citet{2018E&PSL.485..111W} show that reaction stoichiometry controls $\beta_{\rm th}$ values by considering dissolution reactions of individual feldspar minerals and find $\beta_{\rm th} = 0.25$ for albite and K-feldspar. However, we find that the presence of ions produced by the water-bicarbonate system makes $\beta_{\rm th}$ dependent on equilibrium constants of all reactions considered in addition to stoichiometry and hence $\beta_{\rm th} = 0.26$ for albite and K-feldspar (Table~\ref{tab:therm_fit}). The $\beta_{\rm th}$ value of the divalent cation-producing anorthite is 0.69, considerably higher than other divalent cation-producing minerals. This is again attributed to reaction stoichiometry as demonstrated by \citet{2018E&PSL.485..111W}, although their study results in a slightly smaller $\beta_{\rm th} = 0.67$ because of neglecting reactions in the water-bicarbonate system.

The weathering flux produced by the rock with a higher $\beta_{\rm th}$ is higher (Figure~\ref{fig:CAeq_peri}a). However, the choice of secondary minerals produced during weathering influences this result. For instance, if kaolinite is considered instead of halloysite, the weathering flux of basalt is higher by more than an order of magnitude, which is even higher than that of peridotite (see Table~\ref{tab:therm_fit}). The effect of a secondary mineral on the weathering flux of granite is smaller than on basalt. This difference is attributed to the impact of the secondary mineral on weathering reactions of feldspar endmember minerals. Anorthite, a constituent of basalt in our model, exhibits an order of magnitude increase in the weathering flux when kaolinite is used instead of halloysite (Table~\ref{tab:therm_fit}). Whereas, the weathering flux of albite and K-feldspar, constituents of granite in our model, increases by a factor of three for the same change. Moreover, the non-consideration of secondary minerals for peridotite weathering results in high weathering flux at high $P_{\rm CO_2}$, which in reality may be lower \citep[e.g.,][]{2019E&PSL.52415718K}. Thus, the choice of secondary mineral strongly affects $w$ and $\beta_{\rm th}$ of rocks. However, this effect is small for the weathering of individual minerals (Table~\ref{tab:therm_fit}). 

Contrary to observations of the increase in kinetic weathering flux with temperature \citep[e.g.,][]{1981JGR....86.9776W}, the thermodynamic weathering flux for rocks decreases with temperature at a fixed $P_{\mathrm{CO}_2}$ (Figure~\ref{fig:w_rock_therm}b). The fitting parameter, the activation energy of thermodynamic weathering provides a scaling relation between weathering and temperature. Unlike kinetic weathering, $E_{\rm th}$ is negative for weathering of all rocks and minerals except albite and K-feldspar (Table~\ref{tab:therm_fit}). This result is a consequence of the decrease in equilibrium constants as a function of temperature for all mineral dissolution reactions except those of albite and K-feldspar (see Appendix~\ref{app:data}). The choice of secondary mineral influences the magnitude of $E_{\rm th}$ but not its sign. \citet{2018E&PSL.485..111W} observe this effect for plagioclase feldspars which contain anorthite in addition to albite. This effect is discussed further in Section~\ref{sec:disTemp}. 

The fitting parameters provided in Table~\ref{tab:therm_fit} should be used with caution as $\beta_{\rm th}$ depends on $T$ and $E_{\rm th}$ depends on $P_{\mathrm{CO}_2}$. For example, $\beta_{\rm th}$ for peridotite at $P_{\mathrm{CO}_2}$ = 280~$\mu$bar varies between 0.71 at $T$ = 273~K and 0.77 at $T$ = 373~K. Whereas, $E_{\rm th}$ for peridotite at $T$ = 288~K varies between $-$55~kJ~mol$^{-1}$ at $P_{\mathrm{CO}_2}$ = 1~$\mu$bar and $-$46~kJ~mol$^{-1}$ at $P_{\mathrm{CO}_2}$ = 1
~bar. This provides further reason to calculate thermodynamic concentrations consistently by considering all necessary reactions simultaneously, as formulated in this study.

\subsection{Climate Sensitivity of Peridotite Weathering} \label{sec:resultsClimate}

\begin{figure*}[!ht]
  \centering
  \medskip
  \includegraphics[width=\textwidth]{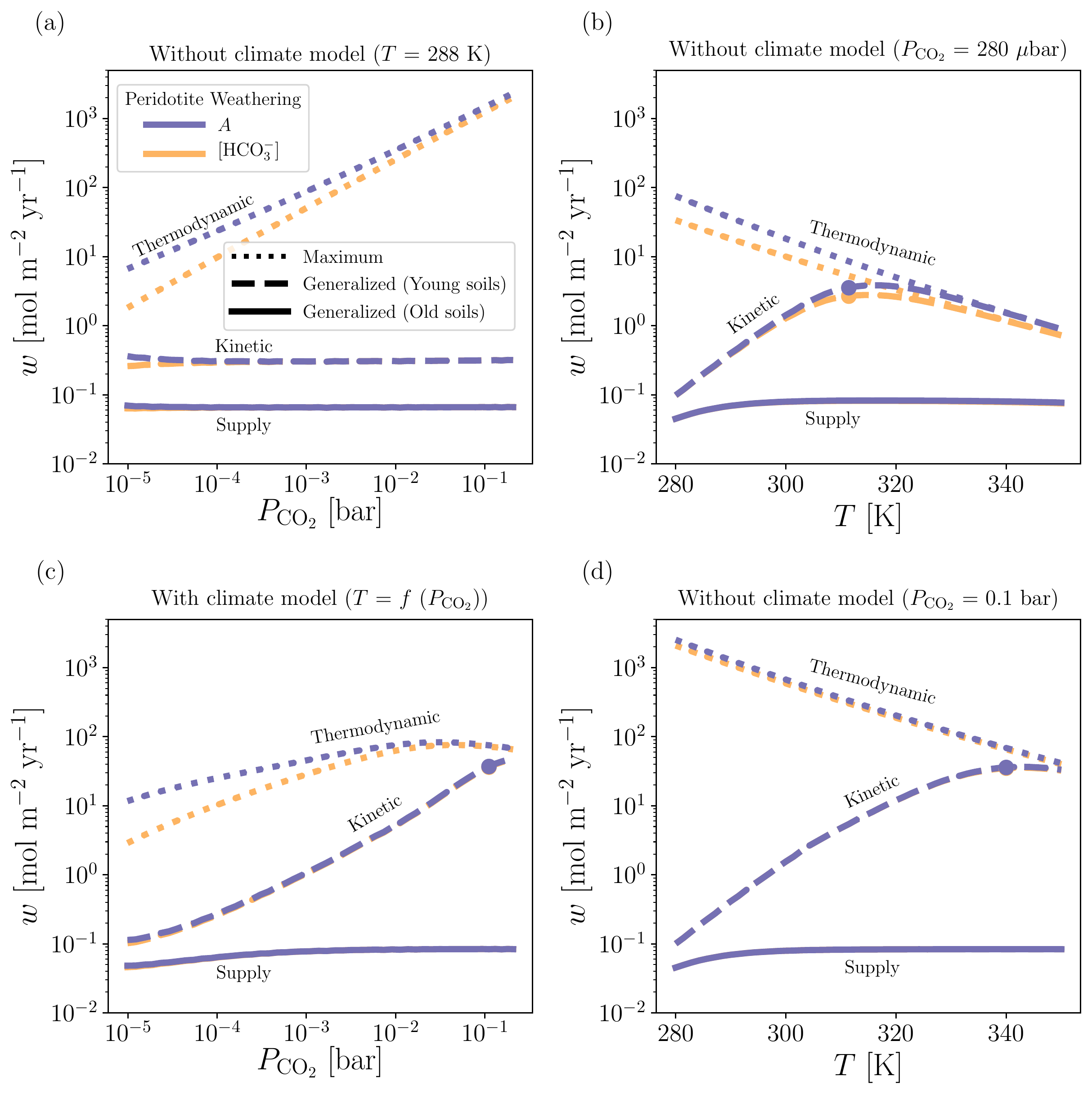}
  \caption{ Sensitivity of peridotite weathering flux of the maximum, generalized young soils ($t_{\mathrm{s}} = 0$) and generalized old soils ($t_{\mathrm{s}} = 100$~kyr) models to climate properties at present-day mean runoff $q$ = 0.3~m~yr$^{-1}$ (other parameters take reference values). (a) Sensitivity to $P_{\mathrm{CO}_2}$ at $T = 288$~K. (b) Sensitivity to $T$ at $P_{\mathrm{CO}_2}$ = 280~$\mu$bar. (c) Sensitivity to $P_{\mathrm{CO}_2}$ at $T = f\,(P_{\mathrm{CO}_2})$ given by the climate model (Section~\ref{sec:methodsClimate}). (d) Sensitivity to $T$ at $P_{\mathrm{CO}_2}$ = 10$^5$~$\mu$bar. Colored disks denote the transition between kinetic and thermodynamic regimes. }
  \label{fig:w_peri_clim}
\end{figure*}

Figure~\ref{fig:w_peri_clim} shows the sensitivity of both maximum and generalized weathering fluxes of peridotite to CO$_2$ partial pressure and surface temperature. The maximum weathering model gives an upper limit to the weathering flux. The generalized weathering flux is either equal to or smaller than the maximum weathering flux depending on if the generalized model encounters the thermodynamic regime or not. The generalized weathering flux in the kinetic regime is lower than that in the thermodynamic regime and higher than that in the supply regime. Since the weathering flux in the thermodynamic regime is an upper limit to weathering, the kinetic flux cannot exceed the thermodynamic flux. To isolate the effect of $P_{\mathrm{CO}_2}$ on weathering, we present the peridotite weathering flux as a function of $P_{\mathrm{CO}_2}$ at a fixed surface temperature (Figure~\ref{fig:w_peri_clim}a). The sensitivity of weathering to temperature is demonstrated at two $P_{\mathrm{CO}_2}$ values (Figure~\ref{fig:w_peri_clim}b,d). In reality, $T$ depends on $P_{\mathrm{CO}_2}$ due to the greenhouse effect of CO$_2$ which is normally modeled using a climate model (Section~\ref{sec:methodsClimate}). The strength of the coupling between $T$ and $P_{\mathrm{CO}_2}$ is sensitive to numerous parameters including incident solar flux and planetary albedo that are fixed to present-day values (Table~\ref{tab:f_params}). Since the solar flux is held constant, the coupling between $T$ and $P_{\mathrm{CO}_2}$ becomes stronger than that during Earth's early history where the solar flux dropped to about 70\% of its present-day value. Figure~\ref{fig:w_peri_clim}(c) demonstrates peridotite weathering under the limit of strong coupling between $T$ and $P_{\mathrm{CO}_2}$.

The maximum (thermodynamic) carbonate alkalinity and bicarbonate ion fluxes increase monotonically with $P_{\mathrm{CO}_2}$ at $T = 288$~K (Figure~\ref{fig:w_peri_clim}a). This monotonic behavior is a direct result of chemical equilibrium calculations with $P_{\mathrm{CO}_2}$ as a free parameter at a fixed temperature. The difference between the maximum $A$ and [HCO$_3^-$] fluxes at low $P_{\mathrm{CO}_2}$ is due to the excess contribution of [CO$_3^{2-}$] to $A$ when the solution pH is basic (see Figure~\ref{fig:CAeq_peri}). The thermodynamic weathering flux decreases with $T$ at a fixed $P_{\mathrm{CO}_2}$ (Figure~\ref{fig:w_peri_clim}b,d). As explained in Section~\ref{sec:resultsThermo} and later discussed in Section~\ref{sec:disTemp}, this decrease is a result of the decrease in equilibrium constants of mineral dissolution reactions as a function of temperature (see Appendix~\ref{app:data}). The thermodynamic flux at $P_{\mathrm{CO}_2}$ = 0.1~bar is higher than at $P_{\mathrm{CO}_2}$ = 280~$\mu$bar by about a factor of 30. Unlike at $P_{\mathrm{CO}_2}$ = 280~$\mu$bar, at $P_{\mathrm{CO}_2}$ = 0.1~bar there is a negligible difference between the thermodynamic $A$ and [HCO$_3^-$] fluxes. When $T$ depends on $P_{\mathrm{CO}_2}$ via a climate model, the behavior of thermodynamic weathering as a function of $P_{\mathrm{CO}_2}$ depends on the trade-off between the individual effects of $T$ and $P_{\mathrm{CO}_2}$ (Figure~\ref{fig:w_peri_clim}c). Up to $P_{\mathrm{CO}_2} = 0.01$~bar, the thermodynamic fluxes increase with $P_{\mathrm{CO}_2}$ because the $P_{\mathrm{CO}_2}$ effect dominates. Whereas, beyond $P_{\mathrm{CO}_2} = 0.05$~bar, there is a small decrease in the thermodynamic fluxes because the effect of $T$ takes over.

When soils are young (soil age is zero, $t_{\mathrm{s}} = 0$), the generalized [HCO$_3^-$] flux at $T = 288$~K, which is in the kinetic regime, is almost constant for the given $P_{\mathrm{CO}_2}$ range (Figure~\ref{fig:w_peri_clim}a). The kinetic rate coefficient depends on $T$ as well as the solution pH which in turn depends on $P_{\mathrm{CO}_2}$ (Section~\ref{sec:methodsGeneralized}). However, the kinetic weathering flux is independent of $P_{\mathrm{CO}_2}$ at a fixed temperature because $k_{\mathrm{eff}}$ of peridotite is independent of $P_{\mathrm{CO}_2}$ when the solution pH is basic (Figure~\ref{fig:w_peri_clim}a). This $k_{\mathrm{eff}}$ is determined by the fayalite dissolution reaction as it is rate limiting among the considered mineral dissolution reactions for peridotite (see Appendix~\ref{app:data}). As a function of $T$ at constant $P_{\mathrm{CO}_2}$, kinetic weathering exhibits a strong dependence on temperature as seen in Figure~\ref{fig:w_peri_clim}(b,d). For the $P_{\mathrm{CO}_2}$ = 280~$\mu$bar case, the generalized ($t_{\mathrm{s}} = 0$) model switches from the kinetic to thermodynamic regime at $\sim$310~K where the limit of maximum weathering is encountered. For the $P_{\mathrm{CO}_2}$ = 0.1~bar case, this transition temperature increases to $\sim$340~K. The transition from kinetic to thermodynamic regime occurs at high $T$ and low $P_{\mathrm{CO}_2}$. The sequence of this transition is in contrast to the switch from the thermodynamic to kinetic regime that occurs at $P_{\mathrm{CO}_2} = 1$~$\mu$bar as a function of $P_{\mathrm{CO}_2}$ (see Figure~\ref{fig:CA_peri}). If the climate model is invoked, the generalized ($t_{\mathrm{s}} = 0$) weathering flux increases steeply with $P_{\mathrm{CO}_2}$ and encounters the thermodynamic regime at about $P_{\mathrm{CO}_2} = 0.1$~bar, a result of the kinetic temperature dependence (Figure~\ref{fig:w_peri_clim}c). When $T$ is strongly coupled to $P_{\mathrm{CO}_2}$, the generalized model may switch from the thermodynamic to kinetic regime at low $P_{\mathrm{CO}_2}$ and from the kinetic regime back to the thermodynamic regime at high $P_{\mathrm{CO}_2}$. 

When soils are old (present-day characteristic soil age, $t_{\mathrm{s}} = 100$~kyr), the [HCO$_3^-$] flux at $T = 288$~K as a function of $P_{\mathrm{CO}_2}$ is constant and in the supply regime (Figure~\ref{fig:w_peri_clim}a). This regime is limited by the supply of fresh rocks and the influence of chemical kinetics on weathering is small compared to when soils are young ($t_{\mathrm{s}}$ is small). This regime is almost independent of $T$ (Figure~\ref{fig:w_peri_clim}b,d). Consequently, unlike the kinetic weathering flux, the supply-limited weathering flux is independent of $P_{\mathrm{CO}_2}$ when $T$-dependence via the climate model is invoked (Figure~\ref{fig:w_peri_clim}c). In the supply regime, the weathering flux depends on the age of soils which is held constant. There is almost no difference between the generalized ($t_{\mathrm{s}} = 100$~kyr) $A$ and [HCO$_3^-$] fluxes between Figure~\ref{fig:w_peri_clim}(a) and (c) because the inclusion of the $T$ effect via the climate model is not strong enough to escape from the supply regime. In contrast, for the generalized ($t_{\mathrm{s}} = 0$) model, invoking the climate model increases the weathering fluxes in the kinetic regime enough to enter the thermodynamic regime at about $P_{\mathrm{CO}_2} = 0.1$~bar. 

\subsection{Endmember Cases of Continental and Seafloor Weathering} \label{sec:resultsContSeaf}

\begin{figure*}[!ht]
  \centering
  \medskip
  \includegraphics[width=\textwidth]{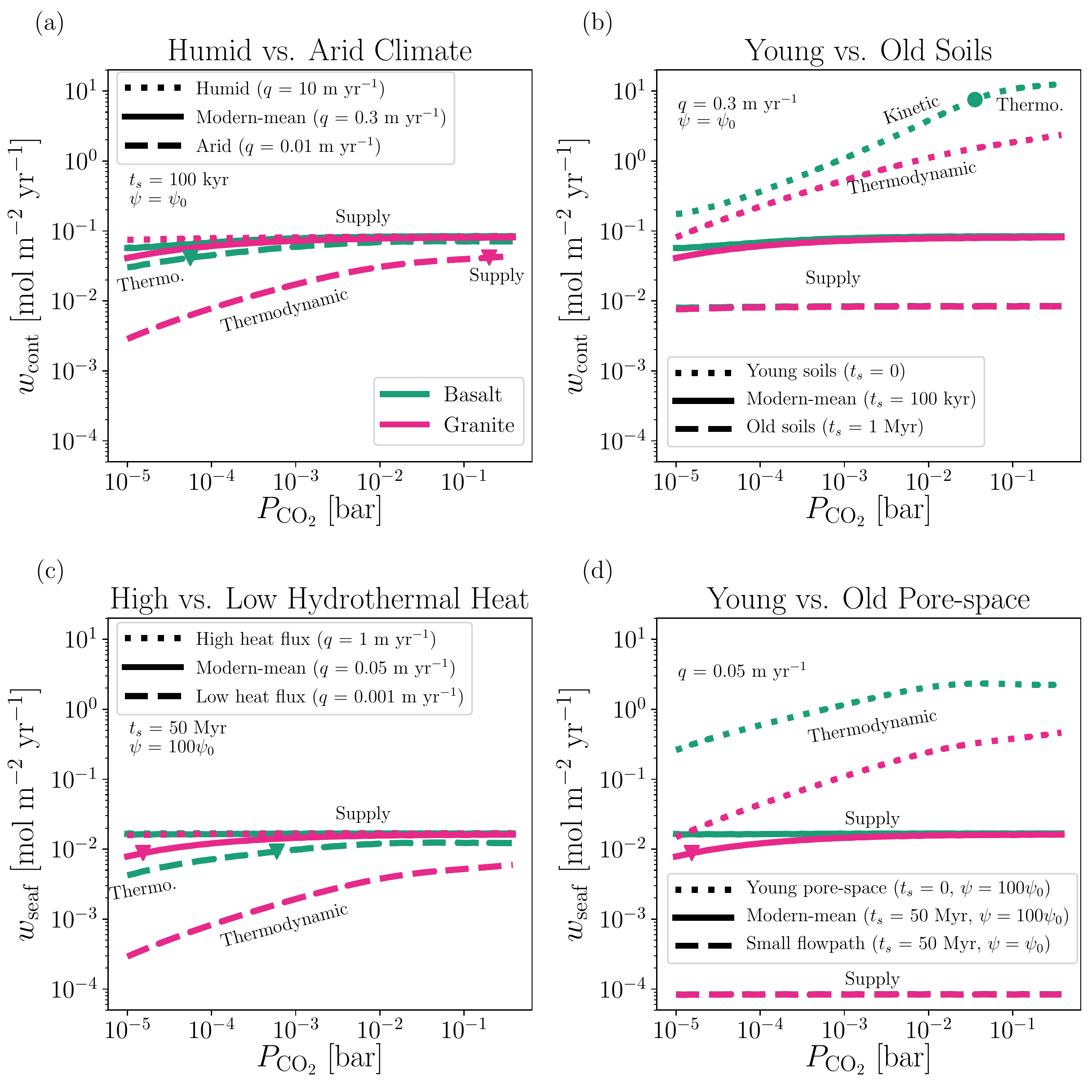}
  \caption{ Carbonate alkalinity flux on continents (a,b) and seafloor (c,d) for basalt and granite as a function of $P_{\mathrm{CO}_{2}}$, where $T = f(P_{\mathrm{CO}_{2}})$ is given by the climate model (Section~\ref{sec:methodsClimate}). Colored disks denote the transition between kinetic and thermodynamic regimes, and inverted triangles denote the transition between thermodynamic and supply regimes. }
  \label{fig:w_xCO2_endmembers}
\end{figure*}

We apply the generalized weathering model to both continental (Figure~\ref{fig:w_xCO2_endmembers}a,b) and seafloor silicate weathering (Figure~\ref{fig:w_xCO2_endmembers}c,d) for diverse cases that may represent weathering scenarios on temperate rocky exoplanets. The climate model is used to couple $T$ with $P_{\mathrm{CO}_2}$ by fixing the stellar flux and planetary albedo to present-day Earth values (Section~\ref{sec:methodsClimate}). This implies a strong coupling between $T$ and $P_{\mathrm{CO}_2}$ such that $T$ varies from 280~K to 350~K when $P_{\mathrm{CO}_2}$ varies from 10~$\mu$bar to 0.5~bar. The lithology and pore-space properties of continents and seafloor on present-day Earth are different from each other. This presents an opportunity to test the generalized weathering model in an extended parameter-space beyond applications to continental weathering. We show the results for basalt and granite in Figure~\ref{fig:w_xCO2_endmembers}. The sensitivity of the carbonate alkalinity weathering flux to $P_{\mathrm{CO}_2}$ is a complex function of climate, fluid flow rate, rock and pore-space properties. This result implies that the weathering flux cannot be simply approximated by Equation~(\ref{eq:w_pCO2}) assuming either kinetic weathering \citep[e.g.,][]{1981JGR....86.9776W,1983AmJS..283..641B,2001JGR...106.1373S,2015ApJ...812...36F} or thermodynamic weathering \citep[][and Section~3.1, this study]{2018E&PSL.485..111W}. 

In Figure~\ref{fig:w_xCO2_endmembers}(a), the continental $A$ flux is calculated for three values of runoff that are representative of arid, modern-mean and humid climates. \citet{1999ChGeo.159....3G} report regional variations in the present-day runoff from 0.01$-$3~m~yr$^{-1}$. Since the choice of humid runoff in our model is higher than the arid runoff by three orders of magnitude, the corresponding weathering fluxes should differ by the same amount if the weathering fluxes are in the thermodynamic regime. However, the differences are smaller in the given $P_{\mathrm{CO}_2}$ range. This is because non-thermodynamic regimes exhibit smaller weathering fluxes than the thermodynamic regime. For example, up to $P_{\mathrm{CO}_2}$ = 0.2~bar, the arid model of granite is in the thermodynamic regime (upper limit for this model), whereas the humid model of granite is in the supply regime (lower limit for this model). Thus, the differences between the two models are smaller than the maximum possible differences. For more arid climates, lithology plays an even more important role as the weathering becomes thermodynamically-limited for the whole $P_{\mathrm{CO}_2}$ range. In contrast, the supply regime is largely independent of lithology. For this reason, lithology has negligible impact on the modern-mean and humid cases. 

The age of soils is a key parameter that determines if the weathering is limited by reaction kinetics or limited by the supply of fresh rocks. Figure~\ref{fig:w_xCO2_endmembers}(b) shows that lithology has no influence on the weathering flux of old soils ($t_{\mathrm{s}} = 100$~kyr and $t_{\mathrm{s}} = 1$~Myr) as opposed to young soils ($t_{\mathrm{s}} = 0$). The old soil models are in the supply regimes. Granite and basalt in young soils are in different weathering regimes. Granite is in the thermodynamic regime for the whole $P_{\mathrm{CO}_2}$ range because the net reaction rate ($D_w$ = 1$-$3~m~yr$^{-1}$) is higher than the fluid flow rate ($q = 0.3$~m~yr$^{-1}$). In contrast, $D_w$ (0.05$-$4.5~m~yr$^{-1}$) of basalt is lower than $q = 0.3$~m~yr$^{-1}$ up to $P_{\mathrm{CO}_2} = 0.5$~bar, implying a transition from kinetic to thermodynamic regime above this $P_{\mathrm{CO}_2}$. 

On Earth, the continental and seafloor pore-space differ in terms of pore-space properties besides the differences in lithology. The dimensionless pore-space parameter $\psi$ depends on the flowpath length $L$ that is normally assumed to be of the order of the regolith thickness (Section~\ref{sec:methodsGeneralized}). Since the thickness of the oceanic crust where seafloor weathering occurs is of the order of 100~m \citep[e.g.,][]{1986JGR....9110309A,2018AREPS..46...21C}, we assume $L = 100$~m and $\psi = 100 \psi_0$. Moreover, the average age of the oceanic crust on Earth at present day is approximately equal to 50~Myr, about 500 times the characteristic age of continental soils. 

In Figure~\ref{fig:w_xCO2_endmembers}(d), we compare the seafloor $A$ flux for characteristic seafloor values of $t_{\mathrm{s}} = 50$~Myr and $\psi = 100 \psi_0$ with two models, one with $t_{\mathrm{s}} = 0$ and another with $\psi = \psi_0$. The present-day seafloor $A$ flux is in the supply regime making it independent of lithology. The supply-limited fluxes of seafloor weathering are smaller than those of continental weathering by a factor of 5 because $w \propto \frac{\psi}{t_{\mathrm{s}}}$ in this regime. When $\psi$ is lowered from $100 \psi_0$ to $\psi_0$, the supply-limited weathering flux decreases by two orders of magnitude at a much lower $P_{\mathrm{CO}_2}$. For the young seafloor pore-space case, the $A$ fluxes of basalt and granite are in the thermodynamic regime, and consequently the impact of lithology is pronounced. Compared to the $t_{\mathrm{s}} = 0$ basalt model in Figure~\ref{fig:w_xCO2_endmembers}(b) that is in the kinetic regime, the $t_{\mathrm{s}} = 0$ basalt model in Figure~\ref{fig:w_xCO2_endmembers}(d) is in the thermodynamic regime. This difference arises due to the choice of $\psi$ that makes the net reaction rate ($D_w$) higher than the fluid flow rate, pushing basalt into the thermodynamic regime. Being in the thermodynamic regime, chemical equilibrium controls the weathering flux of basalt. At high $P_{\mathrm{CO}_2}$, the thermodynamic weathering flux of basalt decreases slightly as the decreasing effect of $T$ takes over the increasing effect of $P_{\mathrm{CO}_2}$. This decreasing effect of $T$ is due to the decrease in equilibrium constants of weathering reactions as a function of temperature (Appendix~\ref{app:data}). The effect of $T$ on the weathering flux of granite is small and hence there is no net decrease in the weathering flux at high $P_{\mathrm{CO}_2}$. 

In Figure~\ref{fig:w_xCO2_endmembers}(c), we calculate the seafloor $A$ flux for three hydrothermal fluid flow rates, where the two extreme $q$ values differ by three orders of magnitude, similar to the strategy in Figure~\ref{fig:w_xCO2_endmembers}(a). Since the fluid flow rates are directly proportional to the hydrothermal heat flux \citep{1994JGR....99.3081S,2013GGG....14.1771C}, a variation in the hydrothermal heat flux implies a variation in the fluid flow rate. Depending on the age of oceanic crust, present-day non-porosity-corrected fluid flow rates are observed between 0.001$-$0.7~m~yr$^{-1}$ \citep{2003E&PSL.216..565J}. The three cases of seafloor weathering shown in Figure~\ref{fig:w_xCO2_endmembers}(c) are broadly similar to their continental counterparts. The modern-mean and high fluid flow rates result in lithology-independent weathering fluxes that are in the supply-limited regimes. The low hydrothermal fluid flow rate causes the granite model to be in the thermodynamic regime for the full $P_{\mathrm{CO}_2}$ range, although the basalt model transition from the thermodynamic to supply regime at $P_{\mathrm{CO}_2} = 0.6$~mbar.

\section{Discussion and Implications} \label{sec:dis}

\subsection{Weathering Regimes and the Role of Lithology} \label{sec:disRegimes}

\begin{figure*}[!ht]
  \centering
  \includegraphics[width=\linewidth]{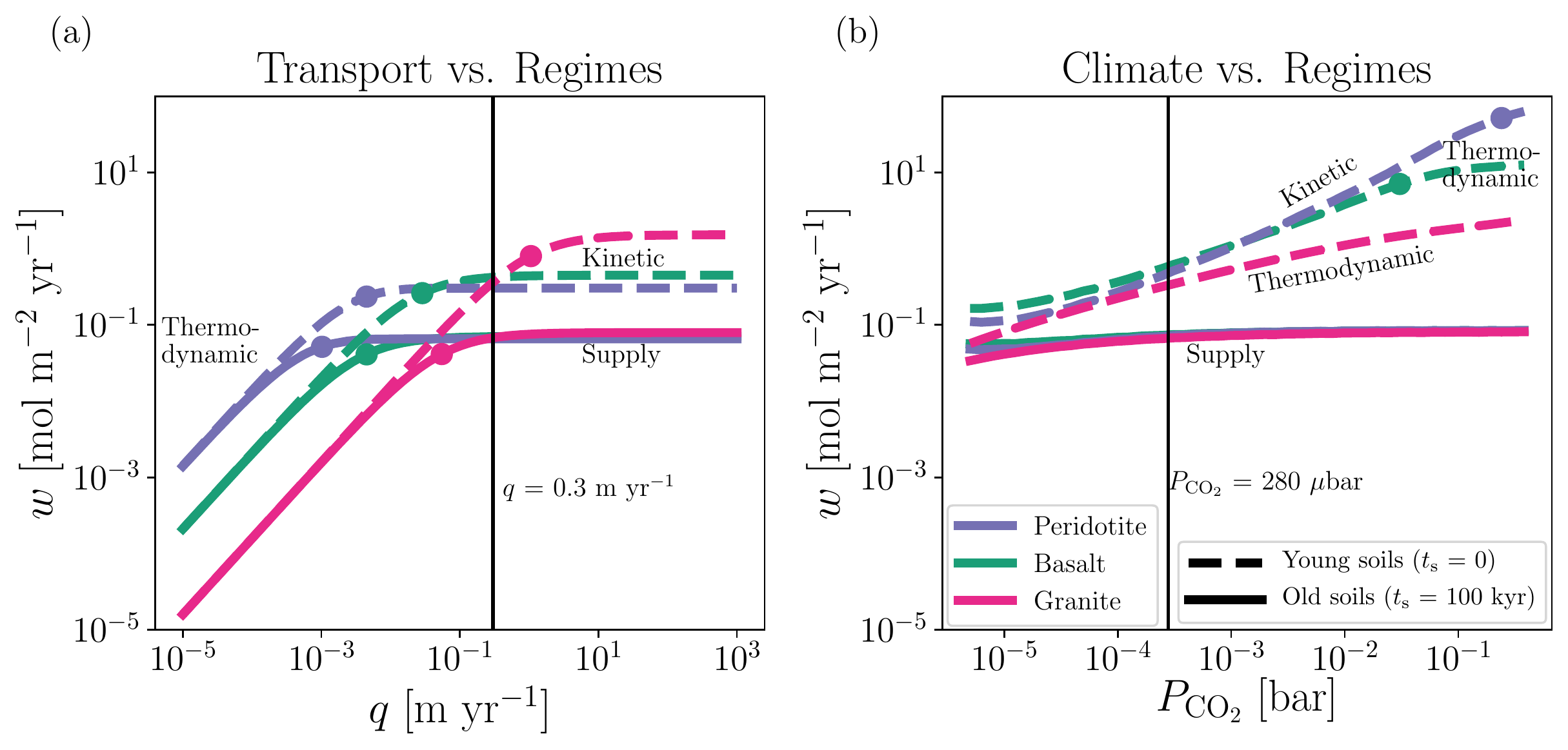}
  \caption{ Weathering regimes with respect to carbonate alkalinity flux of rocks for two generalized weathering models ($t_{\mathrm{s}} = 0$ and $t_{\mathrm{s}} = 100$~kyr). (a) Weathering flux as a function of fluid flow rate at $P_{\mathrm{CO}_2} = 280$~$\mu$bar and $T$ = 288~K. The vertical black line is the modern-mean runoff ($q$ = 0.3~m~yr$^{-1}$). (b) Weathering flux as a function of $P_{\mathrm{CO}_2}$ at modern-mean runoff of $q = 0.3$~m~yr$^{-1}$ where $T = f(P_{\mathrm{CO}_2})$ is given by the climate model (Section~\ref{sec:methodsClimate}). The vertical black line is the pre-industrial CO$_2$ partial pressure ($P_{\mathrm{CO}_2}$ = 280~$\mu$bar). Colored disks mark the transition between thermodynamic and kinetic/supply regimes. }
  \label{fig:w_rock_regimes}
\end{figure*}

In this study, silicate weathering rates are computed by the simultaneous consideration of dissolution reactions of all minerals present in a rock as well as reactions in the water-bicarbonate system. Three common silicate rocks (peridotite, basalt and granite) are examined. We develop the maximum weathering model (Section~\ref{sec:methodsMaximum}) presuming chemical equilibrium and the generalized weathering model (Section~\ref{sec:methodsGeneralized}) that applies to both equilibrium and non-equilibrium conditions. The generalized weathering model allows us to explore weathering in three different regimes (Figure~\ref{fig:w_rock_regimes}). To simulate the transport-based dilution of equilibrium concentrations of weathering products, the solute transport equation of \citet{2014Sci...343.1502M} is implemented. This equation is based on the interplay between the fluid flow rate ($q$) and the net reaction rate ($D_w$). When $q < D_w$, reactions are limited by thermodynamics or transport (runoff) and the weathering is thermodynamically-limited (Figure~\ref{fig:w_rock_regimes}a). In contrast, when $q > D_w$, reactions are limited by kinetics (and independent of runoff) and the weathering is kinetically-limited. Another parameter, the age of soils ($t_{\mathrm{s}}$), is introduced by \citet{2014Sci...343.1502M} to model the effect of limited supply of fresh rocks on the net reaction rate. The higher the $t_{\mathrm{s}}$, the lower is the $D_w$. This gives rise to another regime, supply-limited weathering (Figure~\ref{fig:w_rock_regimes}a). 

As a function of transport and climate properties, the carbonate alkalinity weathering flux shows a strong dependence on lithology in the thermodynamic and kinetic regimes and a weak dependence on lithology in the supply regime (Figure~\ref{fig:w_rock_regimes}). This impact of lithology for the three weathering regimes as a function of $q$ is seen in Figure~\ref{fig:w_rock_regimes}(a). There is approximately an order of magnitude difference between the thermodynamic fluxes of peridotite and basalt as well as those of basalt and granite.  The thermodynamic weathering flux is proportional to the equilibrium carbonate alkalinity which is strongly sensitive to the mineralogy considered for a given rock. In the kinetic regime, the differences are smaller but significant. The kinetic weathering flux is proportional to the effective rate coefficient of a given rock. In contrast, in the supply-limited regime, the $A$ fluxes almost overlap with each other because there is no impact of lithology. The supply regime is strongly sensitive to pore-space space properties that are held constant. 

The climate sensitivity of the weathering flux showcases that the weathering regime may depend on lithology. Figure~\ref{fig:w_rock_regimes}(b) shows that the generalized $t_{\mathrm{s}} = 0$ model of the weathering fluxes of peridotite and basalt increase steeply with $P_{\mathrm{CO}_2}$ up to $\sim 0.1$~bar where $T = f(P_{\mathrm{CO}_2})$ is obtained from the climate model. Beyond this $P_{\mathrm{CO}_2}$ value, peridotite and basalt enter the thermodynamic regime. This is because the weathering flux in the kinetic regime cannot keep increasing indefinitely. As soon as the model hits the thermodynamic upper limit, the model follows the thermodynamic sensitivities of weathering that are independent of pore-space properties and depend on chemical equilibrium and lithology. It is important to note that extrapolations of kinetic weathering expressions (Equation~\ref{eq:w_pCO2}) used in previous studies \citep[e.g.,][]{1981JGR....86.9776W,1983AmJS..283..641B,2001JGR...106.1373S,2015ApJ...812...36F,2017NatCo...815423K} may incorrectly predict the weathering flux to be higher than the upper limit provided by the thermodynamic flux. 

Unlike basalt and peridotite, the generalized young-soils ($t_{\mathrm{s}} = 0$) model of granite is in the thermodynamic regime for the given $P_{\mathrm{CO}_2}$ range (Figure~\ref{fig:w_rock_regimes}b). This is also evident from Figure~\ref{fig:w_rock_regimes}(a) where the vertical line ($q = 0.3$~m~yr$^{-1}$) falls right before the thermodynamic to kinetic regime transition for granite. Thus, the climate sensitivity of the weathering of fresh granite at $q = 0.3$~m~yr$^{-1}$ is determined largely by thermodynamics instead of kinetics. On the other hand, for the generalized models at the present-day characteristic soil age of $t_{\mathrm{s}} = 100$~kyr, the weathering fluxes of the three rocks overlap with each other. This is because this $t_{\mathrm{s}}$ value is so high that the effect of $k_{\mathrm{eff}}$ on the `net reaction rate' $D_w$ is negligible in pushing the model out of the supply regime. In this regime, the weathering flux is independent of the climate and transport properties. 

Since most laboratory measurements of kinetic rate coefficients are available for individual minerals, previous studies discuss weathering of individual minerals instead of rocks \citep[e.g.,][]{1981JGR....86.9776W,1983AmJS..283..641B}. In reality, all minerals in rocks undergo weathering contemporaneously, rendering consideration of individual minerals in isolated systems as less informative. Since the minerals in a rock are in contact with the aqueous solution, solute concentrations are buffered by the dissolution reactions of these minerals. It is essential to consider these reactions simultaneously to solve for solute concentrations. The generalized weathering model shows that the choice of individual minerals or rocks determines the weathering regime. For example, the feldspar endmember minerals (anorthite, albite and K-feldpsar) are in the thermodynamic regime for the $P_{\mathrm{CO}_2}$ range considered, whereas rocks are exhibit both thermodynamic and kinetic regimes (see Figure~\ref{fig:CA_mine}). In their implementation of the fluid transport-controlled model, \citet{2020ApJ...896..115G} find that weathering is largely independent of kinetics because of their choice of oligoclase (a type of plagioclase feldspar mineral) to model weathering, which is in the thermodynamic regime of weathering for a wide range of CO$_2$ partial pressures similar to feldspar endmembers shown in Figure~\ref{fig:CA_mine}.

\subsection{Positive Feedback of Weathering at High Temperature} \label{sec:disTemp}

It is widely accepted that weathering intensifies with surface temperature  \citep[e.g.,][]{1981JGR....86.9776W,1983AmJS..283..641B,2000AREPS..28..611K,2008kwri.book.....B}; but see \citet{1999ChGeo.159....3G} and \citet{2011ApJ...743...41K} for alternate viewpoints. This $T$-dependence of weathering is due to the increase in kinetic rate coefficients of mineral dissolution reactions as a function of $T$ (see Appendix~\ref{app:data}). Laboratory and field measurements of kinetic rate coefficients are fitted to the Arrhenius law, $w \propto \exp{(-E / R T)}$, where $E$ is the activation energy and $R$ is the universal gas constant \citep{palandri2004compilation}. The generalized weathering model based on the fluid transport-controlled approach \citep{2014Sci...343.1502M} captures the diversity of weathering regimes in a generic formulation, which is particularly useful for applications to exoplanets with potentially diverse surface environments. Studies applying the fluid transport-controlled model find that $T$ has a small effect on weathering \citep{2018E&PSL.485..111W,2020ApJ...896..115G}. This statement holds in the thermodynamic regime of weathering for certain plagioclase feldspars including oligoclase which coincidentally exhibit a small $T$ sensitivity, although \citet{2018E&PSL.485..111W} find that the equilibrium [HCO$_3^-$] resulting from plagioclase feldspars decreases with $T$.

\begin{figure}[!ht]
  \centering
  \includegraphics[width=\linewidth]{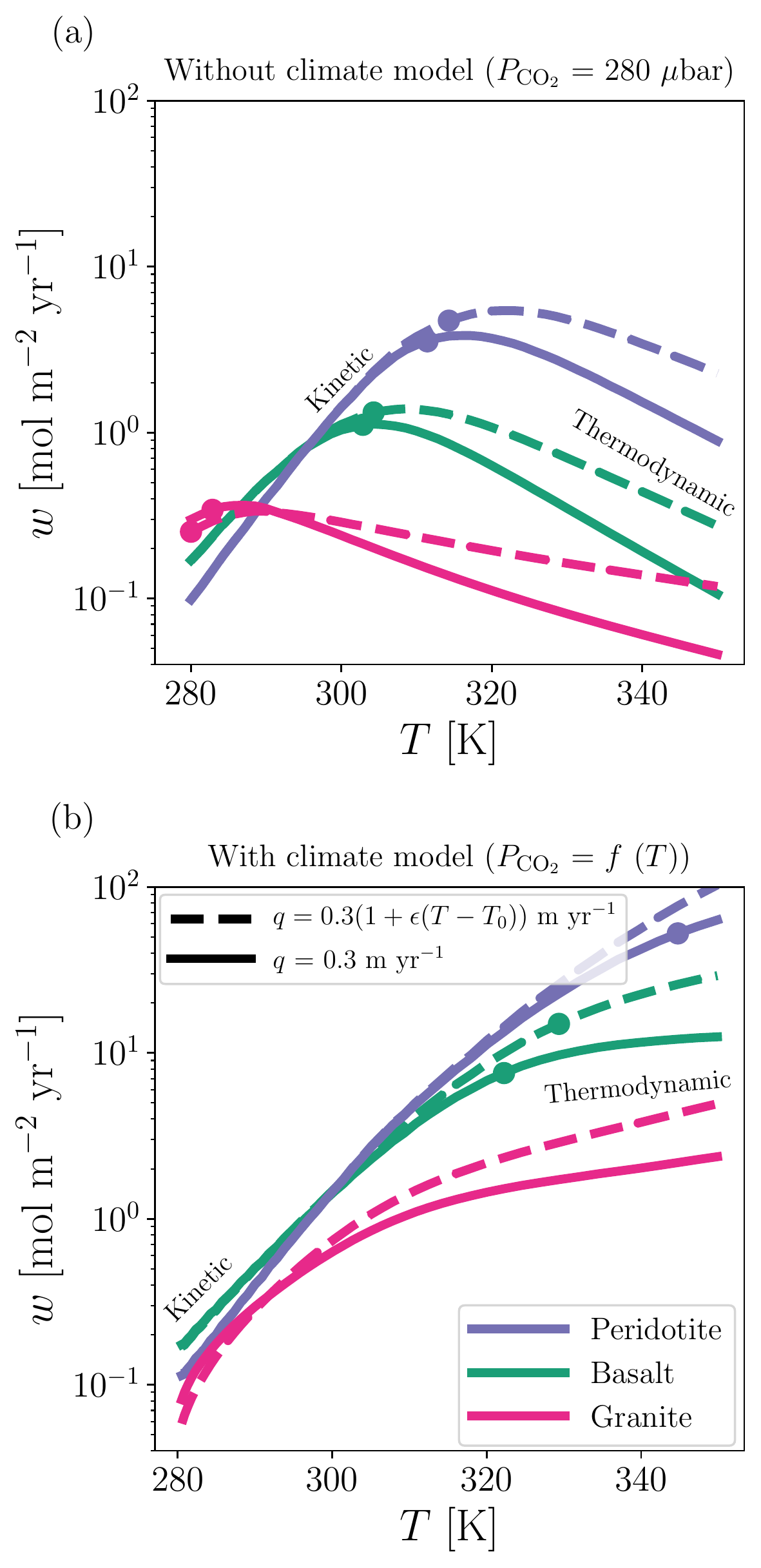}
  \caption{ Generalized carbonate alkalinity flux of rocks as a function of temperature at $t_{\mathrm{s}} = 0$, $P_{\mathrm{CO}_2} = 280$~$\mu$bar for a constant $q = 0.3$~m~yr$^{-1}$ and a $T$-dependent $q$ with $\epsilon = 0.03$~K$^{-1}$ and $T_{0} = 288$~K. Colored disks mark the transition from kinetic to thermodynamic regimes. }
  \label{fig:w_T_rock}
\end{figure}

To isolate the effect of temperature on weathering from that of the $P_{\mathrm{CO}_2}$ effect, we first present weathering as a function of $T$ at a fixed $P_{\mathrm{CO}_2}$ in Figure~\ref{fig:w_T_rock}(a). The generalized weathering model at $t_{\mathrm{s}} = 0$, $q$ = 0.3~m~yr$^{-1}$ and $P_{\mathrm{CO}_2}$ = 280~$\mu$bar (no climate model) shows that the carbonate alkalinity flux of rocks is in the kinetic regime at low temperatures and in the thermodynamic regime at high temperatures. In the kinetic regime, there is a steep increase in the weathering flux of granite up to $\sim$285~K, basalt up to $\sim$310~K, and peridotite up to $\sim$320~K because the effective kinetic rate coefficients show an exponential increase with temperature. Beyond these transition temperatures (where the net reaction rate equals the fluid flow rate), the models enter their respective thermodynamic regimes. This transition implies a switch from the negative feedback of silicate weathering to the carbon cycle to a potential positive feedback. Since the thermodynamic regime gives the maximum possible weathering flux, the kinetic weathering flux cannot exceed the thermodynamic weathering flux in the generalized model. Importantly, the kinetic weathering flux of granite is higher than that of basalt and peridotite because the slowest kinetic reaction among the constituent endmember minerals of granite has a higher kinetic rate coefficient than those of basalt and peridotite. This result is based on the choice of endmember minerals to define rocks and is expected to change with the choice of other endmember minerals or mineral solid solutions. 

The kinetic weathering flux increases with $T$ as expected, however the thermodynamic weathering flux decreases with $T$ (Figure~\ref{fig:w_T_rock}a). As described in Section~\ref{sec:resultsThermo}, if fitted to a kinetic weathering expression (Equation~\ref{eq:w_pCO2}), the thermodynamic model gives a negative value for the activation energy ($E_{\rm th}$) for the weathering of rocks and most minerals. Such a negative value highlights that thermodynamic weathering exhibits a negative slope as a function of $T$. This negative exponential decrease in weathering is a result of the negative slope exhibited by equilibrium constants of mineral dissolution reactions as a function of $T$ except for albite and K-feldspar (see Appendix~\ref{app:data}). This $T$-dependence is a thermodynamic property of mineral dissolution reactions in the aqueous system. For rocks, constituent minerals determine the overall dependence of thermodynamic weathering on $T$. For example, for a rock consisting of only albite and K-feldpsar, the thermodynamic weathering flux is expected to increase with $T$, similar to the prevalent understanding of the $T$-dependence of kinetic weathering. However, apart from these two minerals, all other minerals considered in this study show a negative slope. Since granite consists of both albite and K-feldspar in addition to quartz, phlogopite and annite, granite shows the least negative slope among the three rocks considered (Figure~\ref{fig:w_T_rock}a). Moreover, weathering of basalt in the thermodynamic regime decreases with $T$ more steeply than peridotite because of the presence of anorthite in basalt that exhibits the most negative slope in the equilibrium constant$-$temperature parameter-space. 

When the climate model is invoked assuming present-day solar flux and present-day planetary albedo, $P_{\mathrm{CO}_2}$ increases from 10~$\mu$bar to 0.5~bar as a function of the $T$ varying from 280~K to 350~K and the decreasing effect of $T$ on weathering disappears (Figure~\ref{fig:w_T_rock}b). This is because in the thermodynamic regime the weathering flux shows a positive power-law dependence on $P_{\mathrm{CO}_2}$, which is stronger than the exponential decrease in weathering with temperature. And in the kinetic regime, the exponential increase in weathering with temperature dominates. It is interesting to note that basalt and peridotite enter the thermodynamic regime only at $\sim$325~K and $\sim$345~K respectively, whereas granite is in the thermodynamic regime for the given temperature range. For this climate model, assuming a constant stellar flux results in a strong coupling between $T$ and $P_{\mathrm{CO}_2}$, in contrast to the behavior when $P_{\mathrm{CO}_2}$ is held constant. Depending on the stellar flux and planetary albedo, the strength of coupling between $T$ and $P_{\mathrm{CO}_2}$ is probably in between the two panels shown in Figure~\ref{fig:w_T_rock}, pushing the thermodynamic to kinetic transition temperature given in  Figure~\ref{fig:w_T_rock}(a) to higher values. The supply-limited weathering flux, on the other hand, is independent of $T$ (e.g., Figure~\ref{fig:w_peri_clim}c,d). This is because the supply-limited weathering flux depends on pore-space parameters including $t_{\mathrm{s}}$ and $\psi$ that are assumed to be constant for a given model. In reality, the age of soils may indirectly depend on temperature through the effect of precipitation and runoff on physical erosion rates \citep[e.g.,][]{2005E&PSL.235..211W,2012Geo....40..811W,2015ApJ...812...36F}. 

Previous studies have argued that since runoff depends on precipitation which in turn depends on $T$, the weathering flux should increase even more strongly with $T$ when a $T$-dependent runoff is assumed \citep[e.g.,][]{2001AmJS..301..182B}. Figure~\ref{fig:w_T_rock} shows that this statement does not hold in the kinetic regime of weathering. This is because the kinetic regime is independent of runoff in the fluid transport-controlled model \citep{2014Sci...343.1502M}. Considering a linear dependence of runoff on temperature \citep[e.g., Equations 41, 42 in][]{2020ApJ...896..115G} instead of a constant runoff, there is no impact on weathering because the kinetic regime is independent of runoff in the generalized model.

\subsection{Global Silicate Weathering Rates} \label{sec:disEarth}

During the Archean (2.5$-$4~Ga), the incident solar radiation was about 70$-$80\% of its present-day value,  not high enough to maintain a temperate climate with present-day atmospheric CO$_2$ levels \citep{1972Sci...177...52S,2020SSRv..216...90C}. Although there are no direct measurements of historical weathering rates, the Archean geological record suggests a steady decrease in $P_{\mathrm{CO}_2}$ from the order of 0.1~bar to modern values while maintaining surface temperatures between 280~K and 315~K \citep[and references therein]{2018PNAS..115.4105K}. The lower insolation during the Archean was compensated by the greenhouse effect of CO$_2$. As the insolation increased, climates should have become warmer than the observations suggest. Without a negative feedback of silicate weathering that allowed a decrease in CO$_2$ levels as insolation increased, modern climates would not have been temperate \citep{1981JGR....86.9776W,1983AmJS..283..641B,1993Icar..101..108K}. The extent of silicate weathering during the history of Earth and the contribution of seafloor weathering are debated \citep[e.g.,][]{2001JGR...106.1373S,2001AmJS..301..182B,2015ApJ...812...36F,2018PNAS..115.4105K,2018AREPS..46...21C}. By applying the generalized weathering model to Earth, we lay the foundation for understanding climate regulation on temperate planets. 

\begin{figure*}[!ht]
  \centering
  \medskip
  \includegraphics[width=\textwidth]{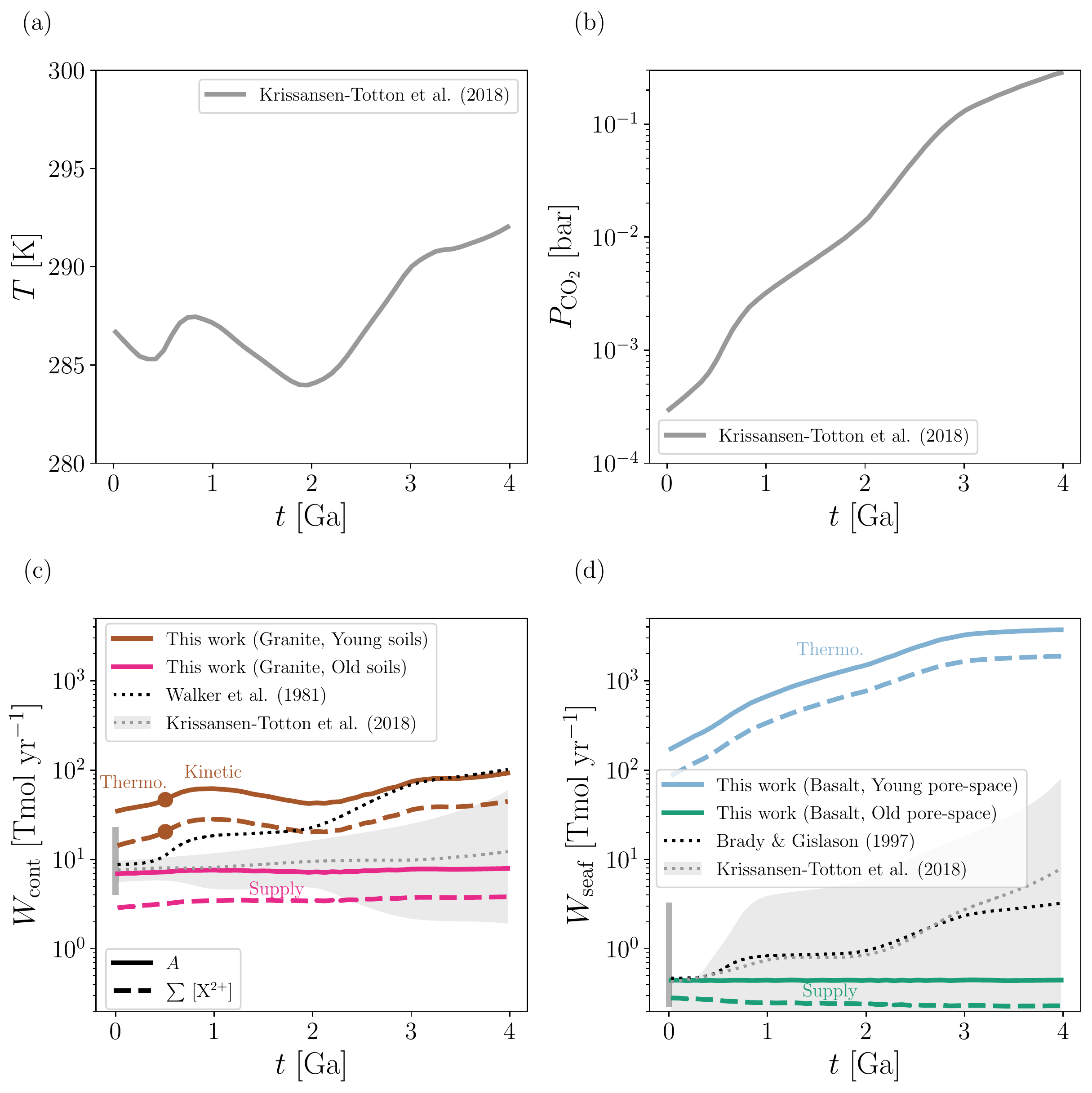}
  \caption{ Continental granite and seafloor basalt weathering rates derived from carbonate alkalinity $A$ and the sum of concentrations of divalent cations $\sum[{\rm X^{2+}}]$ compared with a model from \citet[][Figure~3E,F, hereafter, KT18]{2018PNAS..115.4105K} for the past 4~Ga on Earth. (a) $T$ and (b) $P_{\mathrm{CO}_2}$ are the median values of the same model from KT18 that are used as inputs to the weathering models in (c) and (d). (c) Generalized young soils ($t_{\mathrm{s}} = 0$) and old soils ($t_{\mathrm{s}} = 100$~kyr) granite models take the same values for $\psi = \psi_0$, $q$ = 0.3~m~yr$^{-1}$ (modern mean runoff) and continental weatherable area of 93~Mm$^2$ \citep[][]{2002GBioC..16.1042F}. (d) Generalized young pore-space ($t_{\mathrm{s}} = 0$) and old pore-space ($t_{\mathrm{s}} = 50$~Myr) basalt models take different values for $\psi$ ($\psi = 100 \psi_0$ and $\psi = 18 \psi_0$, respectively) but the same value for $q$ = 0.05~m~yr$^{-1}$ (modern mean hydrothermal fluid flow rate) and seafloor weatherable area of 147~Mm$^2$ \citep[][]{2003E&PSL.216..565J}. Gray shaded regions are 95\% confidence intervals of KT18 models. Vertical gray lines are uncertainties in the estimates of present-day silicate weathering rates given by geological measurements and models. Colored disks denote the transition between thermodynamic and kinetic regimes. }
  \label{fig:earth_weath}
\end{figure*}

Since present-day continents on Earth are largely felsic and the seafloor is mafic, we approximate the lithology of continents by granite and the seafloor by basalt. Continental and seafloor weathering rates on Earth are calculated up to 4~Ga (Figure~\ref{fig:earth_weath}c,d). This is the first application of the fluid transport-controlled weathering model of \citet{2014Sci...343.1502M} to seafloor weathering on Earth. Since there are no direct measurements of historical weathering rates on Earth, we compare the generalized model developed in this study with a model from \citet[][Figure 3, hereafter, KT18]{2018PNAS..115.4105K}.  Instead of using a climate model, time-dependent median models of $P_{\mathrm{CO}_{2}}$ and $T$ from the same KT18 model are used as inputs to the generalized model. The same inputs are used to obtain the continental weathering rate from \citet{1981JGR....86.9776W}  and the seafloor weathering from \citet{1997GeCoA..61..965B}, where the present-day weathering rates of both models are normalized to those of KT18 models. 

To first order, one may expect that continental and seafloor weathering rates depend on the continental surface area ($f\,A_{\rm s}$) and the seafloor surface area ($(1-f)\,A_{\rm s}$), respectively (Equation~\ref{eq:w_rate}). However, not all continental and seafloor surface area undergoes weathering on Earth. \citet{2002GBioC..16.1042F} report the value of continental weatherable area to be equal to 93~Mm$^{2}$, about 60\% of modern continental area ($= 0.6 \, f \, A_{\rm s}$). About 147~Mm$^{2}$ of the seafloor area ($= 0.41 \, (1-f) \, A_{\rm s}$) is expected to contribute to seafloor weathering \citep[given by the exposed area of low-temperature hydrothermal systems,][]{2003E&PSL.216..565J}. Therefore, in Figure~\ref{fig:earth_weath}, the continental and seafloor weathering rates are given by $W_{\mathrm{cont}} = w_{\mathrm{cont}} \times$~93~Mm$^{2}$ and $W_{\mathrm{seaf}} = w_{\mathrm{seaf}} \times$~147~Mm$^{2}$, respectively. A more precise definition of continental weatherable area is the area susceptible to precipitation and runoff, and that of seafloor weatherable area is the area covered by low-temperature hydrothermal systems that are younger than approximately 60~Myr \citep{1994JGR....99.3081S,2003E&PSL.216..565J,2018AREPS..46...21C}. Nonetheless, the definition given in Equation~(\ref{eq:w_rate}) assumes all surfaces are weatherable, thereby giving upper estimates of the global weathering rates on exoplanets. 

The present-day continental weathering rate for young soils ($t_{\mathrm{s}} = 0$) in Figure~\ref{fig:earth_weath}(c) is in the thermodynamic regime (see also Figure~\ref{fig:w_rock_regimes}a). Similarly, the present-day seafloor weathering rate for young pore-space is in the thermodynamic regime (Figure~\ref{fig:earth_weath}d). As the CO$_2$ levels increase (prior to 0.5~Ga), continental weathering for young soils is driven by kinetics (Figure~\ref{fig:earth_weath}c). Whereas, seafloor weathering for young pore-space is in the thermodynamic regime even at high $P_{\mathrm{CO}_2}$. The weathering rate of the young-soils granite model decreases between 1$-$2~Ga. This is because the effective kinetic rate coefficient of granite is almost constant as a function of pH in basic solutions resulting from our calculations and depends mostly on $T$, which decreases between 1$-$2~Ga. Seafloor weathering in the thermodynamic regime increases monotonically because the effect of $P_{\mathrm{CO}_2}$ on weathering wins over the $T$ effect. In the thermodynamic regime, the choice of secondary minerals also affects the weathering rates. For example, on one hand, the granite and basalt weathering rates for $t_{\mathrm{s}} = 0$ models would be higher by a factor of 20 and 3 when kaolinite is used as a secondary mineral instead of halloysite. On the other hand, if secondary minerals incorporating divalent cations are modeled, the weathering rates would be lower because bicarbonate and carbonate ions equivalent to cations locked up in secondary minerals would be removed from the solution \citep[e.g.,][]{2019E&PSL.52415718K}.

The continental weathering rate derived from carbonate alkalinity for young soils and the seafloor weathering rate derived from carbonate alkalinity for young pore-space are respectively about an order of magnitude and up to three orders of magnitude higher than that of KT18 models (Figure~\ref{fig:earth_weath}c,d). The \citet{1981JGR....86.9776W} continental weathering rate and the \citet{1997GeCoA..61..965B} seafloor weathering rate exhibit a monotonic rise ($\beta = 0.3$ and $\beta = 0.23$, respectively). There are multiple reasons for the discrepancy between our $t_{\mathrm{s}} = 0$ models and other models in Figure~\ref{fig:earth_weath}. Our models are designed to provide absolute weathering rates for generality, and are therefore not intrinsically calibrated to present-day weathering rates unlike other studies. The entire exposed planetary surface area may not contain fresh rocks for weathering and continents do not necessarily have a uniform lithology and topography. On Earth's continents, orogeny (mountain building) exposes fresh rocks to the surface that are highly susceptible to weathering (low $t_{\mathrm{s}}$), whereas the contribution of cratons (ancient continental crust, high $t_{\mathrm{s}}$) to weathering is smaller than mountains \citep{2014Sci...343.1502M}. Similarly, ridge volcanism exposes fresh basalt at mid-ocean ridges but the majority of seafloor area is older than a million years.

For a planet with surface conditions where all divalent and monovalent cations react with carbonate and bicarbonate ions to produce carbonates, carbonate alkalinity is a proxy for the flux of CO$_2$ out of the atmosphere-ocean system. However, carbonate alkalinity overestimates the global weathering rate on modern Earth, which is limited by the flux of divalent cations instead of carbonate alkalinity \citep{1997Natur.390...65F}. This is because monovalent cations do not contribute to the carbonate precipitation in the present-day ocean. Figure~\ref{fig:earth_weath}c shows that the weathering rate derived from divalent cations in our granite weathering model is smaller than the carbonate alkalinity weathering rate by up to a factor of five. Alternatively, a planet with no rivers but only coastal springs that are isolated from the atmosphere would produce divalent cations but no carbonate alkalinity and effectively no CO$_2$ flux out of the atmosphere-ocean system. Therefore, care should be taken when extrapolating a single weathering proxy to diverse planetary conditions.

The present-day continental weathering rate derived from carbonate alkalinity matches that of KT18 when the soil age is set to the characteristic soil age (100~kyr). The soil age encapsulates several processes including tectonics, erosion and soil production \citep{1997Natur.388..358H,2003GeCoA..67.4411R,2005E&PSL.235..211W,2014Sci...343.1502M}. For example, the soil age decreases with increasing erosion or increasing mineral supply rates \citep{2014Sci...343.1502M}. On continents, the dimensionless pore-space parameter varies with topography and its mean value for present-day continents is not well-constrained as previous studies adopt values that span an order of magnitude either side of the reference value used in this study \citep{2010E&PSL.294..101M,2011E&PSL.312...48M,2014Sci...343.1502M,2018E&PSL.485..111W}. Given the validity range in parameters and weathering proxies, our model allows the present-day global continental weathering on Earth to either be in thermodynamic, kinetic or supply-limited regimes of weathering.

To match the present-day seafloor weathering rate derived from carbonate alkalinity with that of KT18, the pore-space age is increased to 50~Myr (characteristic seafloor age) and the dimensionless pore-space parameter is decreased from $100 \psi_0$ to $18 \psi_0$.  Since these two pore-space parameters have never been discussed in the context of seafloor weathering, future studies must evaluate their magnitude and extent of validity. For the same parameters, the seafloor weathering rate derived from divalent cations is smaller by less than a factor of two. Both these models are in the supply-limited regime of weathering and therefore exhibit almost constant weathering rates. 

The present-day regional variation in continental runoff \citep[0.01$-$3~m~yr$^{-1}$,][]{1999ChGeo.159....3G,2002GBioC..16.1042F} and seafloor fluid flow rate \citep[0.001$-$0.7~m~yr$^{-1}$,][]{1994JGR....99.3081S,2003E&PSL.216..565J,2013E&PSL.380...12H} is more than two orders of magnitude. At fluid flow rates lower than the mean values (arid climates or low hydrothermal heat flux, see Figure~\ref{fig:w_xCO2_endmembers}), the models might escape kinetic/supply regimes and enter the thermodynamic regime. These models rely on characteristic values of $t_{\mathrm{s}}$, $\psi$ and $q$ to be suitable proxies for calculating global weathering rates, that are not necessarily appropriate for present-day Earth. For example, mountains contribute an order of magnitude more weathering flux than cratons \citep{2014Sci...343.1502M}. Additionally, basaltic regions on continents (omitted in our calculations) may contribute a weathering flux (weathering rate per unit area) higher than that of granitic regions by a factor of five \citep{2016GeCoA.190..265I}. On Earth, local weathering rates can be estimated from data and integrated to compute the global weathering rate for the planetary surface. However, exoplanets lack (Earth independent) data constraints on global properties such as fluid flow rates and pore-space parameters, let alone their potential regional variations. Hence, we propose to study exoplanets by implementing the generalized weathering model to various lithologies using characteristic values of fluid flow rates and pore-space parameters. 

Holding modeling parameters constant throughout Earth's dynamic history is another critical approximation. The uplift of the Himalayan-Tibetan Plateau in the past 40 million years has decreased the characteristic soil age which has likely increased the global weathering rates \citep[][]{2000AREPS..28..611K}. The lithology of continents has also evolved over time and therefore the application of one type of lithology may result in inaccurate weathering rates for certain periods of Earth's history. The continental area fraction (and consequently the weatherable area) was smaller in the Archean than today \citep{2017SedG..357...16D}. Even if the strength of continental weathering flux was similar at the beginning of the Archean compared to today but the continental area was significantly smaller, then the continental weathering rate must have been significantly smaller. The true contribution of the silicate weathering flux to the carbon cycle also depends on the carbonate compensation depth in the oceans \citep{1970GeCoA..34..836P,2005E&PSL.234..299R} as well as reverse weathering \citep{1966AmJS..264..507M,2018Natur.560..471I,2020E&PSL.53716181K} that requires knowledge of ocean salinity and ocean pH, which suggests that an ocean chemistry model should form the basis of future research.

\subsection{Weathering Regimes and the Habitable Zone} \label{sec:disHZ}

It has been proposed that measuring the gaseous abundance of CO$_2$ in the atmospheres of Earth-sized exoplanets will allow one to statistically distinguish between Venus-like and Earth-like climates \citep{2017ApJ...841L..24B,2019BAAS...51c.404C,2020ApJ...896..115G}. A negative feedback associated with weathering is necessary to control the climatic impact of CO$_2$ over geological timescales on an Earth-like planet \citep{1981JGR....86.9776W}. One-dimensional, radiative-convective atmospheric models used to study the boundaries of the classical habitable zone usually assume the presence of the negative weathering feedback to justify the assumption of CO$_2$ being present as a greenhouse gas at the outer boundary of the habitable zone and having only a minor abundance at its inner edge \citep{1993Icar..101..108K,2013ApJ...765..131K}. The effects of CO$_2$ on the runaway greenhouse effect associated with water is debated \citep{1992JAtS...49.2256N,Abe1993Litho..30..223A}. Moreover, if weathering is limited by the supply of fresh rocks, the negative feedback is lost \citep{2012Geo....40..811W,2015ApJ...812...36F}. This is evident from Figure~\ref{fig:w_rock_regimes}, where the supply-limited weathering flux is seen to be insensitive to changes in $T$ and $P_{\mathrm{CO}_2}$. However, if the soil age indirectly depends on $T$ and $P_{\mathrm{CO}_2}$ on geological timescales, the negative feedback may reinstate. Our models of thermodynamic weathering flux show that the weathering flux decreases with $T$ and increases with $P_{\mathrm{CO}_2}$ (Figure~\ref{fig:w_T_rock}). In the parameter-space where the effect of $T$ dominates, it is possible to have a positive climate feedback associated with weathering for lithologies tested in this study. Such a scenario may have implications on the boundaries of the habitable zone. Future studies should combine the climate and weathering models to investigate these possibilities.

\section{Summary and Conclusions} \label{sec:conclusions}

Silicate weathering is a key process in the carbon cycle that transfers CO$_2$ from the atmosphere to the surface of a planet. The intensity of silicate weathering has previously been attributed to the kinetics of fluid-rock reactions \citep[e.g.,][]{1981JGR....86.9776W,1983AmJS..283..641B,2000AREPS..28..611K}. \citet{2011E&PSL.312...48M,2014Sci...343.1502M} show that if the reaction rate exceeds the fluid flow rate, thermodynamics of fluid-rock reactions at chemical equilibrium drives weathering instead of kinetics. This fluid transport-controlled approach models both thermodynamic and kinetic regimes of weathering with a single formulation. Moreover, if there is a limited supply of fresh rocks, the weathering is supply-limited. The applications of this approach to continental weathering on Earth \citep{2018E&PSL.485..111W} and temperate exoplanets \citep{2020ApJ...896..115G} consider weathering reactions of individual minerals. 

In this study, we extend this approach to the weathering of any rock type (lithology) and apply it to seafloor weathering in addition to continental weathering. We find that the simultaneous consideration of weathering reactions of the major minerals present in a rock as well as the reactions in the water-bicarbonate system instead of weathering reactions of individual minerals impacts weathering rates as well as weathering regimes. Moreover, this model allows the calculation of absolute weathering rates instead of weathering rates normalized to present-day values as most previous weathering studies. In addition to climate properties ($T$ and $P_{\mathrm{CO}_2}$) and runoff or fluid flow rate ($q$), this model is mainly sensitive to age of soils ($t_{\mathrm{s}}$) and a dimensionless scaling parameter ($\psi$) based on pore-space and rock properties. The equilibrium constants and kinetic rate coefficients are effectively a function of $T$ and $P_{\mathrm{CO}_2}$. Depending on these five parameters, the weathering for a given lithology is in the thermodynamic, kinetic or supply regimes. Close to the regime transition points, the contribution of both regimes to weathering is similar. 

Weathering reactions at chemical equilibrium give the maximum concentrations of weathering products. We use this approach to calculate the maximum weathering flux for a given lithology. The larger the fraction of divalent cations in rocks, the higher the sensitivity of maximum weathering flux to CO$_2$ partial pressure. This thermodynamic $P_{\mathrm{CO}_2}$ sensitivity (power-law exponent $\beta_{\rm th}$) is 0.71, 0.69 and 0.65 for peridotite, basalt and granite, respectively. These values are subject to change depending on the choice of minerals to define a rock type as well as secondary minerals produced during weathering. For example, the consideration of kaolinite as a secondary mineral instead of halloysite changes $\beta_{\rm th}$ to 0.65 and 0.54 for basalt and granite, respectively.  The thermodynamic $P_{\mathrm{CO}_2}$ sensitivity of these rocks is stronger than the kinetic $P_{\mathrm{CO}_2}$ sensitivity implemented in previous studies \citep[0.22$-$0.55,][]{1981JGR....86.9776W,1991AmJS..291..339B,2013Icar..226.1447D}. However, the combined effect of $P_{\mathrm{CO}_2}$ and $T$ results in a weaker thermodynamic $P_{\mathrm{CO}_2}$ sensitivity. 

The fluid transport-controlled model demonstrates that the weathering flux cannot necessarily be approximated by the kinetic weathering expression (Equation~\ref{eq:w_pCO2}). In our model, planets with arid climates (low runoff) and elevated topography (young soils) are likely in the thermodynamic regime of weathering, exhibiting weathering rates higher than that of modern Earth by three to four orders of magnitude. Moreover, limited supply of fresh rocks mitigates the role of kinetics. The thermodynamic $T$ sensitivity of weathering of rocks is negative, implying that the weathering flux decreases with $T$. This is in contrast with the prevalent understanding that weathering intensifies with an increase in $T$ which is attributed to an increase in kinetic rate coefficients of mineral dissolution reactions with $T$ \citep{1965BdM...88.....2,palandri2004compilation,2008kwri.book.....B}. An important implication of this finding is that when $T$ increases without a strong variation in $P_{\mathrm{CO}_2}$, silicate weathering has the potential to instigate a positive feedback to the carbon cycle.  The focus of future studies should be on applying a generalized weathering model encompassing multiple weathering regimes to model the carbon cycle.


\acknowledgments{We thank Edwin Kite and an anonymous reviewer for their constructive comments that helped to improve this manuscript. We thank Eric Gaidos for stimulating discussions. We acknowledge financial support from the European Research Council via Consolidator Grant ERC-2017-CoG-771620-EXOKLEIN (awarded to KeHe) and Center for Space and Habitability, University of Bern. DJB acknowledges the Swiss National Science Foundation (SNSF) Ambizione Grant 173992. CD acknowledges the SNSF Ambizione Grant 174028. We thank the financial support of the National Centre of Competence in Research PlanetS supported by the SNSF. }

\software{\texttt{astropy} \citep{2013A&A...558A..33A,2018AJ....156..123A},  
          \texttt{CHNOSZ} \citep{2019FrEaS...7..180D}, 
          \texttt{ipython} \citep{2007CSE.....9c..21P},
          \texttt{matplotlib} \citep{2007CSE.....9...90H},
          \texttt{numpy} \citep{2011CSE....13b..22V},
          \texttt{scipy} \citep{2020SciPy-NMeth}
          }


\appendix
\counterwithin{figure}{section}
\counterwithin{table}{section}

\twocolumngrid

\section{Thermodynamics and Kinetics Data}\label{app:data}

\begin{figure*}[!ht]
  \centering
  \medskip
  \includegraphics[width=0.8\textwidth]{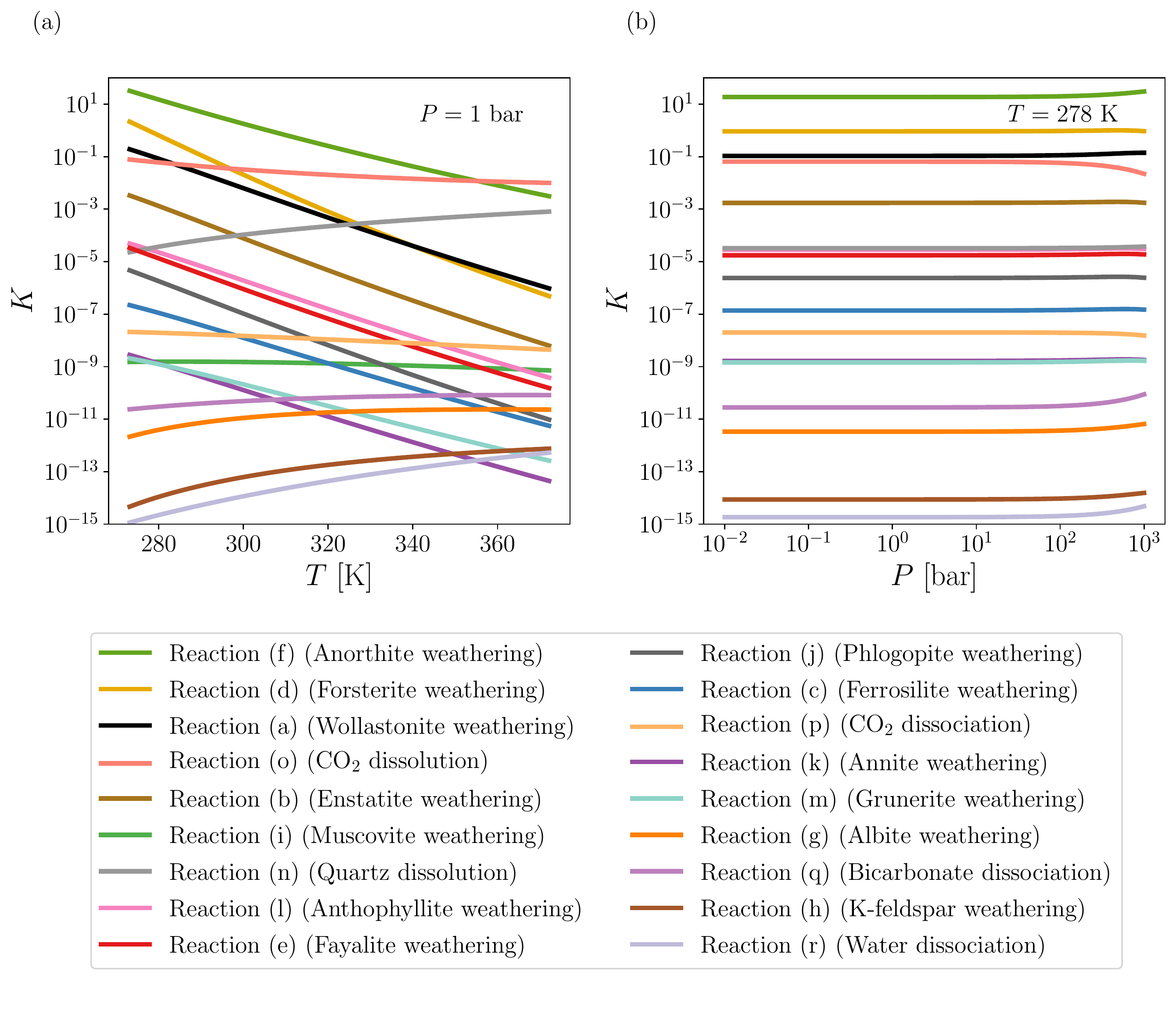}
  \caption{Equilibrium constants for the reactions listed in Table~\ref{tab:reactions} obtained from \texttt{CHNOSZ} \citep{2019FrEaS...7..180D}. (a) $K$ as a function of $T$. (b) $K$ as a function of $P$. Please note that some reactions have fractional stoichiometry in order to ensure high numerical precision. }
  \label{fig:K}
\end{figure*}

The equilibrium constant $K$ of a reaction is given by the difference of the Gibbs energy of formation of products and reactants as follows

\begin{equation}\label{eq:KeqGibbs}
   - R T  \ln{K} = \sum_{i}^{\mathrm{products}} \nu_{i} \Delta_{f} G_{P,T,i} - \sum_{j}^{\mathrm{reactants}} \nu_{j} \Delta_{f} G_{P,T,j}, 
\end{equation} 
where $\Delta_{f} G_{P,T,i}$ is the Gibbs energy of formation of $i^{th}$ species at pressure $P$ and temperature $T$, $\nu_{i}$ is the stoichiometric coefficient and $R$ is the universal gas constant. The Gibbs energy of formation of each species is computed at any $P$ and $T$ in terms of the Gibbs energy of formation at reference pressure $P_0$ and reference temperature $T_0$ by

\begin{equation}\label{eq:GibbsPT}
\begin{split}
   \Delta_{f} G_{P,T} = \Delta_{f} G_{P_{0},T_{0}} - S_{P_{0},T_{0}} (T-T_{0}) \\ + \int_{T_{0}}^{T} C_{P} dT - T \int_{T_{0}}^{T} C_{P} d\ln{T} + \int_{P_{0}}^{P} V dP,
\end{split}
\end{equation} 
where $S_{P_{0},T_{0}}$ is the entropy at the reference pressure and temperature, $C_{P}$ is the heat capacity at constant pressure as a function of temperature, and $V$ is volume as a function of pressure. In this study, $\Delta_{f} G$ values are obtained from the \texttt{CHNOSZ} database \citep{2019FrEaS...7..180D}. The equilibrium constants of reactions given in Table~\ref{tab:reactions} are shown as a function of $P$ and $T$ in Figure~\ref{fig:K}. 

\begin{figure*}[!ht]
  \centering
  \medskip
  \includegraphics[width=0.9\textwidth]{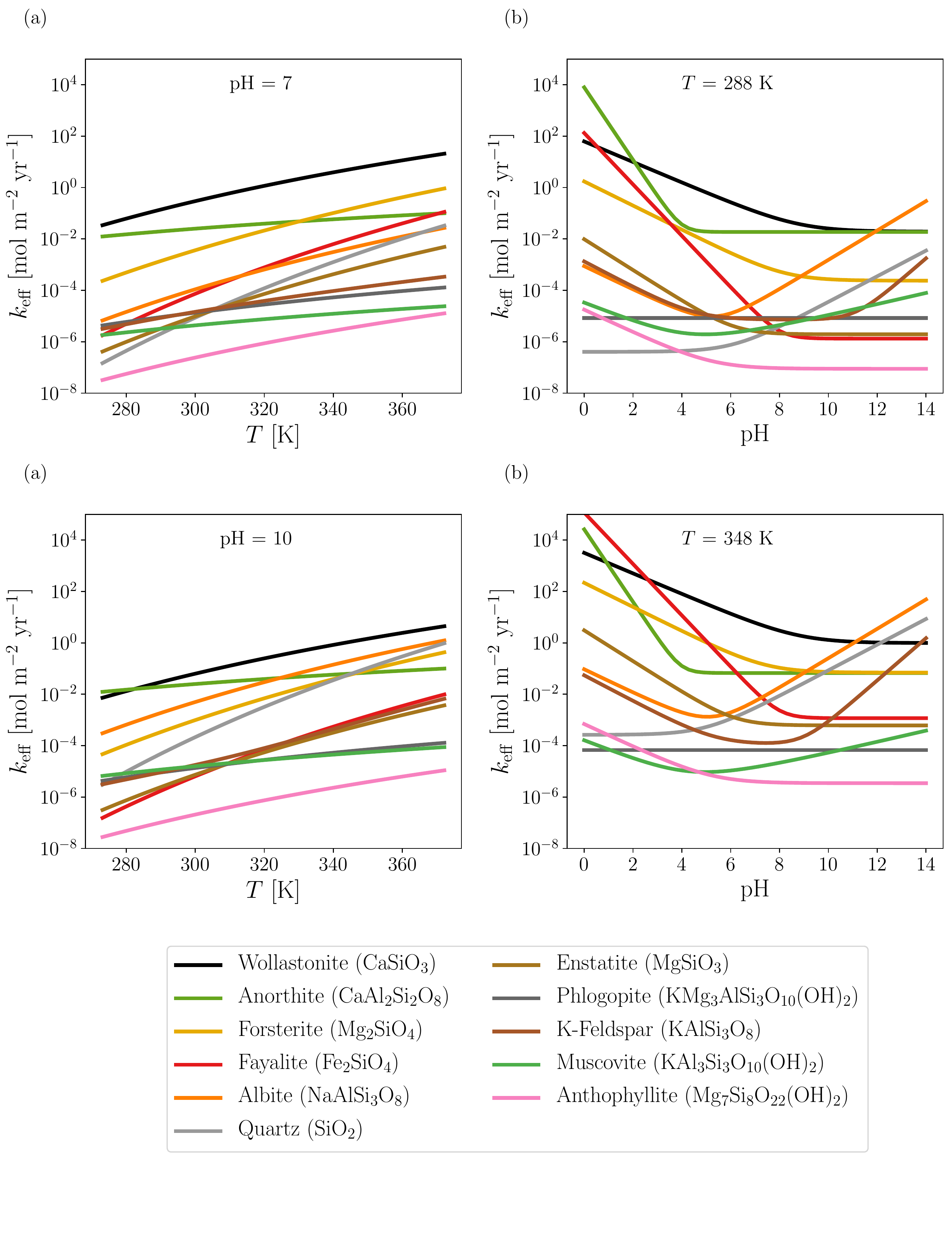}
  \caption{Kinetic rate coefficients for the dissolution of minerals obtained from \citet{palandri2004compilation}. (a) $k_{\mathrm{eff}}$ as a function of $T$ at pH = 7. (b) $k_{\mathrm{eff}}$ as a function of pH at $T = 288$~K. (c) $k_{\mathrm{eff}}$ as a function of $T$ at pH = 10. (d) $k_{\mathrm{eff}}$ as a function of pH at $T = 348$~K. }
  \label{fig:k_eff}
\end{figure*}

The kinetic rate coefficient $k_{\mathrm{eff}}$ of mineral dissolution reactions are obtained from the compilation of \citet{palandri2004compilation}. This compilation (Table~\ref{tab:k_eff}) is based on the fitting of the following equation to experimental kinetics data,

\begin{equation}\label{eq:keff}
\begin{split}
   k_{\mathrm{eff}} = A_{\mathrm{acid}} \exp{ \left( \frac{-E_{\mathrm{acid}}}{R T} \right) } \; 10^{- \mathrm{pH} \; n_{\mathrm{acid}}}
   \\ + A_{\mathrm{neut}} \exp{ \left( \frac{-E_{\mathrm{neut}}}{R T} \right) }
   \\ + A_{\mathrm{base}} \exp{ \left( \frac{-E_{\mathrm{base}}}{R T} \right) } \; 10^{- \mathrm{pH} \; n_{\mathrm{base}}}
\end{split}
\end{equation}
where $E_{\mathrm{acid}}$, $E_{\mathrm{neut}}$ and $E_{\mathrm{base}}$ are the activation energies at acidic, neutral and basic pH, $A_{\mathrm{acid}}$, $A_{\mathrm{neut}}$ and $A_{\mathrm{base}}$ are the preexponential factors at acidic, neutral and basic pH, and $n_{\mathrm{acid}}$ and $n_{\mathrm{base}}$ are the neutral and basic power-law exponents. Figure~\ref{fig:k_eff} shows the variation of $k_{\mathrm{eff}}$ with $T$ and pH for a number of mineral dissolution reactions. For minerals not present in this compilation (ferrosilite, annite, grunerite), the kinetic rate coefficients are obtained from respective endmember minerals of the same mineral group (enstatite, phlogopite, anthophyllite). 

\begin{deluxetable}{lccccc}
\tablecaption{Kinetics data from \citet{palandri2004compilation}. \label{tab:k_eff}}
\tabletypesize{\scriptsize}
\tablehead{
\colhead{Mineral} & \colhead{$E_{\mathrm{acid}}$} & \colhead{$n_{\mathrm{acid}}$} & \colhead{$E_{\mathrm{neut}}$} & \colhead{$E_{\mathrm{base}}$} & \colhead{$n_{\mathrm{base}}$} \\
& kJ mol$^{-1}$ & & kJ mol$^{-1}$ & kJ mol$^{-1}$ & 
}
\startdata
Wollastonite  & 54.7 & 0.400 & 54.7 & $-$  & $-$ \\
Enstatite     & 80.0 & 0.600 & 80.0 & $-$  & $-$ \\
Forsterite    & 67.2 & 0.470 & 79.0 & $-$  & $-$ \\
Fayalite      & 94.4 & $-$   & 94.4 & $-$  & $-$ \\
Anorthite     & 16.6 & 1.411 & 17.8 & $-$  & $-$ \\
Albite        & 65.0 & 0.457 & 69.8 & 71.0 & $-$0.572 \\
K-feldspar    & 51.7 & 0.500 & 38.0 & 94.1 & $-$0.823 \\
Muscovite     & 22.0 & 0.370 & 22.0 & 22.0 & $-$0.220 \\
Phlogopite    & $-$  & $-$   & 29.0 & $-$  & $-$ \\
Anthophyllite & 51.0 & 0.440 & 51.0 & $-$  & $-$ \\
Quartz        & $-$  & $-$   & 90.1 & 108.4 & $-$0.5 \\
\enddata
\tablecomments{ For minerals that are not listed, data from a corresponding endmember mineral from the same mineral group is adopted. }
\end{deluxetable}

\section{Maximum and Generalized Concentrations of Rocks and Minerals} \label{app:Generalized}

Sects.~\ref{sec:methodsMaximum} and \ref{sec:methodsGeneralized} introduce the methods to compute maximum (thermodynamic) and generalized solute concentrations for peridotite weathering. Table~\ref{tab:reactions} lists the mineral dissolution (reactions (a$-$n)) and water-bicarbonate reactions (reactions (o$-$r)), and the relation between equilibrium constants of these reactions and thermodynamic activities. Table~\ref{tab:poly_weath} gives the polynomial equations to calculate the activity of HCO$_3^-$ at chemical equilibrium as a function of $P_{\mathrm{CO}_2}$ for the weathering of all rocks and minerals considered. As an example, the maximum [HCO$_3^-$] for peridotite weathering is obtained as a function of CO$_2$ partial pressure, surface temperature and total pressure in Figure~\ref{fig:HCO3eq_all_peri}. As described in Section~\ref{sec:methodsMaximum}, [HCO$_3^-$]$_{\mathrm{eq}}$ is strongly sensitive to $P_{\mathrm{CO}_2}$ and $T$. However, $P$ has a negligible effect on [HCO$_3^-$]$_{\mathrm{eq}}$ because the equilibrium constants of reactions are largely unchanged up to 1000~bar (Appendix~\ref{app:data}). This figure demonstrates that the total pressure plays a negligible role in determining the solute concentrations of aqueous species. The effect of precipitation of amorphous silica \citep{2018E&PSL.485..111W} is not modeled since this effect changes the weathering flux by less than an order of magnitude (only at high $P_{\mathrm{CO}_2}$) which is smaller than the 5$-$10 orders of magnitude spread in the weathering fluxes discussed in this study.

\begin{figure}[!ht]
  \centering
  \medskip
  \includegraphics[width=0.93\linewidth]{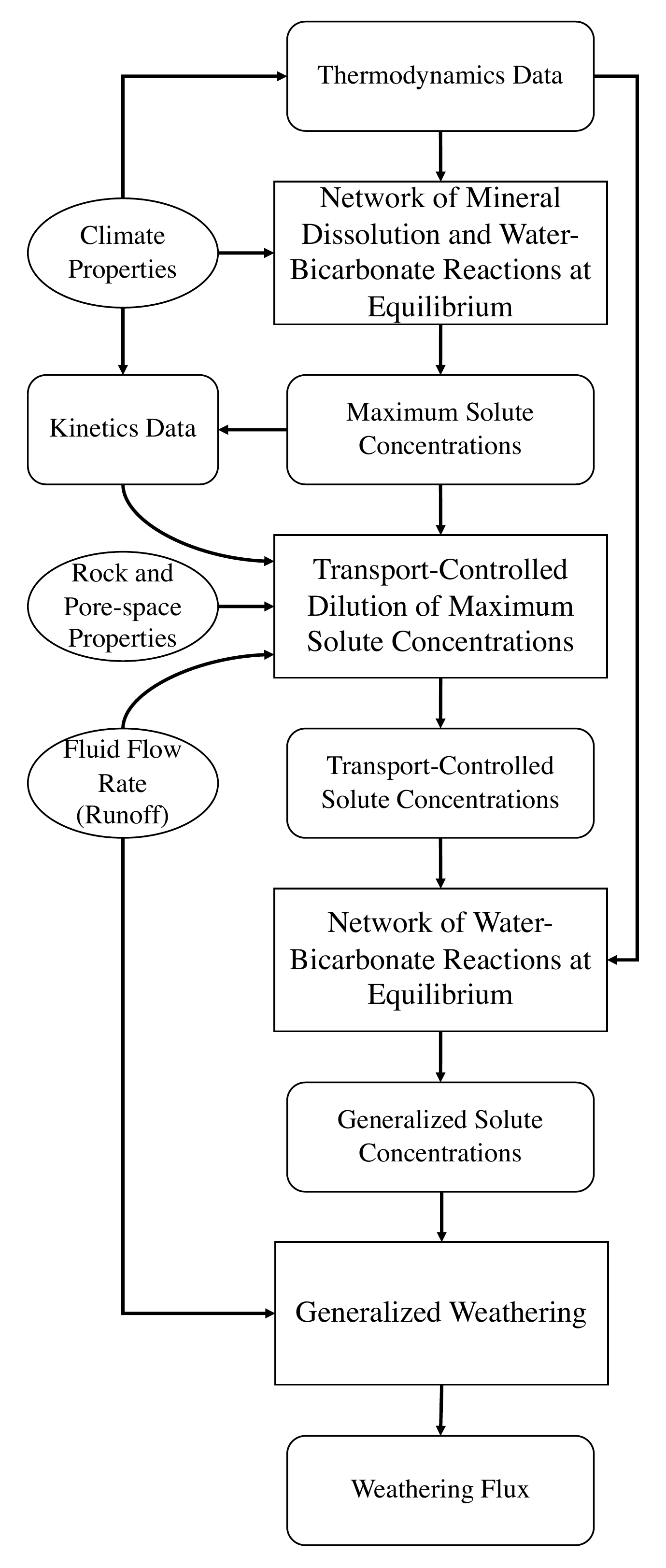}
  \caption{ Schematic describing the methodology of the weathering model \texttt{CHILI}. Square boxes represent software modules, ovals denote parameters (Tables~\ref{tab:params} and \ref{tab:f_params}) and rounded squares represent computed quantities (Table~\ref{tab:variables}). The solute transport equation of \citet{2014Sci...343.1502M} is implemented to calculate diluted solute concentrations. Thermodynamics and kinetics data are obtained from \citet{2019FrEaS...7..180D} and \citet{palandri2004compilation}, respectively (see Appendix~\ref{app:data}). }
  \label{fig:methodology}
\end{figure}

\begin{figure*}[!ht]
  \centering
  \medskip
  \includegraphics[width=0.85\textwidth]{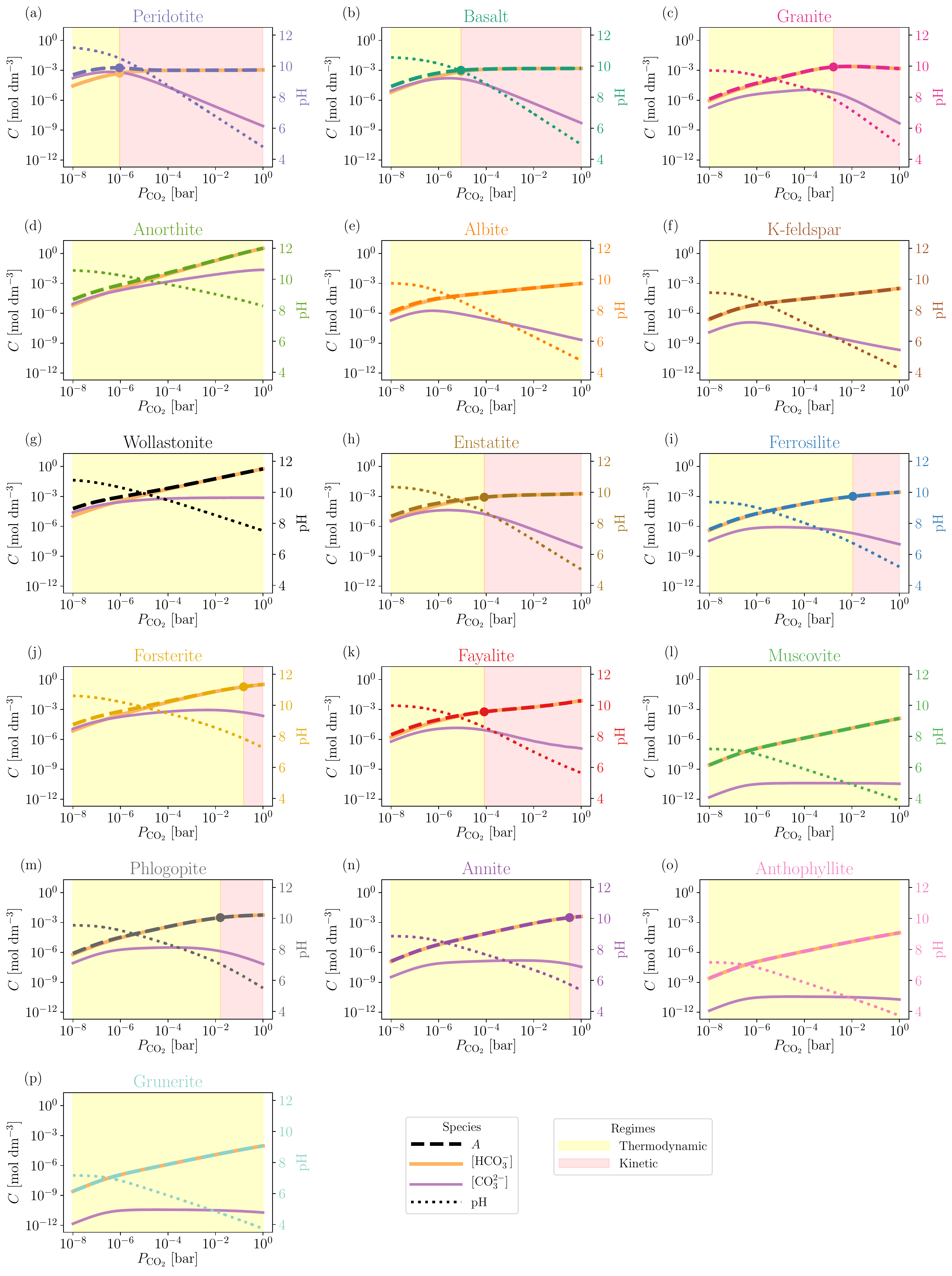}
  \caption{ Carbonate alkalinity and pH as a function of $P_{\mathrm{CO}_{2}}$ at $T = 288$~K (modern surface temperature), $q = 0.3$~m~yr$^{-1}$ (modern mean runoff) and $t_{\mathrm{s}} = 0$ (young soils). From left to right, colored disks mark the transition between thermodynamic and kinetic regimes. }
  \label{fig:CA_mine}
\end{figure*}

\begin{figure}[h]
  \centering
  \medskip
  \includegraphics[width=0.75\linewidth]{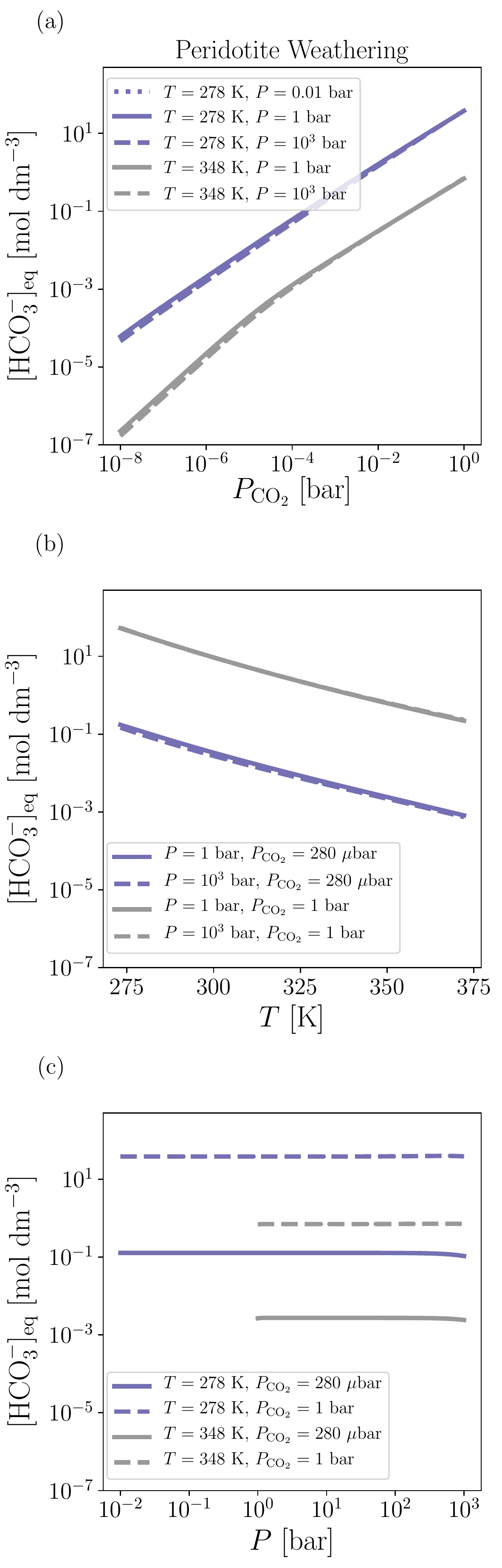}
  \caption{ Impact of climate properties on the equilibrium bicarbonate concentration produced as a result of peridotite weathering. (a) CO$_2$ partial pressure. (b) Temperature. (c) Total Pressure. }
  \label{fig:HCO3eq_all_peri}
\end{figure}

Once the maximum [HCO$_3^-$] is determined for a given rock or mineral, the solute transport equation of \citet{2014Sci...343.1502M} is implemented to dilute the equilibrium value of [HCO$_3^-$] as a function of runoff or fluid flow rate $q$ (Figure~\ref{fig:methodology}). This equation allows to calculate the non-equilibrium concentrations using equilibrium concentrations. Higher the fluid flow rate, more diluted is the resulting [HCO$_3^-$] (Equation~\ref{eq:solute_eq}). This diluted [HCO$_3^-$] is then used as an input to solve for concentrations of other aqueous species such as CO$_3^{2-}$, H$^+$ and OH$^-$ by assuming that the water-bicarbonate reactions obey chemical equilibrium. Figure~\ref{fig:CA_mine} demonstrates that generalized solute concentrations (Section~\ref{sec:methodsGeneralized}) are strongly sensitive to lithology. For example, the transition between thermodynamic and kinetic weathering regimes of peridotite occurs at $P_{\mathrm{CO}_2} = 1$~$\mu$bar for $q = 0.3$~m~yr$^{-1}$. Whereas, this transition occurs at $P_{\mathrm{CO}_2} = 1$~mbar for granite. Once these generalized concentrations are obtained, the generalized weathering flux is calculated using Equation~(\ref{eq:w}).

\begin{rotatetable*}
\begin{deluxetable*}{llLP}
\tablecaption{Reactions and relations between equilibrium constants and activities of reactants and products. \label{tab:reactions}}
\tablewidth{700pt}
\tabletypesize{\tiny}
\tablehead{
\colhead{Index} & \colhead{Symbol} & 
\colhead{Reaction} & \colhead{Equilibrium Constant and Activities} 
} 
\startdata
& & Pyroxene \\
(a) & Wo & \mathrm{CaSiO}_{3}(s) + 2 \mathrm{CO}_{2}(g) + \mathrm{H_{2}O}(l) \xrightleftharpoons{K_{\mathrm{Wo}}}   2 \mathrm{HCO}_{3}^{-} + \mathrm{Ca}^{2+} + \mathrm{SiO}_{2}(aq)         &       K_{\mathrm{Wo}} =  a_{\mathrm{HCO}_{3}^{-}}^{2} \; a_{\mathrm{Ca}^{2+}} \; a_{\mathrm{SiO}_{2}(aq)} \; P_{\mathrm{CO}_{2}}^{-2} \\
(b) & En & \mathrm{MgSiO}_{3}(s) + 2 \mathrm{CO}_{2}(g) + \mathrm{H_{2}O}(l) \xrightleftharpoons{K_{\mathrm{En}}}   2 \mathrm{HCO}_{3}^{-} + \mathrm{Mg}^{2+} + \mathrm{SiO}_{2}(aq)         &       K_{\mathrm{En}} =  a_{\mathrm{HCO}_{3}^{-}}^{2} \; a_{\mathrm{Mg}^{2+}} \; a_{\mathrm{SiO}_{2}(aq)} \; P_{\mathrm{CO}_{2}}^{-2}  \\
(c) & Fs & \mathrm{FeSiO}_{3}(s) + 2 \mathrm{CO}_{2}(g) + \mathrm{H_{2}O}(l) \xrightleftharpoons{K_{\mathrm{Fs}}}   2 \mathrm{HCO}_{3}^{-} + \mathrm{Fe}^{2+} + \mathrm{SiO}_{2}(aq)         &       K_{\mathrm{Fs}} =  a_{\mathrm{HCO}_{3}^{-}}^{2} \; a_{\mathrm{Fe}^{2+}}  \; a_{\mathrm{SiO}_{2}(aq)} \; P_{\mathrm{CO}_{2}}^{-2} \\
\hline
& & Olivine \\
(d) & Fo & \frac{1}{2} \mathrm{Mg_{2}SiO}_{4}(s) + 2 \mathrm{CO}_{2}(g) + \mathrm{H_{2}O}(l) \xrightleftharpoons{K_{\mathrm{Fo}}}   2 \mathrm{HCO}_{3}^{-} + \mathrm{Mg}^{2+} + \frac{1}{2} \mathrm{SiO}_{2}(aq)         &       K_{\mathrm{Fo}} = a_{\mathrm{HCO}_{3}^{-}}^{2} \;  a_{\mathrm{Mg}^{2+}} \; a_{\mathrm{SiO}_{2}(aq)}^{1/2} \; P_{\mathrm{CO}_{2}}^{-2} \\
(e) & Fa & \frac{1}{2} \mathrm{Fe_{2}SiO}_{4}(s) + 2 \mathrm{CO}_{2}(g) + \mathrm{H_{2}O}(l) \xrightleftharpoons{K_{\mathrm{Fa}}}   2 \mathrm{HCO}_{3}^{-} + \mathrm{Fe}^{2+} + \frac{1}{2} \mathrm{SiO}_{2}(aq)         &       K_{\mathrm{Fa}} =  a_{\mathrm{HCO}_{3}^{-}}^{2}  a_{\mathrm{Fe}^{2+}} a_{\mathrm{SiO}_{2}(aq)}^{1/2}P_{\mathrm{CO}_{2}}^{-2} \\
\hline
& & Feldspar \\
(f) & An & \frac{1}{2} \mathrm{CaAl_{2}Si_{2}O}_{8}(s) + \mathrm{CO}_{2}(g) + \frac{3}{2} \mathrm{H_{2}O}(l) \xrightleftharpoons{K_{\mathrm{An}}}   \mathrm{HCO}_{3}^{-} + \frac{1}{2} \mathrm{Ca}^{2+} + \frac{1}{2} \mathrm{Al_{2}Si_{2}O_{5}(OH)}_{4}(s)       &        K_{\mathrm{An}} =  a_{\mathrm{HCO}_{3}^{-}} \; a^{1/2}_{\mathrm{Ca}^{2+}} \; P_{\mathrm{CO}_{2}}^{-1} \\
(g) & Alb & \mathrm{NaAlSi_{3}O}_{8}(s) + \mathrm{CO}_{2}(g) + \frac{3}{2} \mathrm{H_{2}O}(l) \xrightleftharpoons{K_{\mathrm{Alb}}}   \mathrm{HCO}_{3}^{-} + \mathrm{Na}^{+} + \frac{1}{2} \mathrm{Al_{2}Si_{2}O_{5}(OH)}_{4}(s) + 2 \mathrm{SiO}_{2}(aq)       &       K_{\mathrm{Alb}} = a_{\mathrm{HCO}_{3}^{-}} \; a_{\mathrm{Na}^{+}} \; a_{\mathrm{SiO}_{2}(aq)}^{2} \; P_{\mathrm{CO}_{2}}^{-1} \\
(h) & Kfs & \mathrm{KAlSi_{3}O}_{8}(s) + \mathrm{CO}_{2}(g) + \frac{3}{2} \mathrm{H_{2}O}(l) \xrightleftharpoons{K_{\mathrm{Kfs}}}    \mathrm{HCO}_{3}^{-} + \mathrm{K}^{+} + \frac{1}{2} \mathrm{Al_{2}Si_{2}O_{5}(OH)}_{4}(s) + 2 \mathrm{SiO}_{2}(aq)       &       K_{\mathrm{Kfs}} = a_{\mathrm{HCO}_{3}^{-}} \; a_{\mathrm{K}^{+}} \; a_{\mathrm{SiO}_{2}(aq)}^{2} \; P_{\mathrm{CO}_{2}}^{-1} \\
\hline
& & Mica \\
(i) & Ms & \mathrm{KAl_{3}Si_{3}O_{10}(OH)}_{2}(s) + \mathrm{CO}_{2}(g) + \frac{5}{2} \mathrm{H_{2}O}(l) \xrightleftharpoons{K_{\mathrm{Ms}}}   \mathrm{HCO}_{3}^{-} + \mathrm{K}^{+} + \frac{3}{2} \mathrm{Al_{2}Si_{2}O_{5}(OH)}_{4}(s)       &       K_{\mathrm{Ms}} =  a_{\mathrm{HCO}_{3}^{-}}  a_{\mathrm{K}^{+}}P_{\mathrm{CO}_{2}}^{-1} \\
(j) & Phl & \frac{1}{3} \mathrm{KMg_{3}AlSi_{3}O_{10}(OH)}_{2}(s) + \frac{7}{3} \mathrm{CO}_{2}(g) + \frac{7}{6} \mathrm{H_{2}O}(l) \xrightleftharpoons{K_{\mathrm{Phl}}}          &       K_{\mathrm{Phl}} = a_{\mathrm{HCO}_{3}^{-}}^{7/3} \; a_{\mathrm{K}^{+}}^{1/3}  \; a_{\mathrm{Mg}^{2+}} \; a_{\mathrm{SiO}_{2}(aq)}^{2/3} \; P_{\mathrm{CO}_{2}}^{-7/3} \\
& & \frac{7}{3} \mathrm{HCO}_{3}^{-} + \frac{1}{3} \mathrm{K}^{+} + \mathrm{Mg}^{2+} + \frac{1}{6} \mathrm{Al_{2}Si_{2}O_{5}(OH)}_{4}(s) + \frac{2}{3} \mathrm{SiO}_{2}(aq) & \\
(k) & Ann & \frac{1}{3} \mathrm{KFe_{3}AlSi_{3}O_{10}(OH)}_{2}(s) + \frac{7}{3} \mathrm{CO}_{2}(g) + \frac{7}{6} \mathrm{H_{2}O}(l) \xrightleftharpoons{K_{\mathrm{Ann}}}          &       K_{\mathrm{Ann}} = a_{\mathrm{HCO}_{3}^{-}}^{7/3} \; a_{\mathrm{K}^{+}}^{1/3} \; a_{\mathrm{Fe}^{2+}} \; a_{\mathrm{SiO}_{2}(aq)}^{2/3} \; P_{\mathrm{CO}_{2}}^{-7/3}  \\
& & \frac{7}{3} \mathrm{HCO}_{3}^{-} + \frac{1}{3} \mathrm{K}^{+} + \mathrm{Fe}^{2+} + \frac{1}{6}  \mathrm{Al_{2}Si_{2}O_{5}(OH)}_{4}(s) + \frac{2}{3} \mathrm{SiO}_{2}(aq) & \\
\hline
& & Amphibole \\
(l) & Ath & \frac{1}{7} \mathrm{Mg_{7}Si_{8}O_{22}(OH)}_{2}(s) + 2 \mathrm{CO}_{2}(g) + \frac{6}{7} \mathrm{H_{2}O}(l) \xrightleftharpoons{K_{\mathrm{Ath}}}   2 \mathrm{HCO}_{3}^{-} + \mathrm{Mg}^{2+} + \frac{8}{7} \mathrm{SiO}_{2}(aq)       &       K_{\mathrm{Ath}} = a_{\mathrm{HCO}_{3}^{-}}^{2} \; a_{\mathrm{Mg}^{2+}} \; a_{\mathrm{SiO}_{2}(aq)}^{8/7} \; P_{\mathrm{CO}_{2}}^{-2} \\
(m) & Gru & \frac{1}{7} \mathrm{Fe_{7}Si_{8}O_{22}(OH)}_{2}(s) + 2 \mathrm{CO}_{2}(g) + \frac{6}{7} \mathrm{H_{2}O}(l) \xrightleftharpoons{K_{\mathrm{Gru}}}   2 \mathrm{HCO}_{3}^{-} + \mathrm{Fe}^{2+} + \frac{8}{7} \mathrm{SiO}_{2}(aq)       &       K_{\mathrm{Gru}} = a_{\mathrm{HCO}_{3}^{-}}^{2} \; a_{\mathrm{Fe}^{2+}} \; a_{\mathrm{SiO}_{2}(aq)}^{8/7} \; P_{\mathrm{CO}_{2}}^{-2} \\
\hline
& & Quartz \\
(n) & Qz & \mathrm{SiO}_{2}(s)  \xrightleftharpoons{K_{\mathrm{Qz}}}  \mathrm{SiO}_{2}(aq)        &       K_{\mathrm{Qz}} = a_{\mathrm{SiO}_{2}(aq)}  \\
\hline
& & Water-Bicarbonate~System \\
(o) & CO$_2$ & \mathrm{CO}_{2}(g)  \xrightleftharpoons{K_{\mathrm{CO_2}}}    \mathrm{CO}_{2}(aq)        &       K_{\mathrm{CO_2}} = a_{\mathrm{CO}_{2}(aq)} \; P_{\mathrm{CO}_{2}}^{-1}   \\
(p) & Bic & \mathrm{CO}_{2}(g) + \mathrm{H_{2}O} \xrightleftharpoons{K_{\mathrm{Bic}}}    \mathrm{HCO}_{3}^{-} + \mathrm{H}^{+}        &       K_{\mathrm{Bic}} = a_{\mathrm{HCO}_{3}^{-}}  \; a_{\mathrm{H}^{+}} \; P_{\mathrm{CO}_{2}}^{-1}  \\
(q) & Car & \mathrm{HCO}_{3}^{-}  \xrightleftharpoons{K_{\mathrm{Car}}}    \mathrm{CO}_{3}^{2-} + \mathrm{H}^{+}        &       K_{\mathrm{Car}} = a_{\mathrm{CO}_{3}^{2-}} \; a_{\mathrm{H}^{+}} \; a_{\mathrm{HCO}_{3}^{-}}^{-1}  \\
(r) & Wat & \mathrm{H_{2}O} \xrightleftharpoons{K_{\mathrm{Wat}}}    \mathrm{OH}^{-} + \mathrm{H}^{+}       &       K_{\mathrm{Wat}} = a_{\mathrm{OH}^{-}} \; a_{\mathrm{H}^{+}} \\[1mm]
\enddata
\tablecomments{Moles of several species are fractional so that the equilibrium constants do not exceed computational numerical precision. See Figure~\ref{fig:K} for the dependence of $K$ on $P$ and $T$. }
\end{deluxetable*}
\end{rotatetable*}

\begin{rotatetable*}
\begin{deluxetable*}{lL}
\tablecaption{Polynomial equations in $a_{\mathrm{HCO}_{3}^{-}}$ and $P_{\mathrm{CO}_{2}}$ for minerals and rocks considered in this study. \label{tab:poly_weath}}
\tablewidth{700pt}
\tabletypesize{\scriptsize}
\tablehead{
\colhead{Weathering of a mineral/rock} & \colhead{Polynomial in $a_{\mathrm{HCO}_{3}^{-}}$ and $P_{\mathrm{CO}_{2}}$} \\
\colhead{(Reaction indices from Table~\ref{tab:reactions})}
} 
\startdata
Peridotite (a,b,d,e,o,p,q,r) & \!\begin{aligned} (2 K_{\mathrm{Car}} K_{\mathrm{En}}^{2} ) a_{\mathrm{HCO}_{3}^{-}}^{4} + K_{\mathrm{En}}^{2} (K_{\mathrm{Wat}} + K_{\mathrm{Bic}} P_{\mathrm{CO}_{2}}) a_{\mathrm{HCO}_{3}^{-}}^{3} - K_{\mathrm{Bic}}^{2} K_{\mathrm{En}}^{2} P_{\mathrm{CO}_{2}}^{2} a_{\mathrm{HCO}_{3}^{-}} \\ - 2 K_{\mathrm{Bic}} K_{\mathrm{Fo}} (K_{\mathrm{Fo}} K_{\mathrm{En}} +  K_{\mathrm{Fa}}K_{\mathrm{En}} + K_{\mathrm{Wo}} K_{\mathrm{Fo}}) P_{\mathrm{CO}_{2}}^{3}  = 0 \end{aligned} \\
Basalt (a,b,c,f,g,o,p,q,r) & \!\begin{aligned} (2 K_{\mathrm{Car}} K_{\mathrm{Wo}}^{2} ) a_{\mathrm{HCO}_{3}^{-}}^{4} + K_{\mathrm{Wo}}^{2} (K_{\mathrm{Wat}} + K_{\mathrm{Bic}} P_{\mathrm{CO}_{2}}) a_{\mathrm{HCO}_{3}^{-}}^{3} - K_{\mathrm{Bic}} (K_{\mathrm{Bic}} K_{\mathrm{Wo}}^{2} + K_{\mathrm{An}}^{4}  K_{\mathrm{Alb}}) P_{\mathrm{CO}_{2}}^{2}  a_{\mathrm{HCO}_{3}^{-}} \\ - 2 K_{\mathrm{Bic}} K_{\mathrm{An}}
^{2} K_{\mathrm{Wo}} (K_{\mathrm{Wo}} + K_{\mathrm{En}} + K_{\mathrm{Fs}} ) P_{\mathrm{CO}_{2}}^{3}  = 0 \end{aligned} \\
Granite (g,h,j,k,n,o,p,q,r) & \!\begin{aligned} (2 K_{\mathrm{Car}} K_{\mathrm{Qz}}^{2} K_{\mathrm{Kfs}}^{1/3}) a_{\mathrm{HCO}_{3}^{-}}^{4} + K_{\mathrm{Qz}}^{2} K_{\mathrm{Kfs}}^{1/3} (K_{\mathrm{Wat}} + K_{\mathrm{Bic}} P_{\mathrm{CO}_{2}}) a_{\mathrm{HCO}_{3}^{-}}^{3} - K_{\mathrm{Bic}} K_{\mathrm{Kfs}}^{1/3} [K_{\mathrm{Bic}} K_{\mathrm{Qz}}^{2} + (K_{\mathrm{Alb}} \\ +  K_{\mathrm{Kfs}})] P_{\mathrm{CO}_{2}}^{2}  a_{\mathrm{HCO}_{3}^{-}}  - 2 K_{\mathrm{Bic}} K_{\mathrm{Qz}}^{2} ( K_{\mathrm{Phl}} + K_{\mathrm{Ann}}) P_{\mathrm{CO}_{2}}^{3} = 0 \end{aligned} \\
Wollastonite (a,o,p,q,r)  &   \!\begin{aligned} (2 K_{\mathrm{Car}}) a_{\mathrm{HCO}_{3}^{-}}^{3} + (K_{\mathrm{Wat}} + K_{\mathrm{Bic}} P_{\mathrm{CO}_{2}} ) a_{\mathrm{HCO}_{3}^{-}}^{2}  - (K_{\mathrm{Bic}}^{2} + 2 K_{\mathrm{Bic}} \sqrt{K_{\mathrm{Wo}}} ) P_{\mathrm{CO}_{2}}^{2}  = 0 \end{aligned} \\
Enstatite (b,o,p,q,r) &      \!\begin{aligned} (2 K_{\mathrm{Car}}) a_{\mathrm{HCO}_{3}^{-}}^{3} + (K_{\mathrm{Wat}} + K_{\mathrm{Bic}} P_{\mathrm{CO}_{2}} ) a_{\mathrm{HCO}_{3}^{-}}^{2} - (K_{\mathrm{Bic}}^{2} + 2 K_{\mathrm{Bic}} \sqrt{K_{\mathrm{En}}} ) P_{\mathrm{CO}_{2}}^{2}  = 0 \end{aligned} \\
Ferrosilite (c,o,p,q,r)  &   \!\begin{aligned} (2 K_{\mathrm{Car}}) a_{\mathrm{HCO}_{3}^{-}}^{3} + (K_{\mathrm{Wat}} + K_{\mathrm{Bic}} P_{\mathrm{CO}_{2}} ) a_{\mathrm{HCO}_{3}^{-}}^{2}  - (K_{\mathrm{Bic}}^{2} + 2 K_{\mathrm{Bic}} \sqrt{K_{\mathrm{Fs}}} ) P_{\mathrm{CO}_{2}}^{2}  = 0 \end{aligned} \\
Forsterite (d,o,p,q,r) &  \!\begin{aligned} (2 K_{\mathrm{Car}}) a_{\mathrm{HCO}_{3}^{-}}^{10/3} + (K_{\mathrm{Wat}} + K_{\mathrm{Bic}} P_{\mathrm{CO}_{2}} ) a_{\mathrm{HCO}_{3}^{-}}^{7/3}  - (K_{\mathrm{Bic}}^{2} P_{\mathrm{CO}_{2}}^{2}) a_{\mathrm{HCO}_{3}^{-}}^{1/3}  - 2^{4/3} K_{\mathrm{Bic}} K_{\mathrm{Fo}}^{2/3}  P_{\mathrm{CO}_{2}}^{7/3}  = 0 \end{aligned} \\
Fayalite (e,o,p,q,r) &  \!\begin{aligned} (2 K_{\mathrm{Car}}) a_{\mathrm{HCO}_{3}^{-}}^{10/3} + (K_{\mathrm{Wat}} + K_{\mathrm{Bic}} P_{\mathrm{CO}_{2}} ) a_{\mathrm{HCO}_{3}^{-}}^{7/3}  - (K_{\mathrm{Bic}}^{2} P_{\mathrm{CO}_{2}}^{2}) a_{\mathrm{HCO}_{3}^{-}}^{1/3}  - 2^{4/3} K_{\mathrm{Bic}} K_{\mathrm{Fa}}^{2/3}  P_{\mathrm{CO}_{2}}^{7/3}  = 0 \end{aligned} \\
Anorthite (f,o,p,q,r) &  \!\begin{aligned} (2 K_{\mathrm{Car}}) a_{\mathrm{HCO}_{3}^{-}}^{4} + (K_{\mathrm{Wat}} + K_{\mathrm{Bic}} P_{\mathrm{CO}_{2}} ) a_{\mathrm{HCO}_{3}^{-}}^{3}  - (K_{\mathrm{Bic}}^{2} P_{\mathrm{CO}_{2}}^{2}) a_{\mathrm{HCO}_{3}^{-}}  - 2 K_{\mathrm{Bic}} K_{\mathrm{An}}^{2} P_{\mathrm{CO}_{2}}^{3}   = 0 \end{aligned} \\
Albite (g,o,p,q,r) &    \!\begin{aligned} (2 K_{\mathrm{Car}}) a_{\mathrm{HCO}_{3}^{-}}^{3} + (K_{\mathrm{Wat}} + K_{\mathrm{Bic}} P_{\mathrm{CO}_{2}} ) a_{\mathrm{HCO}_{3}^{-}}^{2}  - (2^{-2/3} K_{\mathrm{Bic}} K_{\mathrm{Alb}}^{1/3} P_{\mathrm{CO}_{2}}^{4/3} ) a_{\mathrm{HCO}_{3}^{-}}^{2/3}  - (K_{\mathrm{Bic}}^{2} P_{\mathrm{CO}_{2}}^{2})   = 0 \end{aligned} \\
K-feldspar (h,o,p,q,r) &   \!\begin{aligned} (2 K_{\mathrm{Car}}) a_{\mathrm{HCO}_{3}^{-}}^{3} + (K_{\mathrm{Wat}} + K_{\mathrm{Bic}} P_{\mathrm{CO}_{2}} ) a_{\mathrm{HCO}_{3}^{-}}^{2}  - (2^{-2/3} K_{\mathrm{Bic}} K_{\mathrm{Kfs}}^{1/3} P_{\mathrm{CO}_{2}}^{4/3} ) a_{\mathrm{HCO}_{3}^{-}}^{2/3}  - (K_{\mathrm{Bic}}^{2} P_{\mathrm{CO}_{2}}^{2})   = 0 \end{aligned} \\
Muscovite (i,o,p,q,r) &   \!\begin{aligned} (2 K_{\mathrm{Car}}) a_{\mathrm{HCO}_{3}^{-}}^{3} + (K_{\mathrm{Wat}} + K_{\mathrm{Bic}} P_{\mathrm{CO}_{2}} ) a_{\mathrm{HCO}_{3}^{-}}^{2}  - (K_{\mathrm{Bic}}^{2} + K_{\mathrm{Bic}} K_{\mathrm{Ms}} ) P_{\mathrm{CO}_{2}}^{2}  = 0 \end{aligned} \\
Phlogopite (j,o,p,q,r) &   \!\begin{aligned} (2 K_{\mathrm{Car}}) a_{\mathrm{HCO}_{3}^{-}}^{19/6} + (K_{\mathrm{Wat}} + K_{\mathrm{Bic}} P_{\mathrm{CO}_{2}} ) a_{\mathrm{HCO}_{3}^{-}}^{13/6}  - (K_{\mathrm{Bic}}^{2} P_{\mathrm{CO}_{2}}^{2}) a_{\mathrm{HCO}_{3}^{-}}^{1/6}  - 2^{2/3} 3^{1/2} K_{\mathrm{Bic}} K_{\mathrm{Phl}}^{1/2} P_{\mathrm{CO}_{2}}^{13/6} = 0 \end{aligned} \\
Annite (k,o,p,q,r) & \!\begin{aligned} (2 K_{\mathrm{Car}}) a_{\mathrm{HCO}_{3}^{-}}^{19/6} + (K_{\mathrm{Wat}} + K_{\mathrm{Bic}} P_{\mathrm{CO}_{2}} ) a_{\mathrm{HCO}_{3}^{-}}^{13/6}  - (K_{\mathrm{Bic}}^{2} P_{\mathrm{CO}_{2}}^{2}) a_{\mathrm{HCO}_{3}^{-}}^{1/6}  - 2^{2/3} 3^{1/2} K_{\mathrm{Bic}} K_{\mathrm{Ann}}^{1/2} P_{\mathrm{CO}_{2}}^{13/6} = 0 \end{aligned} \\
Anthophyllite (l,o,p,q,r) &   \!\begin{aligned} (2 K_{\mathrm{Car}}) a_{\mathrm{HCO}_{3}^{-}}^{3} + (K_{\mathrm{Wat}} + K_{\mathrm{Bic}} P_{\mathrm{CO}_{2}} ) a_{\mathrm{HCO}_{3}^{-}}^{2}  - (2^{-3/5} 7^{8/15} K_{\mathrm{Bic}} K_{\mathrm{anth}}^{7} P_{\mathrm{CO}_{2}}^{29/15} ) a_{\mathrm{HCO}_{3}^{-}}^{1/15} - (K_{\mathrm{Bic}}^{2} P_{\mathrm{CO}_{2}}^{2})   = 0 \end{aligned} \\
Grunerite (m,o,p,q,r) &    \!\begin{aligned} (2 K_{\mathrm{Car}}) a_{\mathrm{HCO}_{3}^{-}}^{3} + (K_{\mathrm{Wat}} + K_{\mathrm{Bic}} P_{\mathrm{CO}_{2}} ) a_{\mathrm{HCO}_{3}^{-}}^{2}  - (2^{-3/5} 7^{8/15} K_{\mathrm{Bic}} K_{\mathrm{Gru}}^{7} P_{\mathrm{CO}_{2}}^{29/15} ) a_{\mathrm{HCO}_{3}^{-}}^{1/15}   - (K_{\mathrm{Bic}}^{2} P_{\mathrm{CO}_{2}}^{2})   = 0 \end{aligned} \\
\enddata
\end{deluxetable*}
\end{rotatetable*}

\section{Equilibrium Constant of the CO2 Dissolution Reaction} \label{app:Henry}

Henry's law states that the amount of gas dissolved in the liquid ([CO$_2(aq)$] = $a_{\mathrm{CO}_2(aq)} \times 1 $~mol~dm$^{-3}$) is proportional to its partial pressure above the liquid, $P_{\mathrm{CO}_2}$. For the CO$_2$ dissolution reaction ((o) in Table~\ref{tab:reactions}), the proportionality constant is the equilibrium constant $K_{\mathrm{CO}_2}$ that itself depends on pressure and temperature,

\begin{equation} \label{eq:henry_th}
    a_{\mathrm{CO}_2(aq),\mathrm{thermo}} = K_{\mathrm{CO}_2} P_{\mathrm{CO}_2}. 
\end{equation}
We obtain the dimensionless $K_{\mathrm{CO}_2}$ as a function of $T$ and $P$ from the \texttt{CHNOSZ} thermodynamic database \citep{2019FrEaS...7..180D}.

\citet[Equation~8.14,][]{2010ppc..book.....P} provide an approximate dimensional Arrhenius-type fitting function $K_H$ at any temperature $T$ for Henry's law constant,

\begin{equation} \label{eq:KHenry}
     K_{H} (T) = K^{0}_{H} \exp{ [ - C_{H} (\frac{1}{T} - \frac{1}{T_{0}}) ] },
\end{equation} with empirical factors, $K^{0}_{H}$ = 1600~$\mathrm{\frac{mol~water}{mol~CO_2(aq)}}$ at a reference temperature $T_{0} = 298$~K, and $C_{H} = 2400$~K. The relation between $a_{\mathrm{CO}_2(aq)}$ and $P_{\mathrm{CO}_2}$ using $K_H$ is given by

\begin{equation} \label{eq:henry_ar}
    a_{\mathrm{CO}_2(aq),\mathrm{arrhen}} = \frac{u}{K_{H}} P_{\mathrm{CO}_2},
\end{equation} 
where $u$ = 55.5~$\mathrm{\frac{mol~water}{mol~CO_2(aq)}}$ (1~dm$^{3}$ of water contains 55.5 moles of water) is a conversion factor between the standard states of $K_{H}$ (1~$\mathrm{\frac{mol~CO_2(aq)}{mol~water}}$) and $K_{\mathrm{CO}_2}$ (1~$\mathrm{\frac{mol~CO_2(aq)}{dm^{3}~water}}$).

\begin{figure}[!ht]
  \centering
  \medskip
  \includegraphics[width=\linewidth]{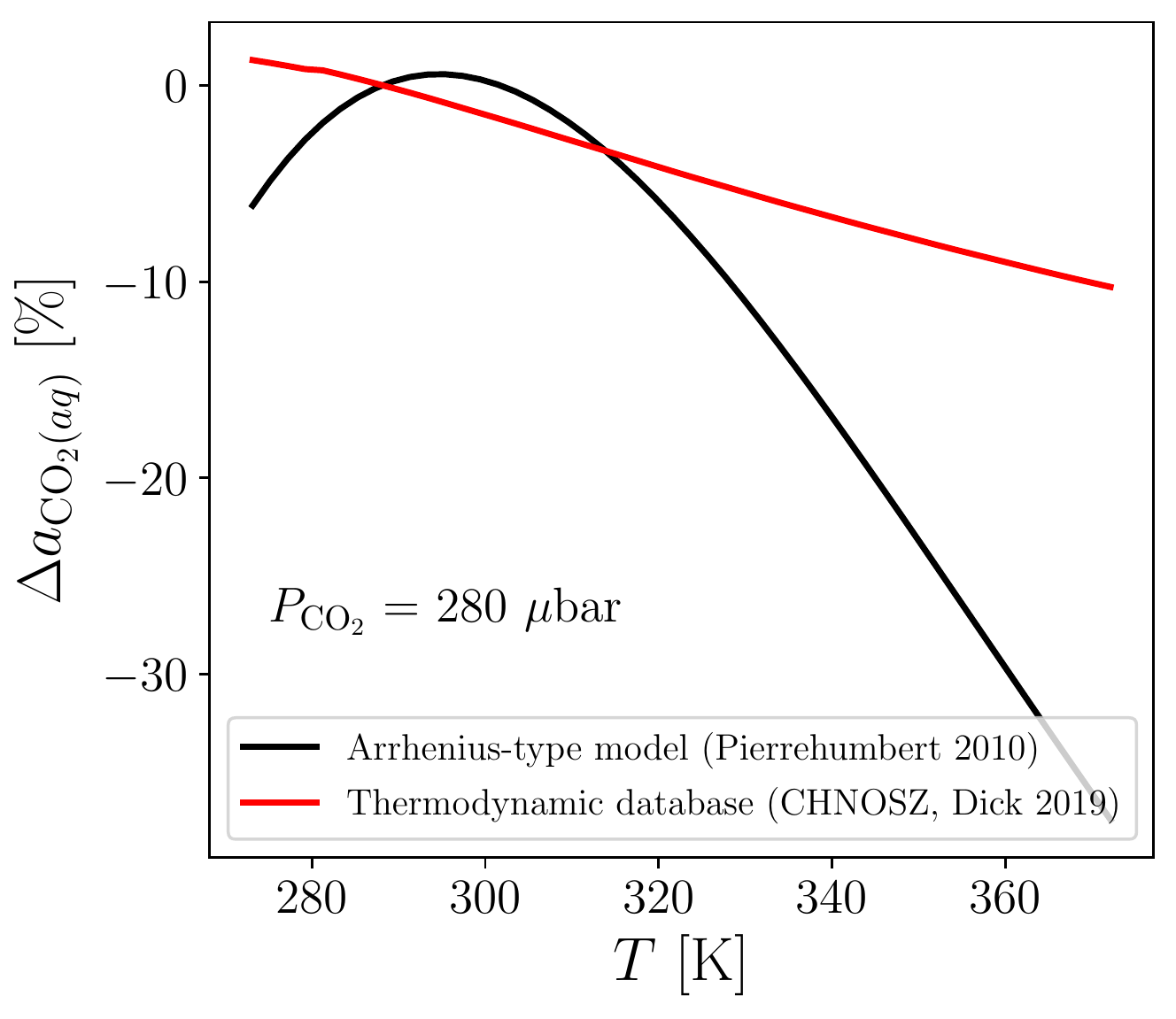}
  \caption{ Difference between model and experimental data for $a_{\mathrm{CO}_2(aq)}$ as a function of $T$. The experimental data is obtained from the compilation of \citet{diamond2003solubility}. The models are given by the Arrhenius-type equation \citep{2010ppc..book.....P} and the thermodynamic database  \citep{2019FrEaS...7..180D}. }
  \label{fig:henry_CO2}
\end{figure}

In Figure~\ref{fig:henry_CO2}, these two models (Equations~\ref{eq:henry_th} and \ref{eq:henry_ar}) are compared with the fit to experimental data on the solubility of CO$_2$ in pure water compiled by \citet{diamond2003solubility}. The Arrhenius-type model is within 6\% of that of the experimental data up to 330~K and deviates by up to 37\% at higher temperatures. The thermodynamic model performs better than the Arrhenius-type model at all temperatures except for 288$-$313~K and is within 10\% of the experimental data at 373~K. For this reason, we use the thermodynamic model to calculate the solubility of CO$_2$ in water instead of the Arrhenius-type model (Equation~\ref{eq:KHenry}). 


\section{Sensitivity of the Damk\"{o}hler Coefficient to Parameters}\label{app:Dw}

\begin{figure*}[!ht]
  \centering
  \medskip
  \includegraphics[width=\textwidth]{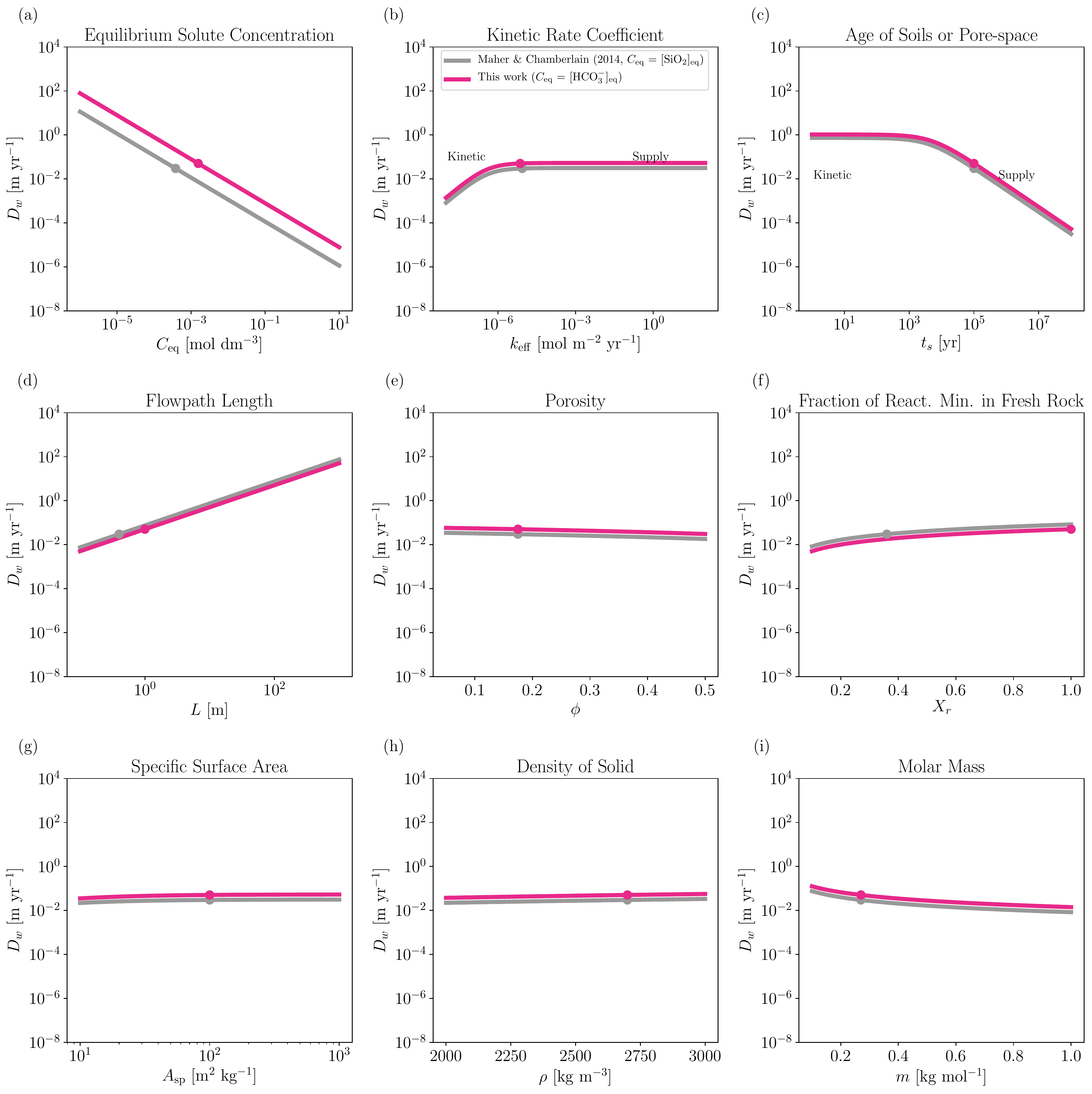}
  \caption{(a$-$i) Sensitivity of the Damk\"{o}hler coefficient to all nine parameters (Equation~\ref{eq:Dw}) and comparison between the Damk\"{o}hler coefficients of granite-like model from this work and the granite model from \citet{2014Sci...343.1502M}. Colored disks represent the default values of $D_w$ and respective parameters. The lines represent the extent of the variation in parameters. The key difference between the two models is in the species of interest: HCO$_3^-$ with [HCO$_3^-$]$_{\mathrm{eq}}$ = 1555~$\mu$mol~dm$^{-3}$ (this work) and SiO$_2$ with [SiO$_2$]$_{\mathrm{eq}}$ = 380~$\mu$mol~dm$^{-3}$ \citep{2014Sci...343.1502M}. Other differences include the  flowpath length and fraction of reactive minerals: $L$ = 1~m  and $X_r = 1$ (this work), and $L$ = 0.4~m and $X_r = 0.36$ \citep{2014Sci...343.1502M}. These two differences result in $D_w$ = 0.05~m~yr$^{-1}$ (this work) and $D_w$ = 0.03~m~yr$^{-1}$ \citep{2014Sci...343.1502M}. Kinetic and supply regimes of weathering are highlighted in (b) and (c). }
  \label{fig:Dw_MACH}
\end{figure*}

The Damk\"{o}hler coefficient $D_w$ depends on seven parameters and two computed quantities (Equation~\ref{eq:Dw}). The two computed quantities, equilibrium solute concentration $C_{\mathrm{eq}}$ (= [HCO$_3^-$]$_{\mathrm{eq}}$ in this study) and effective kinetic rate coefficient $k_{\mathrm{eff}}$, are treated as free parameters in Figure~\ref{fig:Dw_MACH}. All nine parameters are varied for a maximum possible range of their known values (Figure~\ref{fig:Dw_MACH}). $D_w$ is largely sensitive to four quantities, $C_{\mathrm{eq}}$, $k_{\mathrm{eff}}$, $t_{\mathrm{s}}$ (age of soils) and $L$ (flowpath length). The flowpath length is absorbed into the dimensionless pore-space parameter $\psi$ which is a control parameter for the models in the main text (Equation~\ref{eq:Dw}). Figure~\ref{fig:Dw_MACH}(b,c) highlights the interdependence of $k_{\mathrm{eff}}$ and $t_{\mathrm{s}}$. At low $k_{\mathrm{eff}}$ or low $t_{\mathrm{s}}$, $D_w$ is strongly sensitive to $k_{\mathrm{eff}}$ and insensitive to $t_{\mathrm{s}}$, implying the presence of `fast kinetic' regime. At high $k_{\mathrm{eff}}$ or high $t_{\mathrm{s}}$, $D_w$ is independent of $k_{\mathrm{eff}}$ and decreases strongly with $t_{\mathrm{s}}$, implying that weathering is in the `slow kinetic' or supply-limited regime due to insufficient supply of fresh rocks for weathering.  Figure~\ref{fig:Dw_MACH} also compares $D_w$ of our granite model to that of \citet{2014Sci...343.1502M}. These two models show similar trends between $D_w$ and respective parameters. The difference between the two models arise mainly from our assumption of endmember silicate minerals, instead of solid solutions, for the granite model, that results in a higher $C_{\mathrm{eq}}$ in our model by a factor four. 

\section{Climate Models}\label{app:climate}

A climate model provides a relation between the surface temperature $T$, the CO$_2$ partial pressure $P_{\mathrm{CO}_2}$, top-of-atmosphere stellar flux $S$ and planetary albedo $\alpha$. \citet{2019ApJ...875....7K} provide a fitting function to the climate model of \citet{2013ApJ...765..131K,2014ApJ...787L..29K} which is valid for $T$ in the range 150$-$350~K with $P_{\mathrm{CO}_2}$ in the range $10^{-5}-10$~bar at saturated H$_2$O and 1~bar N$_2$. The fitting function is given by

\begin{equation} \label{eq:clim1}
    F_{\mathrm{OLR}}(T,P_{\mathrm{CO}_2}) = I_0 + \mathbf{T \ B \ P^{t}},
\end{equation} 
\begin{equation} \label{eq:clim2}
    \mathbf{T} = [1 \; \xi \; \xi^2 \; \xi^3 \; \xi^4 \; \xi^5 \; \xi^6],
\end{equation}
\begin{equation} \label{eq:clim3}
    \mathbf{P} = [1 \; \chi \; \chi^2 \; \chi^3 \; \chi^4],
\end{equation} 
where \textbf{t} denotes the transpose of the vector, the outgoing longwave radiation $F_{\mathrm{OLR}}$ is a function of $T$ and $P_{\mathrm{CO}_2}$, $I_0 = -3.1$~W~m$^{-2}$ and $\xi = 0.01 \; (T - 250)$. For $P_{\mathrm{CO}_2} < 1$~bar,

\begin{equation} \label{eq:clim4}
   \chi = 0.2 \; \log_{10}{P_{\mathrm{CO}_2}},
\end{equation} 

\begin{gather} \label{eq:clim5}
\small 
\mathbf{B} =
    \begin{bmatrix} 
     87.8373 & -311.289 & -504.408 & -422.929 & -134.611 \\
     54.9102 & -677.741 & -1440.63 & -1467.04 & -543.371 \\
     24.7875 &  31.3614 & -364.617 & -747.352 & -395.401 \\
     75.8917 &  816.426 &  1565.03 &  1453.73 &  476.475 \\
     43.0076 &  339.957 &  996.723 &  1361.41 &  612.967 \\
    -31.4994 & -261.362 & -395.106 & -261.600 & -36.6589 \\
    -28.8846 & -174.942 & -378.436 & -445.878 & -178.948
    \end{bmatrix}.
\end{gather}
For $P_{\mathrm{CO}_2} > 1$~bar,

\begin{equation} \label{eq:clim6}
   \chi = \log_{10}{P_{\mathrm{CO}_2}},
\end{equation} 
\begin{gather} \label{eq:clim7}
\small
\mathbf{B} =
    \begin{bmatrix} 
     87.8373 & -52.1056 &  35.2800 & -1.64935 & -3.42858 \\
     54.9102 & -49.6404 & -93.8576 &  130.671 & -41.1725 \\
     24.7875 &  94.7348 & -252.996 &  171.685 & -34.7665 \\
     75.8917 & -180.679 &  385.989 & -344.020 &  101.455 \\
     43.0076 & -327.589 &  523.212 & -351.086 &  81.0478 \\
    -31.4994 &  235.321 & -462.453 &  346.483 & -90.0657 \\
    -28.8846 &  284.233 & -469.600 &  311.854 & -72.4874
    \end{bmatrix}.
\end{gather}

\begin{figure*}[!ht]
  \centering
  \medskip
  \includegraphics[width=\textwidth]{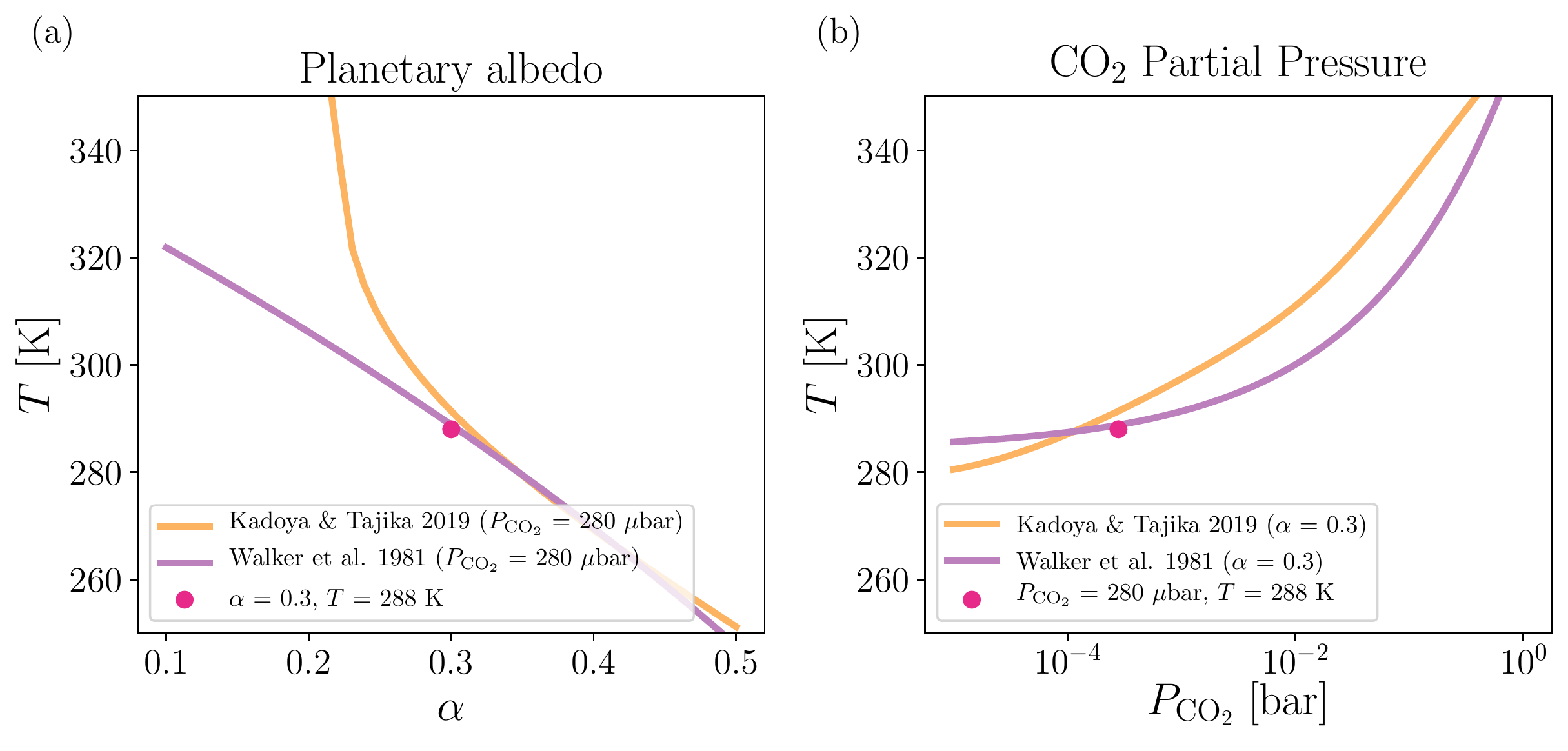}
  \caption{Comparison between the climate models of \citet{1981JGR....86.9776W} and \citet{2019ApJ...875....7K}. (a) The effect of planetary albedo on temperature. (b) The effect of the partial pressure of CO$_2$ on temperature. }
  \label{fig:climate_WHAK_KATA}
\end{figure*}

Equations~(\ref{eq:clim1}$-$\ref{eq:clim7}) are solved by balancing the energy fluxes of the globally-averaged absorbed instellation $S_{\mathrm{avg}}$ and $F_{\mathrm{OLR}}$, where

\begin{equation} \label{eq:clim8}
    F_{\mathrm{OLR}} = S_{\mathrm{avg}},
\end{equation}
and 

\begin{equation} \label{eq:clim9}
    S_{\mathrm{avg}} = \frac{(1 - \alpha)}{4} S,
\end{equation} 
with the geometric factor 4 comes from ratio of the planet surface area to the area of its cross-section. For present-day albedo ($\alpha = 0.3$) and present-day solar flux ($S = 1360$~W~m$^{-2}$), this fit results in $T$ between 280~K and 350~K and $P_{\mathrm{CO}_2}$ between $10^{-5}$~bar and 0.5~bar. 

Another climate model used frequently in carbon cycle studies \citep[e.g.,][]{2015ApJ...812...36F} is the one from \citet{1981JGR....86.9776W}. The relation between $T$ and $P_{\mathrm{CO}_2}$ is given by

\begin{equation} \label{eq:clim10}
    T = T^* + 2(T_e - T_e^*) + 4.6 \left( \frac{P_{\mathrm{CO}_2}}{P^*_{\mathrm{CO}_2}} \right)^{0.346}
\end{equation}
where $T_e$ is the effective temperature given by $T_e = (S_{\mathrm{avg}}/\sigma_{\mathrm{SB}})^{1/4}$ with $\sigma_{\mathrm{SB}} =  5.67\times10^{-8}$~W~m$^{-2}$~K$^{-4}$ as the Stefan-Boltmann constant and the present-day values of temperature, effective temperature and CO$_2$ partial pressure are assumed to be $T^*$ = 285~K, $T_e^*$ = 254~K and $P^*_{\mathrm{CO}_2} = 330 \times 10^{-6}$~bar, respectively \citep{1984Icar...57..335K}.

Figure~\ref{fig:climate_WHAK_KATA} shows the comparison between the models of \citet{2019ApJ...875....7K} and \citet{1981JGR....86.9776W}. Both models show almost the same temperatures for $\alpha$ between 0.3 and 0.5 at $P_{\mathrm{CO}_2} = 280$~$\mu$bar. However, for $\alpha < 0.25$, the \citet{2019ApJ...875....7K} model shows a steep temperature rise with decreasing $\alpha$. As a function of $P_{\mathrm{CO}_2}$ at $\alpha = 0.3$, both models exhibit temperatures within 5\% of each other.


\pagebreak

\bibliography{weathering}{}
\bibliographystyle{aasjournal}



\end{document}